\documentclass[dutch,english]{book}
\usepackage{babel}                                                             
\usepackage{wick}                                                              
\usepackage{fancyheadings}
\usepackage{amsthm}
\usepackage{amssymb}
\usepackage{epsfig}
\usepackage{psfrag}
\usepackage{makeidx}
\renewcommand{\chaptermark}[1]{\markboth{\chaptername\ \thechapter.\ #1}{}}

\newcommand{\be}{\begin{equation}}
\newcommand{\ee}{\end{equation}}
\newcommand{\bq}{\begin{eqnarray}}
\newcommand{\eq}{\end{eqnarray}}
\newcommand{\bqs}{\begin{eqnarray*}}
\newcommand{\eqs}{\end{eqnarray*}}
\newcommand{\ap}{\alpha'}
\newcommand{\Hil}{\mathcal{H}}
\newcommand{\vac}{|0\rangle}
\newcommand{\leftvac}{\langle 0 |}
\newcommand{\state}[1]{| #1 \rangle}
\newcommand{\etats}[1]{\langle #1 |}
\newcommand{\p}{\partial}
\newcommand{\parder}[1]{\frac{\partial}{\partial #1}}
\newcommand{\parders}[2]{\frac{\partial^{#2}}{\partial #1^{#2}}}
\newcommand{\corr}[1]{\left\langle #1 \right\rangle}
\newcommand{\Corr}[1]{\langle\langle #1 \rangle\rangle}

\newcommand{\BigCorr}[1]{\Big\langle\Big\langle #1 \Big\rangle\Big\rangle}
\newcommand{\ep}{\epsilon}
\newcommand{\bpz}{\mbox{bpz}}
\newcommand{\IZ}{\mathbb{Z}}
\newcommand{\IR}{\mathbb{R}}
\newcommand{\IC}{\mathbb{C}}
\newcommand{\oscNO}[1]{\begin{array}{c} \circ\vspace{-1.5ex} \\ \circ \end{array} #1 
\begin{array}{c} \circ\vspace{-1.5ex}\\\circ \end{array}}
\newcommand{\pO}{\mathcal{O}}

\newcommand{\crea}{a^\dagger}
\newcommand{\cre}[1]{a^{\dagger #1}}
\newcommand{\erf}{\mbox{erf}}
\newcommand{\cft}{conformal field theory}
\newcommand{\co}{conformal}
\newcommand{\ct}{conformal transformation}
\newcommand{\cts}{conformal transformations}
\newcommand{\gn}{ghost number}
\newcommand{\refpj}[1]{~(\ref{#1})}
\newcommand{\sfi}{string field}
\newcommand{\sft}{string field theory}
\newcommand{\ssft}{super string field theory}
\newcommand{\st}{string theory}

\newcommand{\trs}{transformations}
\newcommand{\STRUT}{\rule{0in}{4ex}}
\newcommand{\wh}{\widehat}
\newcommand{\hp}{{\wh\Phi}}
\newcommand{\hq}{{\wh Q}}
\newcommand{\he}{{\wh\eta_0}}
\newcommand{\ha}{{\wh{A}}}

\newlength{\mylength}
\newcommand{\kader}[3]
{\ \\[2ex]
  \setlength{\fboxsep}{1ex}
  \setlength{\mylength}{\linewidth}
  \addtolength{\mylength}{-2\fboxsep}
  \addtolength{\mylength}{-2\fboxrule}
  \framebox[\linewidth]{
    \begin{minipage}{\mylength}
      \vspace{1ex}
      \begin{#1}\label{#2}
        {\rm #3}
      \end{#1}
    \end{minipage}
  }\ \\[2ex]}
\newcommand{\kaderzonderrm}[3]
{\ \\[2ex]
  \setlength{\fboxsep}{1ex}
  \setlength{\mylength}{\linewidth}
  \addtolength{\mylength}{-2\fboxsep}
  \addtolength{\mylength}{-2\fboxrule}
  \framebox[\linewidth]{
    \begin{minipage}{\mylength}
      \vspace{1ex}
      \begin{#1}\label{#2}
        #3
      \end{#1}
    \end{minipage}
  }\ \\[2ex]}

\begin{document}
\thispagestyle{empty}
\begin{figure}
  \epsfig{file=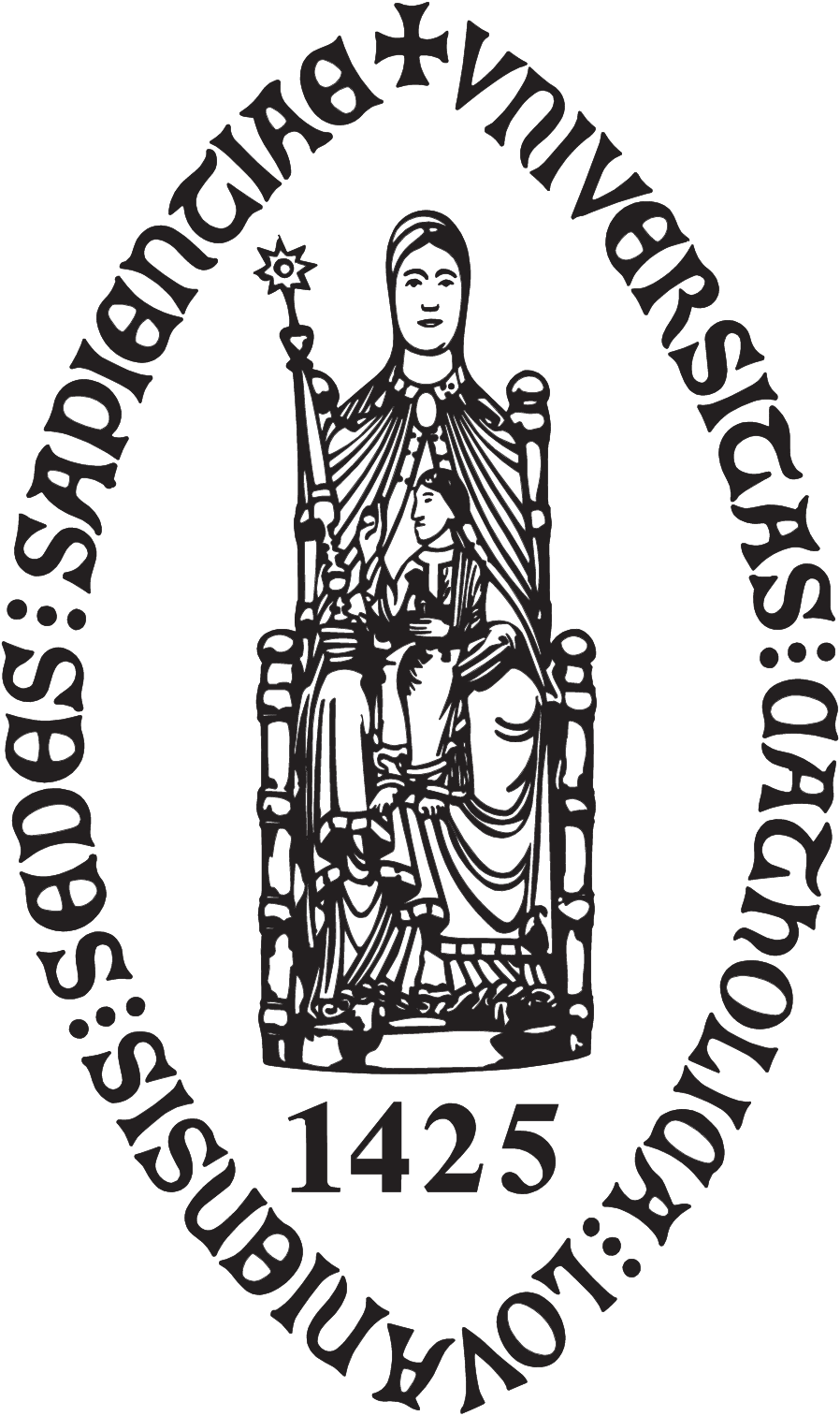,height=2cm,angle=0,trim=-4130 0 0 0}
\end{figure}
\begin{center}
\begin{minipage}{12.5cm}
\vspace{-3.8cm}
\begin{flushleft}
{\large Katholieke Universiteit Leuven\\
 Faculteit der Wetenschappen\\
 Instituut voor Theoretische Fysica\\}
\end{flushleft}
\end{minipage}
\vfill \vfill \vfill \vfill
\begin{minipage}{12.5cm}
\vspace{-1cm}
\begin{center}
{\huge\bf Tachyon Condensation:}
\vskip 5mm
{\huge\bf Calculations in String Field Theory}
\end{center}
\end{minipage}
\vfill \vfill \vfill
\begin{minipage}{12.5cm}
\begin{center}
{\large\bf Pieter-Jan De Smet\\ }
\end{center}
\end{minipage}
\vfill \vfill \vfill \vfill
\begin{minipage}{12.5cm}
\begin{minipage}[t]{6cm}
\begin{flushleft}
Promotor: Prof. Dr. W. Troost
\end{flushleft}
\end{minipage}
\hfill
\begin{minipage}[t]{6cm}
\begin{flushleft}
       Proefschrift ingediend voor\\
       het behalen van de graad van\\
       Doctor in de Wetenschappen\\
\end{flushleft}
\end{minipage}
\end{minipage}
\vfill
\begin{minipage}{12.5cm}
\begin{center}
{\large 2001}
\end{center}
\end{minipage}
\end{center}
\newpage
\thispagestyle{empty}
\selectlanguage{dutch}
\begin{center}
\vspace*{15,5cm}
\begin{minipage}{12,5cm}
\begin{flushleft}
De auteur is Aspirant van het
\\ Fonds voor Wetenschappelijk Onderzoek -- Vlaanderen.
\end{flushleft}
\end{minipage}
\end{center}
\newpage
\selectlanguage{english}


\thispagestyle{empty}
\frontmatter 
\tableofcontents
\index{eta@$\eta$|see{ghosts, $\eta\xi$}}
\index{xi@$\xi$|see{ghosts, $\eta\xi$}}
\index{b@$b$|see{ghosts, $bc$}}
\index{c@$c$|see{ghosts, $bc$}}
\index{beta@$\beta$|see{ghosts, $\beta\gamma$}}
\index{gamma@$\gamma$|see{ghosts, $\beta\gamma$}}
\index{phi@$\phi$|see{ghosts, $\phi$}}
\index{linear dilaton|see{ghosts, $\phi$}}
\index{interaction vertex|see{Witten's vertex}}
\index{open bosonic string field theory|see{string field theory, Witten's sft}}
\index{open super string field theory|see{string field theory, Berkovits' sft}}
\index{Wess-Zumino-Witten|see{WZW-action}}
\index{CFT|see{conformal field theory}}

\parskip 5pt plus 1pt minus 1pt

\mainmatter 
\chapter{Introduction}\label{c:intro}
During the past four centuries, physicists have accumulated mounting evidence
that all interactions in nature can be reduced to
combinations of four fundamental forces:
the gravitational, electromagnetic, strong and weak force.
The last three are compatible with quantummechanics. Indeed, in the
\emph{Standard Model} one can calculate the electromagnetic, weak and strong
interactions between elementary particles. This model is mathematically rather
elegant. Indeed, these three forces all have the same mathematical
structure; they are described by gauge theories in which the force 
carrying particles are spin~1 gauge bosons. The Standard Model has 
been tested thoroughly
in particle colliders\footnote{
Some small cracks are appearing in the Standard Model though. 
Neutrinos are not massless but have an extremely small
mass~\cite{neutrinos}. The anomalous magnetic moment of the muon seems 
to disagree with
theoretical calculations~\cite{moment}.}. 

However, our theory of gravitation -- general relativity -- is not compatible
with quantummechanics. At the moment, nobody can calculate the gravitational
interaction between two elementary particles with all quantummechanical effects
included. For this one needs
\emph{quantumgravity}. Its construction is the main research topic in
theoretical high energy physics. In all probability
the force carrying particle of gravity is a particle of spin~2. This
particle has been given the name ``graviton''. It has not been detected so far.

The last 15 years, most efforts to construct quantumgravity are concentrated in
string theory. 
However, string theory is a theory which is still under construction. In
particular, string theory is only defined \emph{perturbatively}. This means that
string theory is a formalism for calculating scattering amplitudes as a
perturbation series in a small coupling constant. If we compare this with point
particle theories, then perturbative string theory corresponds with a set
Feynman rules. Let us now give a brief review of these elements of string theory which are
necessary to understand the essence of this thesis. For a thorough and
pedagogical discourse we refer to~\cite{Polchinski}.
\section{String theory}
The basic assumption of string theory is that all elementary particles are
different vibration patterns of a small elastic string. Because of this, string theory
generates a unification between the different kinds of elementary particles.
Due to the extended character of a string, string theory also succeeds 
in eliminating infinities which arise when point particles approach one another 
too closely. Some of the vibration patterns generate particles with spin~1.
These massless particles behave exactly like the gauge bosons we met in the
previous section. Another vibration patterns generates a particle with spin~2.
This particle is a graviton! Hence, string theory not only predicts the kind of
particles we actually see in nature, but gravity seems to be an essential part
of the theory as well.

There are several different string
theories.
Firstly, there is one bosonic string theory, in which all particles are bosons. 
If bosonic strings move in a flat background, internal consistency of the theory
demands that the background be 26-dimensional. 
One of
the vibration patterns of the bosonic string in a flat background
is a tachyon. This is a particle
with negative mass squared. The existence of such a particle suggests that the
bosonic string theory is unstable. Because of this and the absence of fermions,
this theory is not considered to be a good theory of quantum gravity.
Secondly, there are string theories having space-time supersymmetry -- super
string theories in short. In these theories there are bosons as well
as fermions but no tachyons. They are considered to be possible candidates for
providing a correct theory of quantum gravity. Internal consistency 
demands that a flat background be 10-dimensional. 
Thirdly, there are some non-supersymmetric heterotic string theories. Most of
these have tachyons. They play no role in the rest of this thesis.
\section{Problems in string theory}
As mentioned in the previous section, string theory predicts that space-time be
10-dimensional. 
Therefore, 6 of these
dimensions must be curled up so tightly that they have escaped detection so far.
This is the main problem of string theory. Is the 10-dimensional Lorentz
invariance broken down to four dimensions? And if so, how are these 6 dimensions
compactified? Posed in a more physical way, 
\begin{center}\label{Question}
$\mathcal{W}${\it hat is the correct vacuum of string theory?}
\end{center}
Here we run up against the inherent limitations of perturbative string theory.
The equations one has to solve to find the
correct vacuum of string theory, are only known perturbatively. They do not
contain enough information to begin the quest. Because the correct vacuum of
string theory is not known, it is impossible to make contact with experiments.
The knowledge of the correct vacuum is a conditio 
sine qua non to make concrete physical
predictions. At this moment, string theory can only make generic predictions,
i.e.~predictions that are valid for all backgrounds. 

In spite of the fact that
string theory is only a  perturbative theory, people have succeeded to obtain
\emph{non-perturbative} results by making use of the huge symmetry structure of
string theory. In particular one can argue that specific superstring theories
are dual to one another. This means that calculations in one 
particular super
string theory at large coupling constant can be translated to different
calculations in another superstring theory at small coupling constant. A key
role in these dualities is played by $D$-branes (1995). 

\begin{figure}[ht]
\begin{center}
\begin{psfrags}
\epsfbox{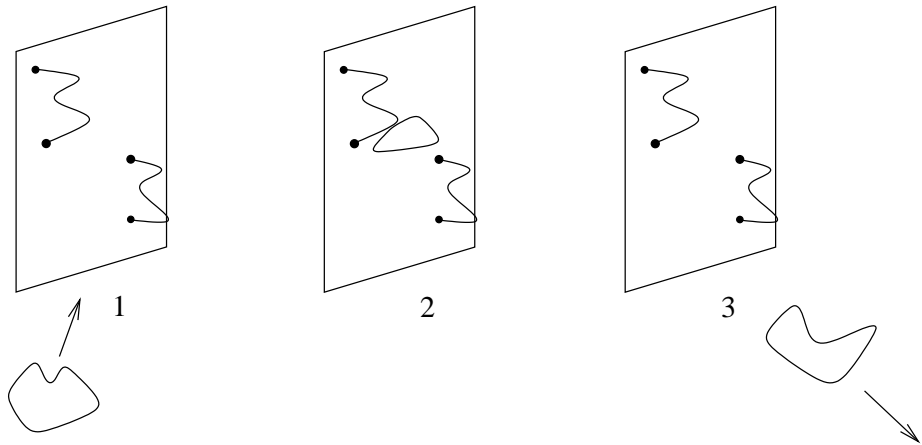}
\end{psfrags} 
\caption{(1) A closed string approaches a $D$-brane. (2) The closed string
interacts with the $D$-brane by way of the (virtual) 
open strings attached to it. (3) After
the collision, the closed string moves away. 
}
\label{fig:Dbrane}
\end{center}
\end{figure}

$D$-branes
were introduced in string theory as hyperplanes on which open strings can end.
The mass of a $D$-brane is inversely proportional to the coupling constant, hence
these are non-perturbative states. Nevertheless, they have a description within
perturbative string theory: the dynamics of the $D$-branes is described by the
open strings ending on them, see figure~\ref{fig:Dbrane}.   

There are a lot of indications that the aforementioned dualities are correct.
Concrete proofs, however, are still lacking. It seems plausible that one needs a
non-perturbative definition of string theory for that. Shortly after the
discovery of perturbative string theory one hoped that string field theories
would provide such a non-perturbative definition. Witten's open bosonic
\sft~\cite{WittenSFT}
seems to do a good job as a non-perturbative definition of open bosonic string
theory. Zwiebach constructed a closed bosonic string field theory~\cite{closed}. However, this
field theory is technically involved and is not suited for concrete
calculations. A lot of problems turned up in the naive attempts to construct
superstring field theories~\cite{Wendt,9108021}. Therefore, most physicists believed that string 
field theory was a too conservative attempt to define strings
non-perturbatively. String field theories were thought not to be able to describe
the rich non-perturbative results found in string theory. Therefore, they
were abandoned as a promising line of research for a decade. After the
discovery of the dualities, it seems that a still mysterious 11-dimensional
theory -- dubbed $\mathcal{M}$-theory could be 
the correct framework to discuss string
theory non-perturbatively. The last couple of years, however, string field
theories have come to the surface again because they seem particularly suitable
to verify Sen's conjectures.
\section{Sen's conjectures}
The origin of these conjectures lies in $D$-branes, as was the case with a lot
of recent discoveries in string theory.
The discovery of $D$-branes gives another point of view on the tachyon of the
open bosonic string theory. Sen suggested in 1999 that this
theory should be looked at as a theory which describes a 
$D25$-brane~\cite{Sen:univ}. Indeed, after the discovery of $D$-branes, it was
realized that open strings have to end on a $D$-brane. The open strings of open
bosonic string theory move across the entire space. Hence, this $D$-brane fills
the entire space -- a $D25$-brane. The space-time of bosonic string theory 
is not empty, it is filled completely by the $D25$-brane!  
Consequently, the existence of the
open string tachyon is a perturbative 
consequence of the instability of the $D25$ 
brane~\cite{9902105}. 
Sen argued that the condensation of the tachyon should correspond to the decay of
the $D25$-brane. In particular, he made three conjectures.
\begin{figure}[ht]
\begin{center}
\begin{psfrags}
\psfrag{A}[][]{potential}
\psfrag{B}[][]{$A$}
\psfrag{C}[][]{$\Psi_0$}
\psfrag{D}[][]{B}
\psfrag{E}[][]{mass of $D25$ brane}
\psfrag{F}[]{$\Psi$}
\epsfbox{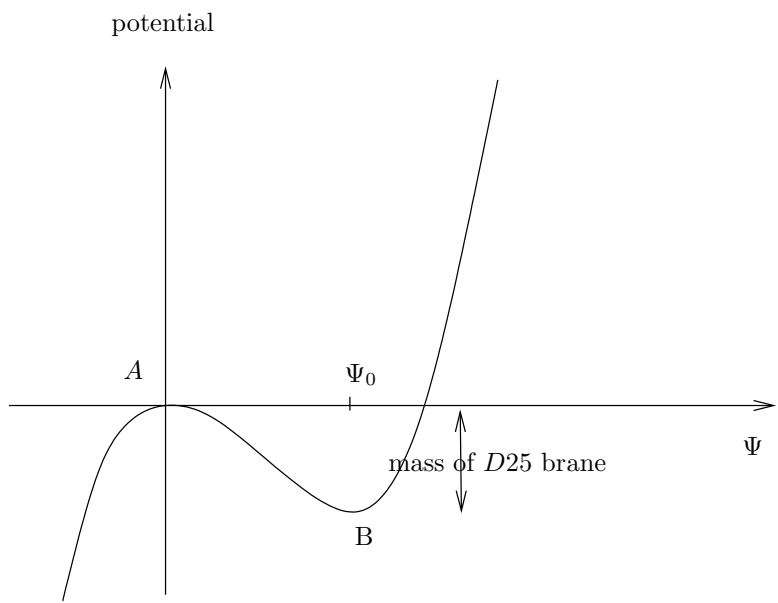}
\end{psfrags} 
\caption{The tachyon potential in the open bosonic string theory. In point~$A$,
the string field $\Psi$ is zero. At this point, the unstable $D25$-brane is
still present. Point~$B$ is the closed string vacuum. According to the
conjecture of Sen, the $D25$-brane decays to this vacuum. Therefore, the
difference in potential energy between these two extrema equals the mass of the
$D25$-brane.}
\label{fig:SenbosE}
\end{center}
\end{figure}
\begin{itemize}
\item[(1)]\label{Senc1}
The difference in the potential between the unstable vacuum and the
perturbatively stable vacuum should be the mass of the $D25$-brane, see
figure~\ref{fig:SenbosE}.
\item[(2)]\label{Senc2} Lower-dimensional $D$-branes should be realized as soliton
configurations of the tachyon and other string fields.
\end{itemize}
After condensation there is no brane left on which open strings could end, hence
\begin{itemize}
\item[(3)]\label{Senc3}
The perturbatively stable vacuum should correspond to the closed string vacuum.
In particular, there should be no physical open string excitations around this
vacuum.
\end{itemize}
There are also unstable branes in superstring theory, e.g.~the $D9$-brane in
type IIA. Analogous conjectures hold in this case~\cite{9805170}, 
see figure~\ref{fig:SensupE}.
\begin{figure}[ht]
\begin{center}
\begin{psfrags}
\psfrag{A}[][]{potential}
\psfrag{B}[][]{$D9$}
\psfrag{C}[][]{$\Psi_0$}
\psfrag{D}[][]{no $D9$}
\psfrag{E}[][]{mass $D9$ brane}
\psfrag{F}[]{$\Psi$}
\epsfbox{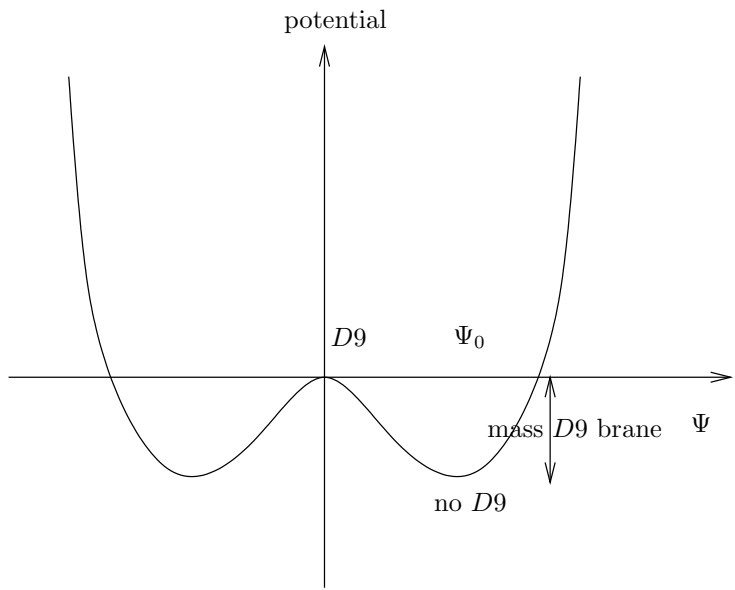}
\end{psfrags} 
\caption{The tachyon potential on the unstable $D9$-brane.}
\label{fig:SensupE}
\end{center}
\end{figure}
Open string field theories are appropriate to verify Sen's
conjectures. Indeed, one has to calculate nothing but the potential to verify the
first conjecture. This kind of calculation fits very well in the framework of a
field theory.

\begin{figure}[ht]
\begin{center}
\begin{psfrags}
\psfrag{a}[][]{(a)}
\psfrag{b}[][]{(b)}
\epsfbox{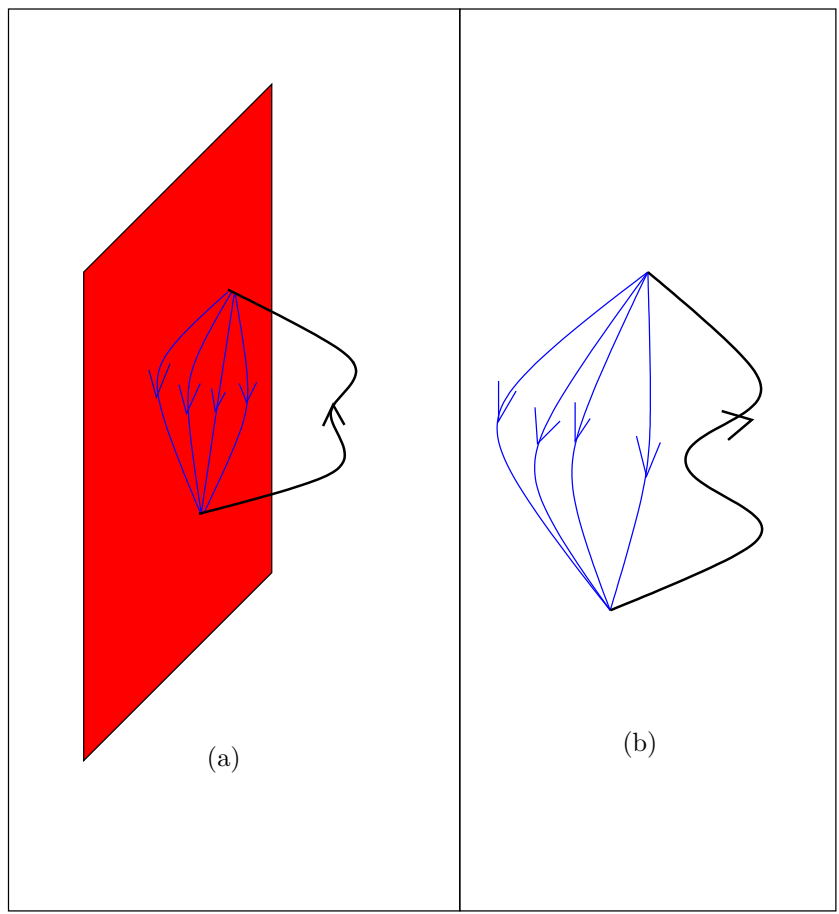}
\end{psfrags} 
\caption{(a) An open string having its endpoints on a non-BPS $D$-brane before
tachyon condensation. The endpoints of the open string are charged under a gauge
field. (b) This configuration after tachyon condensation. The open strings
together with the flux tube form a closed string. }
\label{fig:fluxE}
\end{center}
\end{figure}

The significance of this research is the following. First of all, this kind of
calculations shows that string field theories are able to provide 
non-perturbative information of open string theory. Furthermore, it is
fascinating to speculate that open string field theory gives a \emph{complete}
non-perturbative description of string theory, i.e.~of the open string theory as
well as of the closed string theory. One gets to this speculation in the following
way. According to Sen's third conjecture, the stable vacuum can be identified
with the closed string vacuum. Closed strings will arise as non-perturbative
fluctuations of the stable vacuum. Intuitively, one can see how closed strings 
arise after condensation of a
$D$-brane, see figure~\ref{fig:fluxE},~\cite{0002223,0010240}.

Therefore if $S(\Psi)$ is the open string field
theory action and $\Psi_0$ the stable vacuum (see figures~\ref{fig:SenbosE}
and~\ref{fig:SensupE}), $S(\Psi+\Psi_0)$ is hopefully the action of a closed
\sft.   
If this turns out to be the case, one can study the fate of the closed string
tachyon in the bosonic case. If this turns out to be the case in the 
superstring,
it should be possible to determine the correct vacuum
of superstring theory, see page~\pageref{Question}, and make phenomenological
predictions.
\clearpage
\section{Summary of this thesis}
\subsection*{Chapter 2: Open Bosonic String Field Theory}
In chapter~2 we will use Witten's open \sft\ to study the tachyon of the
open bosonic string. This is the context
in which Sen's conjecture has been most intensely studied. Although our own work
involves the study of the tachyon potential in \emph{super symmetric} theories,
 we
include the bosonic case since it provides a simple setting to discuss various
aspects of the calculation which can then be generalized to the technically more
involved case of the superstring.

We introduce some notation and
technicalities in the first few sections of this chapter. In
section~\ref{s:gluing} we argue that interactions between strings are 
described by gluing their world sheets. This is to be compared with the
interaction of point particles, where world lines are glued together by
interactions in Feynman diagrams. 
We need this
gluing of world sheets to define the central ingredient of Witten's 
action -- the star product -- in section~\ref{s:star}. 

After these 2 sections we have set the stage to treat Witten's action. This
action formally looks like a Chern-Simons action, therefore we give a brief
review of the usual Chern-Simons action in section~\ref{s:CS}. 
Thereafter in section~\ref{s:SFT}, the central part of this chapter, 
we discuss Witten's action. We will elaborate a little bit on this action 
in section~\ref{s:Siegel} where we say how to fix the gauge invariance.

We will use Witten's action to discuss the instability of the open
bosonic \st\ in section~\ref{s:tachpot}. We conclude this chapter with three
sections. In section~\ref{s:Neumann} we give a concrete realization of Witten's
interaction vertex. In
section~\ref{s:proofs} we provide the proofs that have been left over at earlier
stages. In the final section we discuss some other calculations that have been
done in \sft\ in recent months and speculate about further directions research
might take.

\subsection*{Chapter~3: Berkovits' Superstring Field Theory}
In chapter~3 we study the open superstring tachyon in Berkovits' super
string field theory. Therefore this chapter naturally splits into two parts.
\begin{description}
\item[1. Definition of Berkovits' action]
Berkovits' field theory formally looks like a Wess-Zumino-Witten theory.
Therefore, in section~\ref{s:WZW} we give a review of this WZW-action. In
section~\ref{s:BPS} we construct Berkovits' action for a BPS $D9$-brane on this
WZW analog.  
In section~\ref{s:nonBPS} we discuss Berkovits' action for a non-BPS $D9$-brane.
\end{description}
After having set the stage, we can calculate the tachyon potential in this
theory. This part is to a large extent original.  
\begin{description}
\item[2. Study of the tachyon potential] 
In section~\ref{s:level4} we
calculate the tachyon potential up to 
level\footnote{We will explain this notation 
in section~\ref{s:level4}.}~$(2,4)$. We find that at this
approximation the results agree with Sen's prediction up to $89\%$. We want 
to do a higher level calculation to get a more accurate correspondence. However,
the procedure followed in section~\ref{s:level4} turns out to be too tedious to
use to perform higher level calculations. Hence, in section~\ref{s:cons} we
develop a more efficient method. This method uses conservations laws and was
given in the literature for the bosonic string. We extend these laws to the
superstring and use them to do a level~$(4,8)$ calculation in
section~\ref{s:high}. In this approximation we find that the
results agree with Sen's prediction up to $94.4\%$.
\end{description}
In the final section we discuss some other calculations that have been performed
in Berkovits' field theory and speculate about directions research might
take. 
\subsection*{Chapter~4: A Toy Model}
If we want to learn for example
how closed strings arise in the true vacuum of open \sft, it seems plausible
that we need a closed expression for the true vacuum. However this seems to be
very complicated for two reasons. The first reason is that there are an infinite
number of fields acquiring expectation values in the minimum and the second is
the complicated form of the interaction in open \sft. In this 
chapter~4 we try to
learn something about the exact form of the minimum by looking for the
exact form of the minimum in a ``baby version'' of 
\sft. In the baby version of
\sft\ we forget about the ghosts and instead of taking the infinite set of
oscillators $\alpha^\mu_n$ where $\mu: 1,\ldots,26$ and $n: 1,\ldots,+\infty$,
with $[\alpha^\mu_m,\alpha^\nu_n] = m\ \eta^{\mu\nu}\delta_{m+n}$,
we take only one operator $a$ with $[a,\crea]$ = 1. 
\paragraph{}
In section~\ref{TM:action} we will give the action of the toy model. 
In its most general form, the model depends  on some parameters that 
enter in the definition of a  star product and are the analog of 
the Neumann coefficients in bosonic \sft. These parameters
are further constrained if we insist that the toy model star product
satisfies some of the properties that are present in the full \sft.
More specifically, the \sft\ star product satisfies the following properties:
\begin{itemize}
\item
The three-string interaction term is cyclically symmetric, see
equation\refpj{W:prop}.
\item
The star product is associative, see property~\ref{starass}.
\item
Operators of the form $a - a^\dagger$ act as derivations of the star-algebra,
see page~\pageref{der:alpha}.  
\end{itemize}
We impose cyclicity of the interaction term in our toy 
model in
section~\ref{TM:cyc}. We deduce the equations of 
motion in section~\ref{TM:eom}. 
In section~\ref{TM:eom} we define 
the star product for the toy model. 
We discuss the restrictions following from imposing associativity of
the star product in section~\ref{TM:ass}. It turns out
that we are left with 3
different possibilities, hereafter called case I, II and
III. As is the case for the bosonic \sft\ we can also look if there is a
derivation $D = a - a^\dagger$ of the star-algebra. This further restricts the
cases I, II and III to case Id, IId and again Id respectively. This is 
explained
in section~\ref{TM:der}, where we also discuss the existence of an identity
of the star-algebra.

After having set the stage we can start looking for exact solutions. In 
section~\ref{TM:CaseI} we give an exact form for the minimum in case I, in 
section~\ref{TM:othersols} we mention the other exact solutions we have found.
In section~\ref{TM:CaseIId}, we discuss the case IId which perhaps bears
the most resemblance to the full string field theory problem.
In this case, it is possible  to recast the equation of motion in
the form of an ordinary second order nonlinear differential equation.
This equation is not of the Painlev\'{e} type and we have not been able to
find an exact solution. 
Here too, it is possible to get very accurate information about the
stable vacuum using the level truncation method.
We conclude in
section~\ref{TM:end} with some suggestions for further research. 
\section{Conclusions}
The results presented in this thesis have led to more confidence in string field
theories as non-perturbative descriptions of string theory. In the bosonic case,
it is now verified that the $D25$-brane indeed decays to the closed string
vacuum. In the super case, Berkovits' field theory seems to be
singled out as the correct version of superstring field theory. If we can learn
something about closed strings by studying open string field theories, we are 
in the position to determine to correct vacuum of closed string theory. As this
is a very important problem, string field theory will remain an important
research topic in the near future.

\chapter{Open Bosonic String Field Theory}
In this chapter we will use Witten's open \sft\ to study the tachyon of the
open bosonic string. This is the context
in which Sen's conjecture has been most intensely studied. Although our own work
involves the study of the tachyon potential in \emph{supersymmetric} theories, we
include the bosonic case since it provides a simple setting to discuss various
aspects of the calculation which can then be generalized to the technically more
involved case of the superstring.

Witten's \sft\ is thought
to be the correct framework for discussing the field theory of open bosonic
strings. String field theory has the same structure as 
all other field theories, as for example
ordinary $\varphi^4$-theory, Maxwell theory or Yang-Mills theories. The only
difference is the following. Usually when studying field theories, one studies
theories that contain only a \emph{finite} number of space-time fields. The
\sft\ of Witten on the other hand contains an \emph{infinite} number of
space-time fields. Indeed, one has a space-time field corresponding to every
possible oscillation of the open bosonic string. Hence, Witten's action has an
infinite number of kinetic terms and an infinite number of interactions between
all these space-time fields and thus looks rather complicated. Using \cft\ however, Witten's \sft\ can be given 
a very concise formulation.

Before we can give this concise form we need to introduce some notation and
technicalities in the first few sections of this chapter. In
section~\ref{s:gluing} we will explain the gluing of world sheets. We need this
gluing of world sheets to define the central ingredient of Witten's 
action -- the star product -- in section~\ref{s:star}. We
give lots of examples in these 2 sections. We hope this will help the reader to
digest the definitions.

After these 2 sections we have set the stage to give Witten's action. This
action formally looks like a Chern-Simons action, therefore we give a brief
review of this Chern-Simons action in section~\ref{s:CS}. 
Thereafter in section~\ref{s:SFT}, the central part of this chapter, 
we discuss Witten's action. We will elaborate a little bit on this action 
in section~\ref{s:Siegel} where we say how to fix the gauge invariance.

We will use Witten's action to discuss the instability of the open
bosonic \st\ in section~\ref{s:tachpot}. We conclude this chapter with three
sections. In section~\ref{s:Neumann} we give a concrete realization of Witten's
interaction vertex. In
section~\ref{s:proofs} we provide the proofs that have been left over at earlier
stages. In the final section we discuss some other calculations that have been
done in \sft\ in recent months and speculate about further directions research
might take.

\section{Gluing world sheets}\label{s:gluing}
We will first define string field interactions. At this point this definition
comes ``out of the blue''. A motivation for it will be provided in
section~\ref{s:motivation} where we will give a physical interpretation of this
definition. There it will be argued that the string field interactions are 
described by gluing parts of string \emph{world sheets}. This is to be 
compared with the interaction of point particles, where \emph{world lines} are 
glued together by interactions in Feynman diagrams. In the following definition
we will freely use notations and terminology from conformal field theory. A
brief overview of these elements of conformal field theory we use, can be found
in appendix~\ref{CFT}.  
\kader{definition}{dubbra}{
\ \\The \sft\ interaction between $n$ vertex operators $A_1, \ldots, A_n$ is defined
by
$$ 
\Corr{A_1 \cdots A_n} = 
\corr{f_1^{n} \circ A_1(0)\  \cdots\  f_n^{n} \circ A_n(0)} .$$
\ \\[-7ex]
}\index{string field theory!interaction}
The functions $f_k^{n}$ entering in the correlator on the right hand side
denote some appropriate \cts. They are defined by
\be\label{wedges}
f_k^{n}(z) = 
e^{\frac{2 \pi i (k-1)}{n}} \left( \frac{1+ i z}{1-i z}\right) ^ {2/n}
\mbox{ for } n \ge 1,
\ee
they map the unit disk to wedge-formed pieces of the complex plane. In this
chapter we are studying Witten's \sft. Therefore the \cft\ (cft) in use for 
defining the correlator in definition~\ref{dubbra} is the \cft\ of the open
\emph{bosonic} string, i.e.~the $c=26$ matter
cft tensored with the $c = -26$ ghost 
cft\footnote{ 
This will be different in the next chapter. There we discuss 
Berkovits' superstring field theory.
That field theory is thought to describe off-shell dynamics of open 
\emph{superstring} theory. In that chapter we will also use the ``double brackets''
notation. It is obvious that the cft in use for definition~\ref{dubbra} will
then be the cft describing open superstrings, i.e.~the $c=15$ matter super cft
tensored with an appropriate ghost cft. \label{keuzecft}}. 
For a brief review of these conformal
field theories we refer the reader to appendix~\ref{CFT}.

Let us now give some examples to illustrate definition~\ref{dubbra}.
\paragraph{Example 1}\label{W:vb1}
The easiest interaction to write down explicitly is between three ghosts 
$\Corr{c\ c\ c}$\footnote{This example is part of the calculation one needs to
do for the level $(0,0)$ approximation to the tachyon potential in
section~\ref{s:tachpot}.}. Because the ghost $c$ is a primary field with 
weight $-1$, it transforms as
$$ f \circ c(0) = f^{\prime}(0)^{-1}c(f(0)). $$ Therefore we have
\bqs
f^3_1 \circ c(0) &=& {3 \over 4 i}\  c(1),\\
f^3_2 \circ c(0) &=& {3 \over 4 i}{1 \over\omega}\  c(\omega),\\
f^3_3 \circ c(0) &=& {3 \over 4 i}{1 \over\omega^2}\  c(\omega^2),
\eqs
where we used $\omega = \exp 2 \pi i /3$. Using equation\refpj{zeromodes} we find
easily
\bq
\Corr{c\ c\ c} &=& \left({3 \over 4 i}\right)^3 {1 \over\omega\omega^2}
\corr{c(1)\ c(\omega)\ c(\omega^2)}\nonumber\\
&=& \left({3 \over 4 i}\right)^3 
(1-\omega)(1-\omega^2)(\omega-\omega^2)\nonumber\\
&=& - {3^4 \sqrt{3} \over 4^3}\label{vgl:t3}
\eq
\paragraph{Example 2}\label{W:vb2} Here is an example involving 
a non primary field:
$\Corr{\p c,\  c,\ c}$. The vertex operator $\p c$ is non primary and transforms as
\bqs
f \circ \p c(z) &=& \p ( f \circ c)\\
&=& \p \left( f'(z)^{-1} c(f) \right)\\
&=&    \p c(f)-{f'' \over (f')^2} c(f)
\eqs 
We find\footnote{The result of this calculation is zero. This follows from a 
symmetry treated in section~\ref{s:twist}.}
\bqs 
\Corr{\p c,\  c,\ c}&=& \left({3 \over 4i}\right)^2 {1 \over \omega \omega^2}
\corr{\p c(1) \ c(\omega) \ c(\omega^2)} - \corr{c(1) \ c(\omega) \
c(\omega^2)}\\
&=& - {3^2 \over 4^2} \left( (2 - \omega - \omega^2)(\omega-\omega^2) - 
(1-\omega)(1-\omega^2)(\omega-\omega^2)\right)\\
&=&0.
\eqs
\paragraph{Example 3}\label{W:vb3} As a final example\footnote{This example is 
also part of the calculation of
section~\ref{s:tachpot}.} we calculate the interaction
$\Corr{c,\ \p^2 c,\ Tc}$. The \ct\ of the stress-energy tensor is 
well-known~\cite{BPZ}:
$$
f\circ T(z) = f'(z)^2 T(f(z)) + S(f,z), 
$$
where $S(f,z)$ is the Schwarzian derivative
$$S(f,z) = \frac{D}{12}\left[ \frac{3}{2}
\left(\frac{f''}{f'}\right)^2-\frac{f'''}{f'}\right].$$ 
In our case only this anomalous term will contribute. The other \ct\ we need
reads $$ f\circ \partial^2 c = \parders{z}{2} \left[ f'(z)^{-1} c(z) \right].$$
Taking all this together we find:
\bqs
\Corr{c\ \partial^2 c \ Tc } &=& 
\parders{z}{2}\Big\{
\frac{1}{f^{3 \prime}_1(0)f^{ 3 \prime}_2(z) f^{3 \prime}_3(0)} S(f^3_3,0)\\  
& & \qquad\qquad \corr{c(f^3_1(0))\  c(f^3_2(z))\ 
c(f^3_3(0))}\Big\}\Big|_{z=0}\\
&=&- { 11 \cdot 5\cdot 13\over 
2^5\cdot 3\sqrt{3}}
.\eqs

\paragraph{}
As the reader who checked the examples explicitly has noticed, 
the calculations quickly become a bit tedious. They always 
consisted of
two parts. The first part is the calculation of the \ct\ of (non-) primary
fields. The second part is the evaluation of correlators. This latter part is
usually done by performing Wick contractions and can easily be programmed for
computer aided calculations. However, the first part is very difficult to
program\footnote{In appendix~\ref{s:transf} we give detailed information about the
calculation of \cts\ of some non-primary vertex operators.}.

There are easier methods for evaluating string field theory
interactions.
\begin{itemize}
\item Use states instead of vertex operators and translate the \cts\ 
 into Neumann coefficients. This procedure was developed by Gross and
 Jevicki~\cite{GrossJev}. We give a review of this method in
 section~\ref{s:Neumann}.
\item Use conservation laws. These are nicely written down in 
ref.~\cite{conserv}.
We will develop this procedure for the superstring in section~\ref{s:cons}.
\end{itemize}

In the next subsection we will treat the physical interpretation of
definition~\ref{dubbra}.

\subsection{Physical interpretation}\label{s:motivation}
Let us now clarify 
what definition~\ref{dubbra} has to do with gluing 
world sheets by looking at the $n=3$ case. 
At the start, the world sheets of the three
strings are represented as unit half-disks $\{ |z_i| < 1, \Im z \ge 0\}$, $i=1,2,3$
in three copies of the complex plane.

\begin{figure}[h]
\begin{center}
\epsfxsize=\textwidth
\begin{psfrags}
\epsfbox{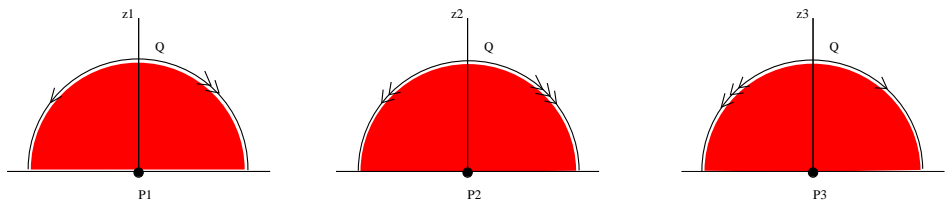}
\end{psfrags} 
\caption{Representation of the cubic vertex as the gluing of three half-disks.
The arrows on the boundaries specify the way these boundaries should be glued.}
\label{Witten:fig1}
\end{center}
\end{figure}

\begin{figure}[ht]
\begin{center}
\epsfxsize=10cm
\begin{psfrags}
\psfrag{P1}[][]{$P_1$}
\psfrag{P2}[][]{$P_2$}
\psfrag{P3}[][]{$P_3$}
\psfrag{T1}[][]{local coordinates around $P_1$}
\psfrag{T2}[][]{local coordinates around $P_2$}
\psfrag{T3}[][]{local coordinates around $P_3$}
\psfrag{Q}[][]{$Q$}
\epsfbox{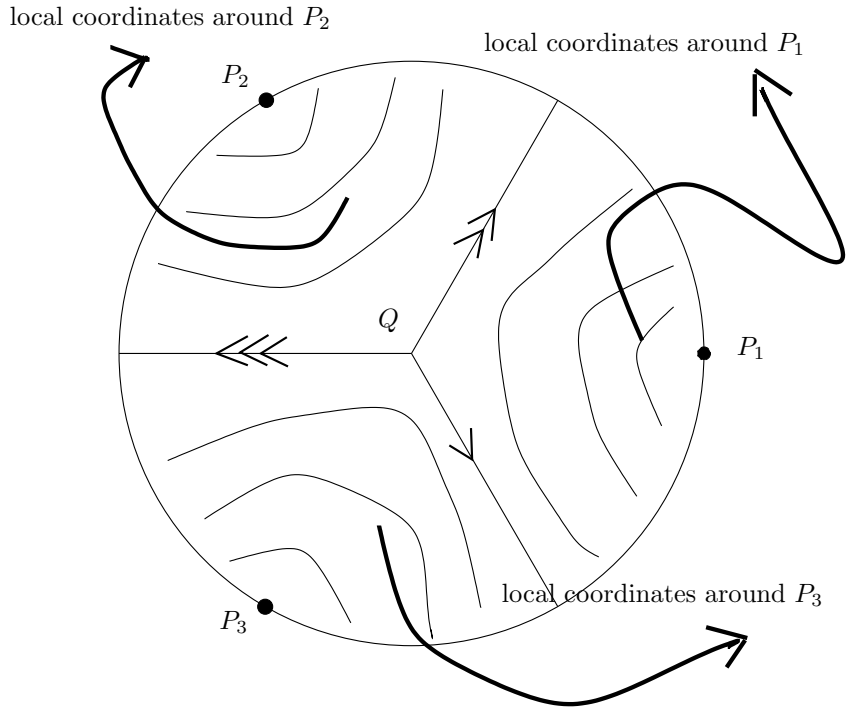}
\end{psfrags} 
\caption{The Riemann surface with three punctures. The three vertex operators
are inserted at the punctures $P_1, P_2$ and $P_3$. The common interaction
point $Q$ is the mid-point of each open string.}
\label{Witten:fig2}
\end{center}
\end{figure}

The three vertex operators $A_1, A_2$ and $A_3$ are inserted in the origin, this
is respectively point $P_1$, $P_2$ and $P_3$.
We glue the boundaries $|z_i|=1$ of the three half-disks with the identification
\bq\label{gluing1} 
z_1 z_3 = -1 && \qquad \mbox{for } |z_1| = 1 
\mbox{ and }\Re z_1 \le 0,\nonumber\\
z_2 z_1 = -1 && \qquad \mbox{for } |z_2| = 1 \mbox{ and }\Re z_2 \le 0,\\
z_3 z_2 = -1 && \qquad \mbox{for } |z_3| = 1 \mbox{ and }\Re z_3 \le
0.\nonumber
\eq
This means that the left half of
the first string is glued to the right hand of the third string and cyclically.
 This construction defines a specific 
``three-punctured disk'', a genus zero Riemann surface with a boundary and 
three punctures. These punctures are marked points on the boundary 
together with a choice of local coordinates
$z_i$ around each puncture. These local coordinates are defined by the
functions\refpj{wedges} which provide the conformal mapping from 
figure~\ref{Witten:fig1} to figure~\ref{Witten:fig2}.

\begin{figure}
\begin{center}
\epsfxsize=12cm
\begin{psfrags}
\psfrag{s1}{string 1}
\psfrag{s2}{string 2}
\psfrag{s3}{string 3}
\epsfbox{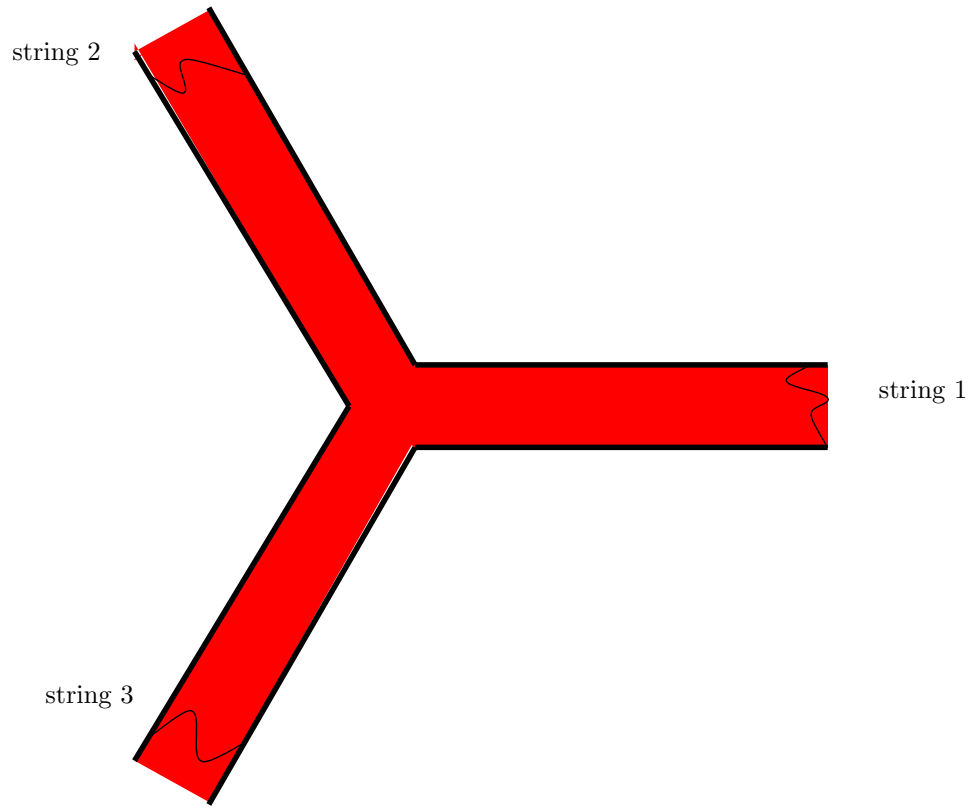}
\end{psfrags} 
\caption{Three strings approach from infinity and interact with one another. 
The right endpoint of the string $i$ joins with
the left endpoint of the string $i+1$.}
\label{Witten:fig3}
\end{center}
\end{figure}

This construction defines the interaction between three strings by gluing their
world sheets. This is nicely illustrated by doing a 
\ct\ on figure~\ref{Witten:fig2}, see
figure~\ref{Witten:fig3}. 
In this picture it is clear that
definition~\ref{dubbra} represents the interaction of three strings. Indeed, by
the state-operator mapping, inserting a vertex operator in the origin of the
world sheet creates a specific state in the string theory Fock space.

\clearpage

\subsection{State formalism}
In definition~\ref{dubbra} we defined the interaction between vertex operators.
As is well known, to every vertex operator corresponds a unique state. This is
the so called state-operator mapping. Let us denote by $\state{A}$ the state
corresponding with the vertex operator $A(z)$:
$$\state{A} = A(0) \vac, $$
where $\vac$ is the $SL(2, \IC)$-invariant vacuum, 
for more information we refer the reader to appendix~\ref{CFT}.
Because the \sft\ interaction defines a multi-linear mapping:
$$\Hil\otimes\cdots\otimes\Hil \to \IC : \state{A_1} \otimes \cdots \otimes
\state{A_n} \to \Corr{A_1 \cdots A_n}, $$
where $\Hil$ is the Hilbert space of states in the open bosonic string
theory, we can define the following.
\kader{definition}{Vertexn}{Witten's vertex $\etats{V_n} \in \Hil^{\ast} \otimes
\cdots \otimes \Hil^{\ast}$  is defined by
$$ \etats{V_n} \state{A_1} \otimes \cdots \otimes \state{A_n} = \Corr{A_1 \cdots
A_n}$$
for all states $\state{A_i}$ with corresponding vertex operators 
$A_i(z)$.}\index{Witten's vertex!abstract}
Of course this is a rather abstract way of defining the bra $\etats{V_n}$. We
will give a concrete form of $\etats{V_3}$ in section~\ref{s:Neumann}.
\paragraph{A special case} The case $n=2$ is nothing but the bpz inner
product treated in appendix~\ref{s:bpz}.
\kader{prop}{prop:V2}{
$$\etats{V_2} \state{A}\state{B} 
= \Bigl( \state{A}, \state{B} \Bigr)_{\mbox{\scriptsize{\bpz}}} 
\equiv \langle A \state{B}$$
}\index{bpz inner product}
\begin{proof}
The left hand side is by definition 
$\corr{f_1^2 \circ A(0) \ f_2^2 \circ B(0)}$ with
$$f_1^2(z) = { 1 + i z \over 1 - i z} \quad\mbox{and}\quad
f_2^2(z) = - { 1 + i z \over 1 - i z}.$$   
This correlator is invariant under the $SL(2,\IC)$ transformation 
$\displaystyle{g(z) = {1+z\over 1-z}i}$. Hence it is equal to 
$$\corr{g \circ f_1^2 \circ A(0) \quad g\circ f_2^2 \circ B(0)}
= \corr{I \circ A(0)\ B(0) }\ ,$$
where $I(z) = -1/z$. This is exactly the bpz inner product as defined in
appendix~\ref{s:bpz}.
\end{proof}

\section{The star product}\label{s:star}
We will now define the star product. This star product is a crucial element of
Witten's open string field theory.
\kader{definition}{starprod}{
Witten's $\star$-product of two states $\state{A}$ and $\state{B}$ is\\ the
state $\state{A}\star \state{B}$ for which for all states $\state{\psi}$: 
$$\quad \etats{\psi}\left( \state{A}\star\state{B}\right) = 
\etats{V_3} \state{\psi} \state{A}\state{B}$$
\ \\[-7ex]
}\index{star product!in bosonic string field theory}
In the left hand side of the definition we have used the bpz inner product, see
appendix~\ref{s:bpz}. It is clear from this definition that the 
ghost number (gn) is a
grading for the star product in the sense that the following holds: 
gn$(A\star B) =$ gn$(A) + $gn$(B)$.
\paragraph{An example}
We will calculate the
following star product: $c_1\vac\star c_1\vac$. This product has ghost number~2,
therefore we have the following expansion:
\bqs
c_1\vac\star c_1\vac&=& x_1\ c_0c_1\vac + x_2\ c_{-1}c_1\vac+ x_3\ c_{-2}c_1\vac\\
& & + x_4\ c_{-1}c_0\vac + x_5\ L_{-2}c_{0}c_1\vac + O(\mbox{level~3})
\eqs
Here $O(\mbox{level~3})$ denotes that we have omitted states of level~3 and
higher\footnote{The level of a state is defined to be the weight of the state
$+1$. Hence $c_0c_1\vac$ has level~0, $c_{-1}c_1\vac$ has level~1, etc.}.
We will now fix these coefficients $x_1, \ldots,x_5$ by imposing the 
definition~\ref{starprod} for all states $\state{\psi}$. This 
definition holds trivially if $\state{\psi}$ has ghost number
different from 1, indeed the left and the right hand side in the definition 
are then both
zero. Therefore we restrict ourselves to states $\state{\psi}$ having ghost
number 1:
\bqs
\state{\psi}&=& \psi_1\ c_1\vac + \psi_2\ c_0\vac+ \psi_3\ c_{-1}\vac\\
& & + \psi_4\ c_{1}L_{-2}\vac + \psi_5\ b_{-2}c_{0}c_1\vac + O(\mbox{level~3})
\eqs

\emph{Left hand side of definition~\ref{starprod}}\\
As explained in appendix~\ref{s:bpz} the bpz conjugate of $\state{\psi}$ is
\bqs
\etats{\psi}&=& \leftvac (\psi_1 c_{-1}- \psi_2 c_0\vac+ \psi_3 c_{1}+\\
& & + \psi_4 c_{-1}L_{2}\vac + \psi_5 b_{2}c_{-1}c_0) + O(\mbox{level~3})
\eqs
and we can calculate with the normalization\refpj{bc:norm}
\bq\label{star:vb1}
\etats{\psi}\ c_1\vac\star c_1\vac&=& \psi_1x_1 + \psi_2 x_2 + \psi_3 x_4\\
&&+  \psi_4x_5 \corr{L_2 L_{-2}} + \psi_5 x_3 \nonumber
\eq
Because we are working in the bosonic string we have $[L_2,L_{-2}] = 4 L_0 +
26/12 \cdot 6$, hence the coefficient of $\psi_4x_5$ is 13.

\emph{Right hand side of definition~\ref{starprod}}\\
We need to calculate
$$\etats{V_3} \left(\psi_1 c_1 + \psi_2 c_0+\psi_3 c_{-1}+\psi_4 c_1L_{-2}
+  \psi_5 b_{-2}c_{0}c_1\right) \vac \otimes c_1\vac\otimes c_1\vac.$$  
This calculation is rather tedious to do by hand, however we have explained all
necessary details in section~\ref{s:gluing}. The result is:
$$-{81 \sqrt{3} \over 64} \psi_1 + 0\ \psi_2 - {33 \sqrt{3} \over 64}\ \psi_3 
+{195 \sqrt{3} \over 64}\psi_4 +{3 \sqrt{3} \over 4}\psi_5$$ 
Therefore comparing with equation~\refpj{star:vb1} we can fix $x_1,\ldots,x_5$ and we find
\bqs
c_1\vac\star c_1\vac&=& -{81 \sqrt{3} \over 64} \ c_0c_1\vac + 0\ c_{-1}c_1\vac
+{3 \sqrt{3} \over 4} \ c_{-2}c_1\vac\\
& & - {33 \sqrt{3} \over 64}\ c_{-1}c_0\vac + {15 \sqrt{3} \over 64}\ L_{-2}c_{0}c_1\vac + 
O(\mbox{level~3})
\eqs
We refer the interested reader to ref.~\cite{conserv} where a closed form expression
of this star-product as
well as numerous others can be found.

\paragraph{}
We conclude this section with two crucial properties of the star product.

\kader{prop}{starass}{ \ 
\begin{enumerate}
\item[(i)] the $\star$-product is associative\footnotemark
\item[(ii)] for all states $A_1,A_2, \ldots, A_n$:\\
$$
\etats{V_n}\state{A_1} \cdots \state{A_k}\state{A_l}\cdots \state{A_n}=
\etats{V_{n-1}}\state{A_1} \cdots (\state{A_k}\star\state{A_l})\cdots \state{A_n}
$$
\end{enumerate}
\ \\[-7ex]}
\footnotetext{One should be very careful about the space 
$\Hil$ one is working with, otherwise there are associativity 
anomalies~\cite{assanom}.}
We refer to ref.~\cite{GrossJev} for a proof of~(i). The difficult proof of the
second statement can be found in ref.~\cite{gluing}.

\section{Short review of the Chern-Simons action}\label{s:CS}
\index{Chern-Simons action|(}
Witten's action has the same structure as a Chern-Simons action. Let us
therefore give a short review of this action.

Consider a $3$-manifold $M$ and let $A$ be a 1-form. 
The Chern-Simons action reads
\be\label{CS:action}\index{Chern-Simons action}
S(A) = \frac{1}{2} \int_M A \wedge dA + \frac{1}{3} \int_M A \wedge A \wedge A\
.
\ee
In general the gauge
field $A$ is a Lie-algebra valued $1$-form: $A = A_a T^a$ where the generators
$T^a$ are the generators of a Lie-algebra:
$$[T^a,T^b] = f^{ab}_{\ \ c}T^c.$$
In the action\refpj{CS:action} the integration symbol is understood to include 
a trace
over the Lie-algebra generators. If we decorate all the symbols with their
Lie-algebra indices the action becomes
$$
S(A) = \frac{1}{2} \int_M A_a \wedge dA_a + 
\frac{1}{6}  f^{ab}_{\ \ c}\int_M A_a \wedge A_b \wedge A_c.
$$
In this formula we have normalized the generators to $\mbox{tr} T^a T^b =
\delta^{ab}$.
\paragraph{Gauge invariance}
This action is invariant under
the following gauge transformation
\be\label{CS:gtr}
\delta A = d\epsilon + A \wedge \epsilon - \epsilon \wedge A 
\mbox{\quad ( $\epsilon$ is a 0-form )}.
\ee
If we write out the Lie-algebra indices explicitly, 
the gauge transformation is
$$
(\delta A)_c = d\epsilon_c +  f^{ab}_{\ \ c}A_a \wedge \epsilon_b .
$$
Witten's action has a kind of ``gauge invariance'' as well. Let us therefore
see which properties we need to prove the gauge invariance\refpj{CS:gtr}.
Tracking down the proof of the gauge invariance, we see that we need the
following five properties:
\be\label{CS:prop}
\left\{ \begin{array}[h]{ll}
1.& d \mbox{ is nilpotent}:\\
&\qquad \displaystyle{d^2 \omega = 0 \mbox{  for all differential forms }
\omega.}\\
2.&d \mbox{ is a derivation:}\\
&\displaystyle{\qquad d(\omega \wedge \eta) = 
d\omega \wedge \eta+(-1)^{|\omega|}\omega \wedge d\eta,} \\
&\qquad \mbox{where } \displaystyle{|\omega| \mbox{ is the degree of } \omega.}\\
3.&\mbox{cyclic symmetry: }\\ 
&\qquad 
\displaystyle{\int \omega\wedge\eta 
=(-1)^{|\omega| | \eta|}\int \eta\wedge\omega}\\
4.&\mbox{Stokes:}\\  
&\displaystyle{\qquad \int d \omega = 0}\\
5.&\mbox{the wedge product is associative}
\end{array}\right.
\ee
We will make sure that the analogs of\refpj{CS:prop} hold in Witten's theory.
\index{Chern-Simons action|)}
\section{Witten's action}\label{s:SFT}
\index{string field theory!Witten's sft|(}
Starting from the Chern-Simons action we will now 
construct Witten's action.

Since \sft\ corresponds to second quantized \st, a point in the
\textit{classical} configuration space of \sft\ corresponds to a specific
\textit{quantum} state of the first quantized theory. As was shown
in ref.~\cite{WittenSFT}, in order to describe a gauge invariant \sft\ we must 
include the full Hilbert space of states of the first quantized open \st\ 
including the $b$ and $c$ ghost fields. Therefore the analog of a 
\emph{differential form} in the
Chern-Simons action is taken to be a 
state in the cft of open bosonic string theory. This is
the motivation of the first entry in the table~\ref{CStoW}.

\begin{table}[h]
\begin{center}
\begin{tabular}{|l|l|p{30ex}|}
\hline
& Chern-Simons& Witten's open sft\\
\hline
\hline
\STRUT
1&differential form&state in CFT\\
\hline
\STRUT
2 &wedge product $\wedge$ & star product $\star$\\ 
\hline\STRUT
3& degree of a differential form & ghost number of a state\\
\hline\STRUT
4 & gauge field $A$ & string field $\Psi$ with ghost number~1\\
\hline\STRUT
5 & gauge parameter $\epsilon$ & state in the CFT with ghost number~0\\
\hline\STRUT
6& exterior derivative $d$& BRST operator $Q$\\
\hline\STRUT
7& integration $\int$ & Witten's vertex $\etats{V}$\\
\hline
\end{tabular}\caption{Dictionary: the elements of Chern-Simons theory and their analogs in
Witten's open \sft.\label{CStoW}}
\end{center}
\end{table}

For the analog of the \emph{wedge product} we take the star product defined in
section~\ref{s:star}. For this analogy to make sense, let us see if the
properties we know of the wedge product hold for the star product. The most
important property of the wedge product is its associativity. This also holds
for the star product, see property~\ref{starass}. A more trivial property of
the wedge product is that it adds degrees:
$$p-\mbox{form}\  \wedge\  q-\mbox{form} =  (p+q)-\mbox{form}.$$
We have a similar property in Witten's theory. Indeed, we can define the
analog for the \emph{degree} of a differential form to be the \gn\ of the
state. As remarked below definition~\ref{starprod} the \gn\ is a grading for the
star product. This is the motivation for the second and the third entries in
table~\ref{CStoW}.

From this assignment and the fact that the \emph{gauge field} $A$ is 
a 1-form, it follows that its analog in Witten's action, the
\sfi\ $\Psi$, must have \gn\ 1. Since
the gauge parameter in the Chern-Simons action is a 0-form, 
it's analog will have ghost number zero. 

The fact that the exterior derivative squares to zero $d^2 =0$, suggests that
we use the BRST-charge $Q$~\refpj{BRST:bos} as a substitute. When we apply the exterior
derivative on a differential form its degree increases with one unit. This is
compatible with the fact that the BRST-charge has \gn\ one.

The \emph{integration} $\int$ is a linear functional on the differential forms. It is
zero when the form has degree different from three. This is nicely compatible
if the analog of the integration $\int$ is Witten's vertex $\etats{V}$. Indeed
$$ \etats{V_n} \state{A_1} \otimes \cdots \otimes \state{A_n} = 0$$
if the total \gn\ of the states $\state{A_i}$ is not three.

Taking this all together, Witten's action reads
\be
\label{W:action}
\fbox{$ \qquad \displaystyle 
S(\state{\Psi}) = - 2 \pi^2 M \left(\frac{1}{2} \etats{V_2}\state{\Psi} 
Q\state{\Psi} 
+ \frac{1}{3} \etats{V_3}
\state{\Psi}\state{\Psi}\state{\Psi}\right), \qquad
$}
\ee
where $\state{\Psi}$ is a state in the open bosonic cft carrying \gn\ one and 
$M$ is the mass of the D-25 brane\footnote{For a justification of this 
normalization, see ref.~\cite{Sen:univ,9912249}.}. 
We
have the following gauge invariance
\be\label{W:gtr}
\delta \Psi  = Q \epsilon + \Psi \star \epsilon - \epsilon \star \Psi, 
\ee
with $\epsilon$ a state in the open bosonic cft of ghost number~0. 
Of course at this point we have only given some motivation for\refpj{W:action}
and\refpj{W:gtr}. We still have to prove the analogs 
of\refpj{CS:prop} if we want to verify that\refpj{W:gtr} is 
a gauge transformation.
For all states $A$ and $B$ we need
\be\label{W:prop}
\left\{ \begin{array}[h]{ll}
1.&$Q$ \mbox{ is a derivation}\\
&\displaystyle{\qquad Q(A \star B) = 
Q_A \star B+(-1)^{\mbox{\scriptsize{gn}} (A)}A  \star QB, } \\
&\qquad \mbox{where } \displaystyle{\mbox{gn}(A) 
\mbox{ is the ghost number  of the state} A.}\\
2.&\mbox{cyclic symmetry: }\\ 
&\qquad 
\displaystyle{\etats{V_2} \state{A}\state{B} 
=(-1)^{\mbox{\scriptsize{gn}} (A)\mbox{\scriptsize{gn}} (B)}
\etats{V_2}\state{B}\state{A}}\\
3.&\mbox{Stokes:}\\  
&\displaystyle{\qquad \etats{V_1} Q  \state{A} = 0}\\

\end{array}\right.
\ee
For a proof of the first and third property, we refer to ref.~\cite{Joris,BSZ}. We
will give an alternative proof of these two properties in the context of the
superstring in sections~\ref{s:ders} and~\ref{s:cons}. The proof of the cyclicity is provided in 
section~\ref{s:proofs}.
\index{string field theory!Witten's sft|)}
\section{Comments on the gauge invariance}\label{s:Siegel}
In this section we will first prove at the linearized 
level\footnote{The Siegel gauge seems to hold at the non-linear level 
as well~\cite{Siegnonlin1,Siegnonlin2,Siegnonlin3}. 
} 
that we can fix the gauge by imposing the 
Feynman-Siegel gauge:
\be\label{W:Siegel}\index{Siegel gauge!in Witten's sft}
b_0 \state{\psi}=0.
\ee In the second part of this section we will show that in this gauge the
quadratic part in\refpj{W:action} reduces to the standard kinetic terms of field
theory.
\paragraph{The Feynman-Siegel 
gauge\protect\footnote{We follow ref.~\cite{9912249} closely.}}
\begin{proof}
We
first show that the Siegel slice intersects each orbit at least once. Suppose we
have a field $\state{\psi}$ with weight $h$ ( see picture~\ref{fig:Siegel}).

\begin{figure}[ht]
\begin{center}
\epsfxsize=\textwidth
\begin{psfrags}
\psfrag{p1}[][]{$\state{\tilde\psi}$}
\psfrag{p2}[][]{$\state{\psi}$}
\psfrag{A}[][]{$\state{A}$}
\psfrag{B}[][]{$\state{B}$}
\psfrag{SL}[][]{Siegel slice}
\psfrag{go}[][]{gauge orbits}
\epsfbox{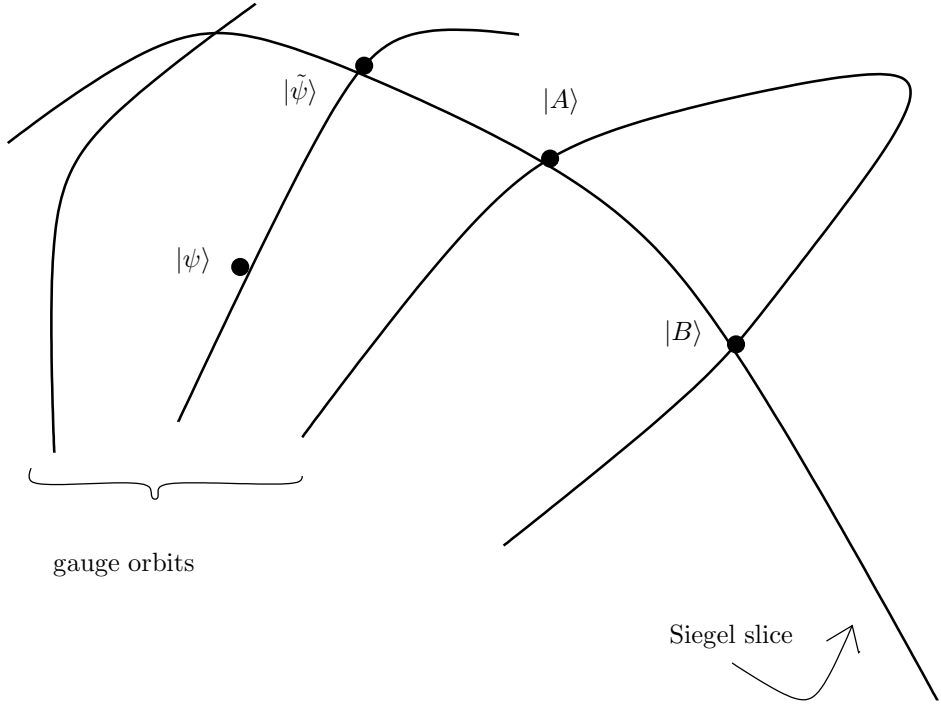}
\end{psfrags} 
\caption{The slice defined by the Siegel gauge intersects every orbit exactly
once.}
\label{fig:Siegel}
\end{center}
\end{figure}

If
we define 
$$\state{\tilde\psi} = \state{\psi}-{1 \over h} Q b_0\state{\psi},$$
then $\state{\tilde\psi}$ clearly belongs to the same gauge orbit as
$\state{\psi}$. Upon using $\{ Q,b_0 \}= L_0^{\mbox{\scriptsize tot}}$ it is easy
to see that  $\state{\tilde\psi}$ satifies the Siegel gauge~\refpj{W:Siegel}.
Notice that this part of the proof breaks down when $h=0$, so we should not
impose the Siegel gauge on fields with weight zero. 

To finish the proof we need to show that the Siegel slice intersects each orbit
only once. Suppose $\state{A}$ and $\state{B}$ lie on the same orbit
$$\state{A}=\state{B}+ Q \state{\epsilon}.$$
Both satisfy the Siegel condition, hence
we have $b_0 Q \state{\epsilon}=0$. We can calculate
\bqs
L_0^{\mbox{\scriptsize{\mbox{tot}}}}\left(\state{A}-\state{B}\right) &=&
L_0^{\mbox{\scriptsize{\mbox{tot}}}}Q \state{\epsilon}\\
&=& \{ Q,b_0\} Q \state{\epsilon}\\
&=& 0\ .
\eqs
We have restricted ourselves to fields with non-zero weight. 
Therefore, the last equation implies $\state{A}-\state{B}=0$, as we 
wanted to show.
\end{proof}

\paragraph{The quadratic part in the Feynman-Siegel gauge}\ \\
In the Siegel gauge the quadratic terms in Witten's action reduce to
\be\label{W:Sieg2}
\etats{\psi_1} Q \state{\psi_2} =  
\etats{\psi_1} c_0 L_0^{\mbox{\scriptsize{tot}}} \state{\psi_2}
.\ee
Indeed, we can do the following calculation
\bqs
\etats{\psi_1} Q \state{\psi_2}&=&\etats{\psi_1} Q\ \{b_0,c_0\}\state{\psi_2}\\
&=&\etats{\psi_1} Q\  b_0 c_0\state{\psi_2}
\qquad \left(\mbox{Siegel gauge on }\state{\psi_2}\right)\\
&=&\etats{\psi_1} \{Q ,b_0\} c_0\state{\psi_2}\qquad\left(\mbox{Siegel gauge on
}\state{\psi_1}\right)\\
&=&\etats{\psi_1}  L_0^{\mbox{\scriptsize{tot}}} c_0\state{\psi_2}\\
&=&\etats{\psi_1}  c_0 L_0^{\mbox{\scriptsize{tot}}} \state{\psi_2} 
\quad \left( [ L_0^{\mbox{\scriptsize{tot}}},c_0 ] =0 \right) \\
\eqs  
Because the oscillator $L_0^{\mbox{\scriptsize{tot}}}$ looks roughly as 
$p^2 + m^2$, this shows
that in this gauge the kinetic terms are the familiar ones.

As an example we will 
now calculate all the quadratic terms needed in
section~\ref{s:tachpot}. We have 
$ \state{\psi} = \left( t c_1 + u c_{-1} + v L_{-2} c_1\right) \vac,$ and
according to definition~\ref{bpzosc} 
$$\bpz(\state{\psi}) = 
\leftvac \left( (-1)^{1-1}t c_{-1} +(-1)^{-1-1} u c_{1} +
(-1)^{-2+2+1-1} v L_{2} c_{-1}\right). $$
We find 
\bqs
\etats{\psi_}  c_0 L_0^{\mbox{\scriptsize{tot}}} \state{\psi_}&=&
t^2 \leftvac c_{-1} c_0 L_0^{\mbox{\scriptsize{tot}}}c_1 \vac + 
u^2 \leftvac c_{1} c_0 L_0^{\mbox{\scriptsize{tot}}}c_{-1} \vac \\
& & +
v^2 \leftvac L_2 c_{-1} c_0 L_0^{\mbox{\scriptsize{tot}}} L_{-2} c_{1} \vac 
  \\
&=& - t^2 \leftvac c_{-1} c_0 c_1 \vac + 
u^2 \leftvac c_{1} c_0 c_{-1} \vac \\
& & +
v^2 \leftvac L_2 L_{-2}\vac 
\leftvac c_{-1} c_0  c_{1} \vac. 
\eqs
Using $[ L_2 ,  L_{-2} ] =  4 L_{0} +
\frac{26}{12} \cdot 6 $ and $\corr{ c_{-1} \ c_0 c_1 } = 1$, we have for the
kinetic term 
\be\label{vgl:t2}
\frac{1}{2} \left( -t^2 - u^2 + 13 v^2 \right)\ .
\ee

We have explained in chapter~\ref{c:intro} that if we want to study the fate of
the open string tachyon we need to find a minimum of the tachyon potential.
Therefore we have to solve the equations of motion
\be\label{W:eom}
Q\Psi + \Psi \star \Psi = 0.
\ee
This is a very
complicated task because an infinite number of space-time fields will acquire 
vacuum expectation values. An exact solution is not known\footnote{See 
however ref.~\cite{0008252} where a recursive solution 
was formulated.}. This is 
probably the most important unsolved problem in this field.
One can simplify the search for a solution of\refpj{W:eom} by
noticing that there is a discrete $\IZ_2$-invariance in Witten's action. This 
twist-invariance will be explained in the next section.
Due to the lack of exact solutions one uses an approximation method to
solve\refpj{W:eom}. We will explain this approximation method -- the
\emph{level truncation} method -- in section~\ref{s:tachpot}. 
\section{Twist invariance}\label{s:twist}
Witten's action is invariant under the following $\IZ_2$-twist:
\be\label{W:Z2}\index{twist invariance!in Witten's sft}
\state{\Phi} \to (-1)^{\mbox{\scriptsize{level}}(\Phi)} \state{\Phi}
.\ee
From this invariance follows in particular that no odd level state can couple
linearly to the tachyon. Indeed, the tachyon has level~0 and hence is a 
twist even state.
Therefore it is consistent with the equations of motion to put the odd level
fields to zero. It is easy to see that the twist symmetry holds for the kinetic
terms in Witten's action. For the interaction piece, twist invariance follows
from the following commutation relation\footnote{This commutation relation
forces the result of example~2 on page~\pageref{W:vb2} to be zero. 
Indeed, the field $\p c$ has 
level~1 and the
fields $c$ have level~0, hence $\Corr{\p c, c, c} = -\Corr{\p c, c, c}$. }
holding for string fields with ghostnumber~1
$$
\Corr{\Phi_1 \Phi_2 \Phi_3} = (-1)^{\sum L_i} \Corr{\Phi_1 \Phi_3 \Phi_2}. 
$$
Here $L_i = h_i+1 $ is the level of the field $\Phi_i$. 
We will provide a proof of this
property in section~\ref{s:proofs}.

There are also other invariances of
Witten's vertex. Zwiebach constructed an $SU(1,1)$
symmetry of the action~\cite{0010190}. This symmetry can be used to trim the tachyon string
field further down to $SU(1,1)$ singlets. We will not use this symmetry 
however.\footnote{An analog of this $SU(1,1)$-symmetry is as far as I know not known
for Berkovits' superstring field theory. In view of the Herculean task to extend tachyon
calculations in Berkovits' action much beyond what is done in this thesis, 
it may pay to look for 
such a symmetry in order to restrict the tachyon string field as tightly as
possible.}

\section[The tachyon potential]
{Level truncation approximation to the tachyon potential}\label{s:tachpot}
\subsection*{Level truncation explained}\label{s:leveltrunc}
\index{level truncation!in Witten's sft}
The level truncation method was first used by Kostelecky and Samuel~\cite{KS}.
The idea is simple. Instead of carrying the infinite number of space-time
fields along during the calculation, one truncates the \sfi\ at a finite
number of terms. The more space-time components one keeps, the better the
approximation. In practice it seems that this level truncation procedure
converges pretty well. 

To do a level~$(m,n)$ approximation, one has to keep all fields up to level~$m$
and all interactions up to level~$n$. The level of a field is simply the
conformal weight of the state, shifted by  a constant in such a way that the
field with lowest weight -- the tachyon with weight $-1$ -- has level~0. Hence
the level of a state $= 1\ + $ the weight of the state. The level of an
interaction is defined to be the sum of the levels of all fields entering into
it.

The level truncation method is thus rather similar to the variational method in
quantum mechanics. There the expectation value of any normalized function $\psi$
can also be used to estimate the ground-state energy. 
\subsection*{The stable vacuum at level $(0,0)$}
There is only one space-time field at level~$0$, the tachyon
$$\Phi =  t\ c_1\vac.$$
Evaluating Witten's action we find for the 
potential
\bqs
V(\Phi )&=& -S(\Phi)= 2 \pi^2 M V_0 \qquad\mbox{with}\\
V_0&=&  -0.5\, t^2-0.730709\,t^3\ .\nonumber\\
\eqs
We have calculated the 
coefficients of $t^2$ and $t^3$ before, see equation\refpj{vgl:t2} and
equation\refpj{vgl:t3} respectively.
At level~$(0,0)$, we have a stationary point in $t_0 = -0.456178$ 
at which $V(\Phi_0)\simeq -0.684616 M$. This is about $ 68 \%$ 
of the exact answer according to Sen's first 
conjecture, see page~\pageref{Senc1}. 
\subsection*{The stable vacuum at level $(2,4)$}
Due to the twist invariance we do not have to include the level~1 components of
the string field. The string field up to level~2 reads
$$\Phi =  t\ c_1\vac + u\ c_{-1}\vac +v\ L_{-2}c_1 \vac. $$  
We give the potential with coefficients evaluated numerically up to 6 significant
digits\footnote{We have calculated the 
quadratic terms on page~\pageref{W:Sieg2} and the coefficient of $tuv$ in 
example 3 on
page~\pageref{W:vb3}. With respect to example 3, there is an additional factor
of $3!$ for the different possible permutations (use\refpj{twistcomm}), an additional factor
of $1/3$ which comes directly from the conventions in Witten's action and
finally a factor of 2 which comes from the state-operator mapping.}
.
\bqs
V(\Phi )&=& -S(\Phi)= 2 \pi^2 M(V_0 + V_2 + V_4 )\\
V_2&=& -0.5\, u^2 + 6.5\, v^2-0.893089\,t^2\,u+5.27734\,t^2\,v\\
V_4&=&-0.171401\,t\,u^2-22.7121\,t\,v^2+4.30006\,t\,u\,v
\eqs
At level~$(2,4)$ we again
find a stationary point at 
$t_0 = -0.541591$, $u_0 = -0.173264$, $v_0 =-0.0518987$ which gives 
$ 0.948553 \%$ 
of the exact answer.

At level~$(2,4)$ it is still easy to eliminate all the higher level fields.
This gives for the effective tachyon potential
$$V_{\mbox{\scriptsize{effective}}}(t) = 
-32.1555\ \frac{
    \left( -4.58883 + t \right) \,
    \left( -0.368932 + t \right) \,
    \,t^2\,
    \left( 0.807429 + t \right) }{\left( -13.71 + 
        t \right) \,
    \left( -0.324841 + t \right) }M.$$
    
\begin{center}
\begin{figure}[ht]
\epsfxsize=\textwidth
\begin{psfrags}
\psfrag{t}[][]{$t$}
\psfrag{V}[][]{$V_{\mbox{\scriptsize{effective}}}(t)$ }
\epsfbox{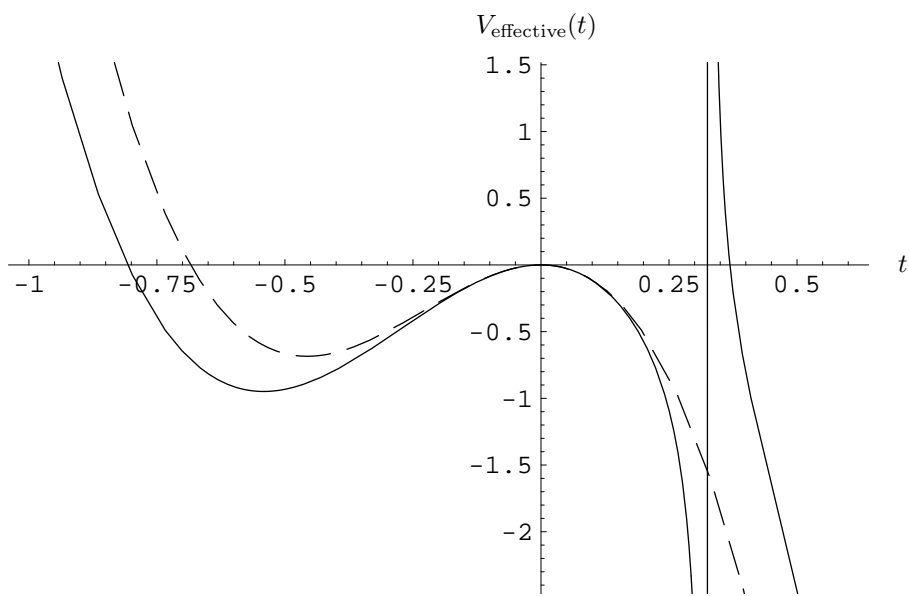}
\end{psfrags} 
\caption{The effective potentials at level $(0,0)$ (dashed line) and at level
$(2,4)$ (full line).}
\label{fig:Veffbos}
\end{figure}
\end{center}
\subsection*{The stable vacuum at higher levels}
Let us calculate the number of states at every level to get an idea of the work
needed to do a higher level calculation.
It is easy to write down the partition function\index{partition function!for bosonic
string field theory}  if we let all oscillators act on the
vacuum $\state{\Omega} = c_1 \vac$, where $\vac$ is the $SL(2,\IC)$-invariant 
vacuum. The vacuum $\state{\Omega}$ is
equivalently defined by
\bqs
b_n \state{\Omega}&=& 0 \qquad\mbox{if    } n \ge 0,\\
c_n \state{\Omega}&=& 0 \qquad\mbox{if    } n \ge 1,\\
L_n \state{\Omega}&=& 0 \qquad\mbox{if    } n \ge -1.\\
\eqs
Notice that $c_0$ does not annihilate this vacuum. However, due to the
Siegel-gauge, fields with non-zero conformal weight will not contain the $c_0$
oscillator. 
Defining $N_{n,k}$ to be the number of fields after gauge fixing 
at level~$n$ with ghost number~$k$, we have
\bqs
\sum_{n\ge 0,\ k \in\mathbb{Z}} N_{n,k}\ q^n g^k &=& g \cdot 
(b\ '\mbox{s}) (c\ '\mbox{s})
(L^{\mbox{m}}\ '\mbox{s})\\
&=& g\cdot\prod_{n=1}^{\infty}(1+ q^n g^{-1}) \prod_{n=1}^{\infty}(1+ q^n g)
 \prod_{n=2}^{\infty}{1\over 1-q^n } 
\\
\eqs 
We expand these products and keep only the ghost number~1 terms. Hence we find
for the number of space-time fields in the string field after gauge fixing 
at every level\footnote{In table~\ref{t:bosstates} there is a field at level~1, which is
missing in this partition function. Indeed,
we should not have applied the Siegel condition on this field, see
section~\ref{s:Siegel}.}:
\bq\label{W:partition}\index{partition function!in Witten's sft}
 \sum_{n \ge 0} N_{n,1} q^n &=& 
1 + 2\,q^2 + 3\,q^3 + 6\,q^4 + 
  9\,q^5 + 17\,q^6\\
  && + 25\,q^7 + 
  43\,q^8 + 64\,q^9 + 102\,q^{10}+O(q^{12})\nonumber
\eq

\begin{table}[h]
\begin{center}
\begin{tabular}{|c|c|p{0.5cm}|}
\hline
Level\STRUT & State & $\mathbb{Z}_2$\\
\hline 
\hline    
0\STRUT& $c_1 \vac $ & +\\ 
\hline
1\STRUT&$c_0 \vac$ &$-$\\ 
\hline  
2\STRUT&$c_{-1} \vac, \   L_{-2} c_1\vac$&$-$\\
\hline
3\STRUT&$c_{-2} \vac, \ b_{-2} c_{-1} c_1 \vac$, $L_{-3} c_1 \vac$&$-$\\ 
\hline
\STRUT&$c_{-3} \vac$,  $b_{-2} c_{-2} c_1 \vac$,  $b_{-3} c_{-1} c_1 \vac$&+\\
4&$L_{-2}c_{-1} \vac$, $L_{-2} L_{-2} c_1 \vac$,  $L_{-4} c_1 \vac$ &\\
\hline 
\end{tabular}\caption{The states after gauge fixing up to level four.
\label{t:bosstates}}
\end{center}
\end{table}

Sen and Zwiebach have calculated the level~$(4,8)$ approximation to the tachyon
potential~\cite{9912249}. From the above partition function\refpj{W:partition}
we see that they needed to include $1+2+6=9$ states, see
table~\ref{t:bosstates}. They got $98.6 \%$ of the conjectured value. Moeller and
Taylor have done the level~$(10,20)$ calculation~\cite{0002237}.
They obtained an accuracy of $99.91 \%$.
From\refpj{W:partition} we see that they needed to include 
171~fields\footnote{In fact, they had 252 fields because they built the states
with the matter oscillators $\alpha^{\mu}_n$ instead of the Virasoro generators
$L_n$.}. 
\subsection*{Final remarks}
It is interesting to see that the level truncation method overestimates the
correct ground-state energy and decreases monotonously to the correct
value\footnote{This is not the case if one fixes the gauge invariance by a gauge
condition other than the Siegel gauge~\cite{Siegnonlin1}. It would be interesting to shed some
light on this issue.}. This is analogous to the behavior of the variational method
in quantum mechanics. The difference is of course that in quantum mechanics one
can prove this behavior whereas in string field theory case one can't even
prove the convergence of the procedure.

At this moment one relies heavily on the level truncation method to get some
results out of string field theory. Therefore, it would also be interesting 
to provide some evidence that the level truncation
method converges to exact answers and discuss its possible limitations.
This is one of the motivations of the study of a toy model in
chapter~\ref{c:TM}.\footnote{In this toy model, it is possible to write 
down some exact solutions in some particular cases and prove the 
convergence of the level truncation method in these cases~\cite{art:TM,Joris}.}

\section{The interaction vertex: a concrete realization}\label{s:Neumann}
\begin{center}
\begin{figure}[ht]
\epsfxsize=\textwidth
\begin{psfrags}
\psfrag{sf}[][]{\begin{minipage}{3cm}{with string functionals}\end{minipage}}
\psfrag{N}[][]{\begin{minipage}{3cm}{with Neumann coefficients}\end{minipage}}
\psfrag{cft}[][]{\begin{minipage}{3cm}{Using \cft}\end{minipage}}
\psfrag{J}[][]{\begin{minipage}{3cm}{\begin{flushleft}see the PhD - thesis 
of Joris Raeymaekers~\cite{Joris}\end{flushleft} }\end{minipage}}
\psfrag{G}[l][]{\begin{minipage}{3cm}{\begin{flushleft}
see Gross and Jevicki~\cite{GrossJev}\end{flushleft}}\end{minipage} }
\psfrag{lc}[t][]{\begin{minipage}{3cm}{see LeClair et
al.~\cite{LeClair}}\end{minipage} }
\psfrag{ts}[][]{this section }
\epsfbox{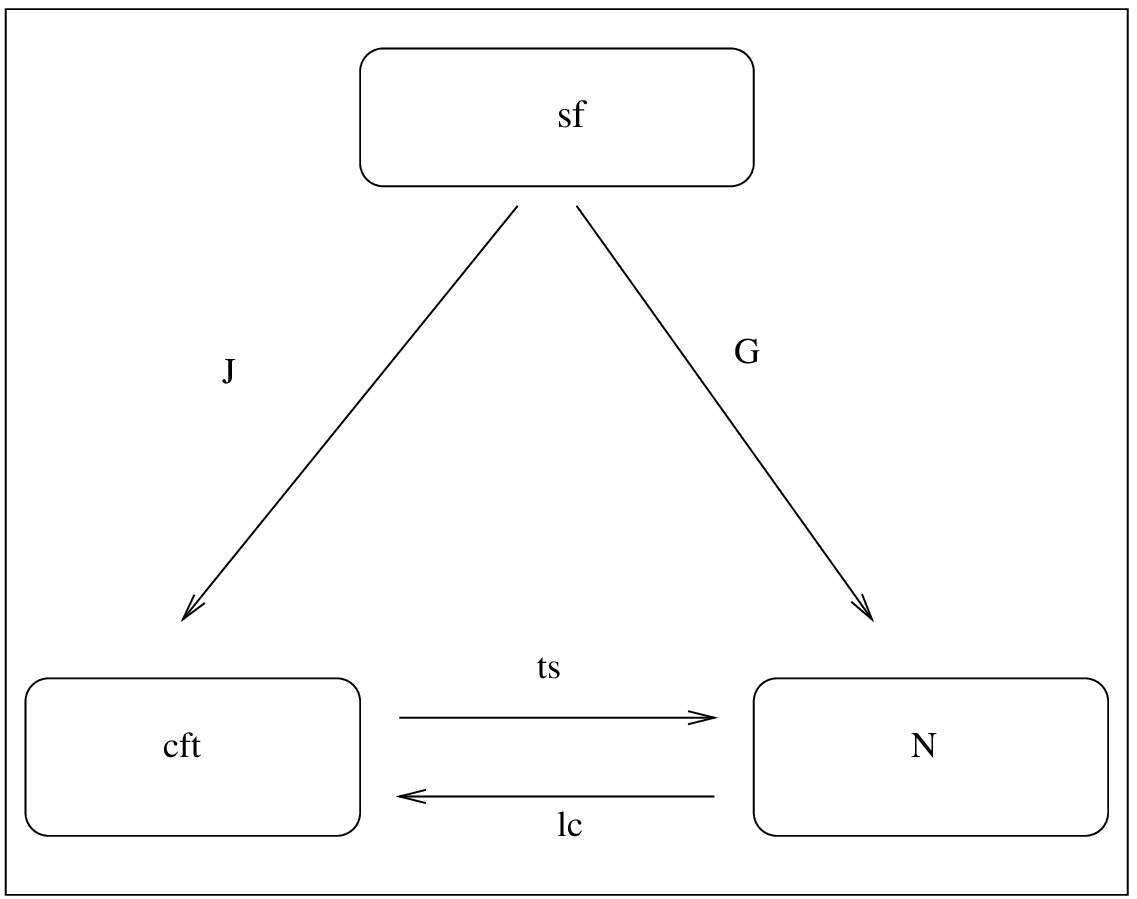}
\end{psfrags} 
\caption{The different realizations of Witten's open \sft.}
\label{Witten:fig4}
\end{figure}
\end{center}

\index{Witten's vertex!concrete realization|(}
We will now discuss a concrete realization of Witten's vertex $\etats{V_3}$. We
use an analog of this realization to construct a toy model for tachyon
condensation in chapter~\ref{c:TM}. Historically, this 
realization
was known before the realization with conformal transformations acting on vertex
operators that we discussed in section~\ref{s:gluing}~\cite{GrossJev}. 
It was constructed directly
from Witten's intuitive formulation of \sft\ involving string
functionals~\cite{WittenSFT}, see figure~\ref{Witten:fig4}.

Because the matter sector is decoupled from the ghost sector, it follows that
Witten's vertex splits
$$\etats{V_3}=\ 
_{\mbox{\scriptsize{matter}}}\etats{V_3}
\otimes\  _{\mbox{\scriptsize{ghost}}}\etats{V_3}.$$

In the following we restrict ourselves to the matter part of the vertex. We are
not interested in the ghost part at this moment because the toy model of
chapter~\ref{c:TM} does not contain a ghost sector. To simplify the discussion we
set the momenta to zero and put $\alpha^{\prime}=2$.
  
Suppose we want to calculate the
following correlator:
\be\label{Neumann1}
\corr{f_1 \circ \mathcal{O}_1(0)\ f_2 \circ \mathcal{O}_2(0)
\ f_3 \circ \mathcal{O}_3(0)}
,\ee
with 
$$\left\{\begin{array}{ll}
 \mathcal{O}_1& = \displaystyle{\frac{i}{(m-1)! }\partial^m X}\\
 \mathcal{O}_2& = \displaystyle{\frac{i}{(n-1)! }\partial^n X}\\
 \mathcal{O}_3& = \mbox{ not specified at the moment}
 \end{array}\right.
$$ 
Then of course\refpj{Neumann1} equals
\bq\label{Neumann2}
\lefteqn{\left. \frac{i}{(m-1)!} \parders{z_1}{m-1}\right|_{z_1 = 0} 
 \left. \frac{i}{(n-1)!} \parders{z_2}{n-1}\right|_{z_2 =
 0}}\nonumber\\
& &  \corr{f_1'(z_1)\ \partial X (f_1( z_1))\  f_2'(z_2)\ 
     \partial X (f_2( z_2))\ f_3 \circ \mathcal{O}_3(0)}
 \eq 
By Wick's theorem, the correlator in\refpj{Neumann2} is a sum of contractions
between the fields $\partial X$. The 3 vertex operators $\mathcal{O}_1$, 
$\mathcal{O}_2$ and $\mathcal{O}_3$ correspond to the 3 states
$\alpha_{-m}^{(1)} \vac$, $\alpha_{-n}^{(2)} \vac$ and $\state{\mathcal{O}_3}$.
The extra superscript denotes in which Hilbert space the oscillators belong.
The interaction $\etats{V} \alpha_{-m}^{(1)} \vac  \otimes 
\alpha_{-n}^{(2)} \vac  \otimes \state{\mathcal{O}_3}$ can then be calculated by
doing contractions between the oscillators $\alpha$ with
$$
\wick{1}{<1 \alpha_{-m}^{(i)}\  >1\alpha_{-n}^{(j)}} = N^{ij}_{mn} 
$$
where 
$$N^{ij}_{mn} = \left. \frac{i}{(m-1)!} 
\frac{\partial^{m-1}}{\partial z^{m-1}}\right|_{z = 0} 
 \left. \frac{i}{(n-1)!} \frac{\partial^{n-1}}{\partial w^{n-1}}\right|_{w =0}
 \frac{- f'_i(z) f'_j(w)}{(f_i(z) -f_j(w))^2}
,$$
the objects $N^{ij}_{mn}$ are called the Neumann 
coefficients\index{Neumann coefficients}. We have
just derived that their generating function is 
\be\label{Neumann3}
\sum_{m,n=1}^{\infty} N^{ij}_{mn}z^{m-1}w^{n-1} = 
 \frac{f'_i(z) f'_j(w)}{(f_i(z) -f_j(w))^2}
.\ee
If the contraction is between oscillators belonging to the same vertex operator, a
singularity develops and we have to be more careful.
Suppose for example that we take the state
$$ \state{\mathcal{O}_3} = \alpha^{(3)}_{-k} \alpha^{(3)}_{-l}\vac, $$
with corresponding normal ordered vertex operator
$$\mathcal{O}_3 = \frac{i}{(k-1)!}\frac{i}{(l-1)!}: \partial^k X \partial^l X:,$$
then the \ct\ reads
\bqs 
\lefteqn{f_3\circ \mathcal{O}_3(w) = \frac{i}{(k-1)!}\frac{i}{(l-1)!} \lim_{z\to w} 
\parders{z}{k-1} \parders{w}{l-1} }\\
& & \left\{ f'(z) f'(w)\ \partial X (f(z))\ 
\partial X ( f(w)) + \frac{1}{(z-w)^2}
 \right\}
\eqs
If we want to calculate the contraction between $\alpha^{(3)}_{-k}$ and 
$\alpha^{(3)}_{-l}$, we have $N^{33}_{kl}$ with
$$N^{33}_{kl} =  -  \frac{1}{(k-1)!}\frac{1}{(l-1)!}\left. \parders{z}{k-1}\right|_{0}  
\left. \parders{w}{l-1}\right|_{0}
\left\{ -\frac{f'_3(z) f'_3(w)}{(f_3(z) -f_3(w))^2}+ \frac{1}{(z-w)^2} 
\right\},
$$
or in general
\be\label{Neumann4}
\sum_{m,n=1}^{\infty} N^{ii}_{mn}z^{m-1}w^{n-1} = 
 \frac{f'_i(z) f'_i(w)}{(f_i(z) -f_i(w))^2}-\frac{1}{(z-w)^2}
.\ee
The reader can easily verify that the right hand side of\refpj{Neumann4} is
finite for $z \to w$. 

Taking all this together, the matter part of the vertex is
$$\etats{V_3}
\state{A}\state{B}\state{C} =\  _{123} \leftvac \exp
\frac{1}{2} N^{ij}_{mn} \alpha^{(i)}_m \alpha^{(j)}_n\  
\state{A}_1\state{B}_2\state{C}_3 , $$
where the indices $i,j$ run from 1 to 3 and the indices $m,n$ from 1 to
$\infty$. The oscillators $\alpha^{(i)}_n$ satisfy the following commutation
relations:
$$ [\alpha^{(i),\mu}_m, \alpha^{(j),\nu}_n] = m\ \delta^{ij} \eta^{\mu\nu}
\delta_{m+n}. $$  
The extra subscript on the states $\state{A}_1$, $\state{B}_2$ and 
$\state{C}_3$ denote that you have to replace the operator $\alpha^{\nu}_n$ by 
$\alpha^{(i),\nu}_n$. 
Including the space time indices Witten's vertex reads
\be\label{V3con}
\fbox{
\ \\[1ex]
$\displaystyle{
_{\mbox{\scriptsize{matter}}}\etats{V_3} = \  _{123} \leftvac \exp
\frac{1}{2} N^{ij}_{mn} \eta_{\mu\nu}\alpha^{(i)\ \mu}_m \alpha^{(j)\ \nu}_n}$
\ \\[1ex]}
\ee
As discussed at the end of section~\ref{s:gluing}, it is this method that is 
easiest to
program. Calculating \sft\ interactions is a matter of doing Wick contractions
with the correct Neumann coefficient. The ghost part of the vertex can be
treated along similar lines~\cite{GrossJev,Tayloreff}. 
For closed formulas of the Neumann coefficients, see ref.~\cite{GrossJev}. 

A table of the Neumann coefficients can easily
found by expanding the generating function~(\ref{Neumann3},\ref{Neumann4}).
As an example we give the following table (a bigger table can be found 
in ref.~\cite{Tayloreff})
\begin{center}\index{Neumann coefficients}
\begin{tabular}{| | c | c | | c | c | c | | }
\hline
\hline
  $n$ &  $m$
 & $N^{11}_{nm}$ & $N^{12}_{nm}$ & $N^{13}_{nm}$
\\
\hline
\hline
1 & 1
 & 
$-5/27$
 & 
$16/27$
 & 
$16/27$
\\ \hline
\hline
1 & 2
 & 
0
 & 
$32\,{\sqrt{3}}/243$
 & 
$-32\,{\sqrt{3}}/243$
\\ \hline
2 & 1
 & 
0
 & 
$-32\,{\sqrt{3}}/243$
 & 
$32\,{\sqrt{3}}/243$
\\ \hline
\hline
1 & 3
 & 
$32/729$
 & 
$-16/729$
 & 
$-16/729$
\\ \hline
2 & 2
 & 
$13/486$
 & 
$-64/243$
 & 
$-64/243$
\\ \hline
3 & 1
 & 
$32/729$
 & 
$-16/729$
 & 
$-16/729$
\\ \hline
\hline
1 & 4
 & 
0
 & 
$-64\,{\sqrt{3}}/2187$
 & 
$64\,{\sqrt{3}}/2187$
\\ \hline
2 & 3
 & 
0
 & 
$-160\,{\sqrt{3}}/2187$
 & 
$160\,{\sqrt{3}}/2187$
\\ \hline
3 & 2
 & 
0
 & 
$160\,{\sqrt{3}}/2187$
 & 
$-160\,{\sqrt{3}}/2187$
\\ \hline
4 & 1
 & 
0
 & 
$64\,{\sqrt{3}}/2187$
 & 
$-64\,{\sqrt{3}}/2187$
\\ \hline
\hline
1 & 5
 & 
$-416/19683$
 & 
$208/19683$
 & 
$208/19683$
\\ \hline
2 & 4
 & 
$-256/19683$
 & 
$128/19683$
 & 
$128/19683$
\\ \hline
3 & 3
 & 
$-893/59049$
 & 
$10288/59049$
 & 
$10288/59049$
\\ \hline
4 & 2
 & 
$-256/19683$
 & 
$128/19683$
 & 
$128/19683$
\\ \hline
5 & 1
 & 
$-416/19683$
 & 
$208/19683$
 & 
$208/19683$
\\ \hline

\hline
\end{tabular}
\end{center}
\index{Witten's vertex!concrete realization}
\index{Neumann coefficients|seealso{Witten's vertex, concrete realization}}
\section{Some proofs}\label{s:proofs}
\begin{enumerate}
\item (cyclicity) for all vertex operators $A_i$: 
$\Corr{ A_1 \cdots A_n} = \Corr{ A_2 \cdots A_n A_1}$ 
\item For all string fields with ghost number~1
\be\label{twistcomm}
\Corr{\Phi_1 \Phi_2 \Phi_3} = (-1)^{\sum L_i} \Corr{\Phi_1 \Phi_3 \Phi_2} 
\ee
\end{enumerate}
where $L_i = h_i+1 $ is the level of the field $\Phi_i$.
\paragraph{Proof of (1).}
We will only give a proof for zero-momentum states. 
A full proof can be found in ref~\cite{BSZ}. 
The cyclicity properties of the conformal field theory correlation functions
are analyzed by using the property:  
\bqs
T\circ f^{n}_i\circ A &=& f^{n}_{i+1} \circ A
\quad \hbox{for} \quad 1\le i\le
(n-1)\\
T\circ f^{n}_n\circ A
&=& T^n\circ f_1\circ A \equiv R\circ f_1\,\circ A ,
\eqs
for any vertex operator A. Here  
$T(w)=e^{2\pi i/n} w$, and $R=T^n$ denotes rotation by $2\pi$. 
While the transformation $R$ acts trivially on the complex plane, 
it
must be viewed in general as the composition $T^n$ of $n$ transformations
by $T$. Thus $R$ affects the 
transformation of fields with non-integer dimension. However, because we
restrict ourselves to zero-momentum states, all fields have integer dimension
and $R$ acts as the identity.
Since $T$
maps the unit disk to itself in a one to one fashion, it corresponds to an
$SL(2,\IR)$ transformation. 
Using
$SL(2,\IR)$ invariance of the correlation functions on the disk, we can
write
$$
\langle (f^{n}_1\circ A_1) \cdots (f^{n}_{n-1}\circ A_{n-1})
(f^{n}_n\circ A_n) \rangle
= \langle (f^{n}_2\circ A_1) \cdots (f^{n}_{n}\circ A_{n-1}) 
(R\circ f^{n}_1\circ A_n) \rangle
$$
The product of all
the operators inside the
correlation function must be Grassmann odd 
in order to get a non-vanishing
correlator. Thus we pick up no minus sign in moving
the $R\circ f^{n}_1\circ A_n$
factor on the right hand side to the first place. Hence the right hand
side may be written as
$$ 
\langle (f^{n}_1\circ A_n) (f^{n}_2\circ A_1) \cdots (f^{n}_{n}\circ
A_{n-1})
\rangle
.$$
\paragraph{Proof of (2).}
Let $M(z) = -z$ and $\tilde I(z) = 1/z$, then it is easy to see that $f^3_k
\circ M = \tilde I \circ f^3_{3-k+2}$. So we have -- restricting ourselves again
to zero-momentum case, so that we do not run into complications with branch cuts --
$f^3_k \circ M \circ \Phi  = \tilde I \circ f^3_{3-k+2} \circ \Phi$. We
calculate
\bqs
\Corr{\Phi_1\ \Phi_2\ \Phi_3} &=& \corr{f^3_1\circ\Phi_1\
f^3_2\circ\Phi_2\ f^3_3\circ \Phi_3}\\
&=& (-1)^{\sum h_i} \corr{f^3_1\circ M\circ\Phi_1\
f^3_2\circ M \circ \Phi_2\ f^3_3\circ M \circ\Phi_3}\\
&=& (-1)^{\sum h_i} \corr{\tilde I \circ f^3_1\circ\Phi_1\
\tilde I \circ f^3_3\circ \Phi_2\ \tilde I \circ f^3_2\circ M \circ\Phi_3}\\
\eqs 
The $SL(2,\IR)$-transformation $\tilde I$  is a combination of the $SL(2,\IR)$
transformation $I(z) = -1/z$ and the world sheet parity transformation
$P(z)=-z$. These are both symmetries in the restricted Fock space we are using.
The string fields have ghost number~1
and so are Grassmann odd, therefore
\bqs
\Corr{\Phi_1 \Phi_2 \Phi_3} &=& - (-1)^{\sum h_i} \Corr{\Phi_1 \Phi_3 \Phi_2}\\
&=&  (-1)^{\sum (h_i+1)} \Corr{\Phi_1 \Phi_3 \Phi_2}
\eqs
This concludes the proof. 
\section{Conclusions and topics for further research}
In this chapter we have treated the basics of Witten's cubic \sft. We have
developed this theory a little bit in an axiomatic way. For a more physical
derivation of Witten's action we refer the reader to Witten's original
paper~\cite{WittenSFT} and
to the PhD-thesis of Joris Raeymaekers~\cite{Joris}. We have chosen 
the axiomatic way
because we wanted to show that Witten's string field theory is defined in a very
precise way, there are absolutely no ambiguities when doing  classical
calculations with this action.

In our discussion of tachyon condensation, we have limited our attention to
those aspects that are necessary as background for the calculations described in
the next chapter. In doing this, we have left out important related
developments, on some of which we will now briefly comment. For a more complete
review of the literature on this subject, see ref.~\cite{Ohmori:thesis}.
\paragraph{}
Witten's string field theory is not background independent. The
background has to be on-shell, i.e.~the world sheet theory of the open strings
moving in this background has to be conformal invariant. A background
independent open string field theory has been proposed~\cite{9208027}. In this
theory, Sen's first and second conjecture have been proved
exactly~\cite{0009103,0009148}.
\paragraph{}
It is believed that a bosonic D-brane with a space-dependent tachyon profile in
the form of a localized ``lump'' of energy describes a D-brane of lower
dimension~\cite{9902105}. The existence of lump solutions describing
lower-dimensional D-branes has been shown in the level truncation
method~\cite{tolumps}. Their tensions have been found to agree with the known
tensions of the lower-dimensional branes.
\paragraph{}
String field theory around
the stable vacuum $\Psi_0$ is described by performing a shift 
$\Psi \to \Psi_0+ \Psi$ in
Witten's action\refpj{W:action}. The shifted action reads
\be\label{W:shifted}
S(\state{\Psi}) = \frac{1}{2} \etats{V_2}\state{\Psi} \tilde Q \state{\Psi} 
+ \frac{1}{3} \etats{V_3}
\state{\Psi}\state{\Psi}\state{\Psi},
\ee
where $\tilde Q$ acts on a general state $A$ as a covariant derivative
\be
\tilde Q A= Q A + \Psi_0 \star A - (-1)^{\mbox{\scriptsize{gn}} A} A \star
\Psi_0.
\ee
If Sen's conjecture is correct and there are no open string excitations around
the stable vacuum, we expect the operator $\tilde Q$ to have vanishing
cohomology. In the level truncation approximation, a study of the
perturbative spectrum around the stable vacuum has been 
performed~\cite{0103085},
supporting the expectation that $\tilde Q$ has vanishing cohomology.
\paragraph{}
For various reasons as discussed in chapter~\ref{c:intro}, it would be extremely 
interesting to
have Witten's theory expanded around the stable vacuum. For this expansion to
be made, it seems likely that one needs the 
exact form of the minimum $\Psi_0$.
Unfortunately, 
despite many efforts, this exact solution remains
elusive. This has not hindered people to make some guesses about the form of 
this shifted action. This shifted \sft\ is called vacuum \sft\index{vacuum sft}. This theory is 
formally 
of the same form as Witten's theory but with another BRST operator that is made
solely out of world sheet ghosts fields. We refer the interested reader 
to the rapidly lengthening series of papers~\cite{Vacuumsft}.

A concrete starting point to find the exact form of the minimum may be as
follows~\cite{Schnabl:B}.
In the Chern-Simons theory the equation
of motion simply says that the field strength $F$ is zero. Such a field 
is trivially given by taking the gauge potential $A$ to be pure gauge:
$$A = g^{-1}\ d g.$$
In Witten's action one can take a similar ansatz:
\be\label{solSch}
\Psi_0 = U \star QV,
\ee
where the states $U$ and $V$ have both ghost number zero and satisfy
\bqs
&& V \star U = I, \mbox{  where } I \mbox{ is the identity element,  and}\\
&& Q(U \star V) = 0.
\eqs
As the reader can verify by plugging\refpj{solSch} into the equation of 
motion\refpj{W:eom}, it is not necessary to have $U \star V = I$.
That the solution can be written in the form\refpj{solSch} was noticed 
by Schnabl by comparing with the large $B$-field 
case~\cite{Schnabl:B}\footnote{
One should be careful with these formal manipulations 
because the identity string field 
is not an
identity for all states~\cite{conserv,Ellwood:I}.}. It would be nice to pursue
this line of research.

It seems likely that one needs this closed form to discuss the
following:
\begin{enumerate}
\item
the absence of open strings around the stable vacuum
\item 
the fate of the $U(1)$ gauge field which lived on the $D$-brane
\item
and perhaps the most important: the emergence of closed strings in this stable
vacuum~\cite{0002223,0010240}. Indeed, if one understands if and how Witten's open string field theory
is able to describe closed strings as well, it would be possible to start
studying the fate of the \emph{closed} string tachyon\footnote{For some
speculations, see for example ref.~\cite{clostach}.}. 
\end{enumerate}

\chapter{Berkovits' Superstring Field Theory}
In this chapter we will study the open superstring tachyon in Berkovits' 
superstring field theory. Therefore this chapter naturally splits into two parts.
\begin{description}
\item[1. Definition of Berkovits' action]
Berkovits' field theory formally looks like a Wess-Zumino-Witten theory.
Therefore, in section~\ref{s:WZW} we give a review of this WZW-action. In
section~\ref{s:BPS} we construct Berkovits' action for a BPS $D9$-brane on this
WZW analog. 
In section~\ref{s:nonBPS} we discuss Berkovits' action for a non-BPS $D9$-brane.
\end{description}
After having set the stage, we can calculate the tachyon potential in this
theory.
\begin{description}
\item[2. Study of the tachyon potential] 
A closed form expression of the string field at the minimum of the potential is
unknown. Therefore, as in the previous chapter, one resorts to the level
truncation method to get approximate results. In section~\ref{s:level4} we
calculate the tachyon potential up to level~$(2,4)$. We find that at this
approximation the results agree with Sen's prediction up to $89\%$. We want 
to do a higher level calculation to get a more accurate correspondence. However,
the procedure followed in section~\ref{s:level4} turns out to be too tedious to
use to perform higher level calculations. Hence, in section~\ref{s:cons} we
develop a more efficient method. This method uses conservations laws and was
given in the literature for the bosonic string. We extend these laws to the
superstring and use them to do a level~$(4,8)$ calculation in
section~\ref{s:high}. In this approximation we find that the
results agree with Sen's prediction up to $94.4\%$.
\end{description}
In the final section we discuss some other calculations that have been performed
in Berkovits' field theory and speculate about directions research might
take. Throughout this chapter, 
technical details and unwieldy formulas have been
referred to appendix~\ref{c:TEX}.
\section{Review of the Wess-Zumino-Witten action}\label{s:WZW}
\index{WZW-action|(}
We will now give a brief review of the Wess-Zumino-Witten action~\cite{artWZW}.
This conformally invariant action plays an important role in string
theory, it describes a string propagating in a
group manifold $G$. A more complete review of this action can be found
in chapter~$15$ of ref.~\cite{Polchinski} and in ref.~\cite{Walton}. 

Suppose $G$ is a compact connected Lie-group and $g$ is a $G$-valued function.
The Wess-Zumino-Witten (WZW) action is written as 
$$S_{\mbox{\scriptsize{WZW}}}= \int_M \mbox{Tr}\ d_z g^{-1} \wedge d_{\bar z} g
+{1 \over 3} \int_N \mbox{Tr}\ \tilde\omega^3.
$$ 
Here $M$ is a
two-dimensional manifold, it is parameterized by
the complex coordinates $(z, \bar z)$. In the quadratic part of the action we
have split the exterior derivative into two parts, $d =  d_z + d_{\bar z} $,
where for all functions $\alpha$
$$ d_z \alpha = { \p \alpha \over \p z} dz \quad \mbox{and}\quad  
d_{\bar z} \alpha = { \p \alpha \over \p \bar z} d \bar z
.$$

\begin{figure}[ht]
\begin{center}
\epsfxsize=7cm
\epsfysize=7cm
\begin{psfrags}
\psfrag{M0}[][]{$M_0$}
\psfrag{M1}[][]{$M = M_1$}
\psfrag{M2}[][]{$M_t$}
\psfrag{c}[][]{$t$}
\psfrag{a}[l][]{$t=0$}
\psfrag{b}[l][]{$t=1$}
\epsfbox{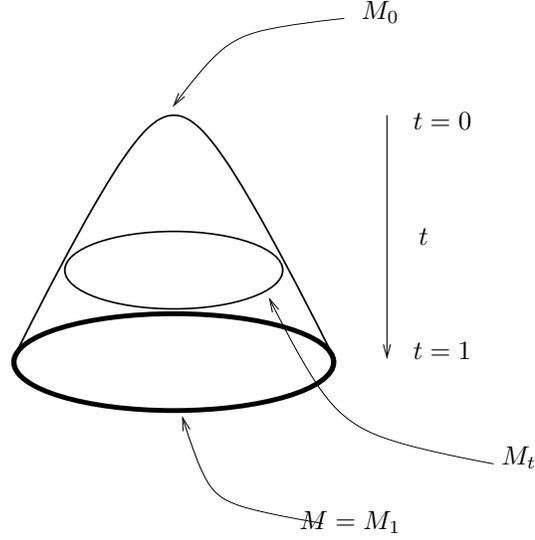}
\end{psfrags} 
\caption{The manifold $N$ used in the Wess-Zumino term. The boundary $M$ is the
set points for which the coordinate $t=1$.}
\label{Berko:fig1}
\end{center}
\end{figure}

The second term is known as the \emph{Wess-Zumino term}. It involves a
three-dimensional manifold $N$  
whose boundary is $M$. In this term we have to extend $g:M\to G$ to 
the three-dimensional manifold,
$\tilde g : N \to G.$
If $G$ is simple, it can be shown that there exists such an extension and that
the WZW-theory is independent of the choice of extension. We will make the
following choice. We extend the manifold $M$ using an additional coordinate $t
\in [0,1]$, see figure~\ref{Berko:fig1}

We extend the function $g$ in the following way
\bqs
&\mbox{if}&\quad g : M \to G: (z,\bar z) \mapsto e^{X(z,\bar z)},\\
&\mbox{then}&\quad\tilde g : N \to G: (t,z,\bar z) \mapsto e^{t X(z,\bar z)}.
\eqs
The Maurer-Cartan form $\tilde\omega$ is defined to be 
$\tilde g^{-1} \tilde d\tilde g$. Here
the exterior derivative contains the
additional coordinate $t$ as well, $\tilde d = d_t + d_z + d_{\bar z}$. 

If we make the
coordinates more manifest in the Wess-Zumino part, the WZW action reads
\kaderzonderrm{equation}{B:WZW}{
S_{\mbox{\scriptsize{WZW}}}= \int_M \mbox{Tr}\ d_z g^{-1} \wedge d_{\bar z} g
+\int_{t=0}^{t=1} dt\ \int_M \!\!\!\mbox{Tr}\ \  \tilde g^{-1}{ \p \tilde g \over \p t} 
\left\{ \tilde g^{-1} d_z \tilde g,\tilde g^{-1} d_{\bar z} \tilde g\right\}.
}
It is this form of the WZW-action that will give motivation for the structure of
Berkovits' action in the next section. 
\paragraph{The equation of motion}
It is well known but a bit tedious to verify that the equation of motion is 
\be\label{WZW:eom} d_{\bar z} \left( g^{-1} d_z g\right) =0. 
\ee
\paragraph{Gauge invariance}
The WZW-action is invariant under
$$g(z,\bar z) \to h_L(\bar z) g(z,\bar z) h_R (z).$$
The infinitesimal form of this gauge transformation is
\be\label{WZW:gt}
\delta g(z,\bar z) = \epsilon_L(\bar z) g(z,\bar z) +  
g(z,\bar z)\epsilon_R(z).
\ee 
Tracking down the derivation of these results one sees that one has used the
following properties.
\be\label{WZW:prop}
\left\{ \begin{array}[h]{ll}
1.& d 1 = 0\\
2.& d \mbox{ is nilpotent}\\
3.&d \mbox{ is a graded derivation:}\\
&\displaystyle{\qquad d(\omega \wedge \eta) = 
d\omega \wedge \eta+(-1)^{|\omega|}\omega \wedge d\eta,} \\
&\qquad \mbox{where } \displaystyle{|\omega| \mbox{ is the degree of } \omega.}\\
4.&\mbox{The integration is cyclically symmetric: }\\ 
&\qquad 
\displaystyle{\int \omega\wedge\eta 
=(-1)^{|\omega| | \eta|}\int \eta\wedge\omega}\\
5.&\mbox{The Stokes property holds:}\\  
&\displaystyle{\qquad \int d_z  \omega = 0} \mbox{  and } 
\displaystyle{\ \int d_{\bar z}  \omega = 0}\\
6.&\mbox{The wedge product is associative.}
\end{array}\right.
\ee
The strategy will then be to prove that the analogs of these properties hold in 
Berkovits' theory.
Then we can conclude that the analogs of the equation of motion\refpj{WZW:eom} 
and of the gauge invariance\refpj{WZW:gt} hold as well.
\index{WZW-action|)}
\section{Berkovits' action for a BPS $D$-brane}\label{s:BPS}
\index{string field theory!Berkovits' sft for a BPS $D$-brane|(}

We now discuss Berkovits' action for a BPS
$D$-brane~\cite{9503099,9912121}. We only treat the
Neveu-Schwarz part of the action\footnote{It is not yet known how to extend
this action to the Ramond sector in a ten-dimensional Lorentz-covariant manner.
However, it can be generalized to a four-dimensional super-Poincar\'e covariant
action which includes both the NS and R sectors, see for
example ref.~\cite{0105230}.} because we are mainly interested in the process of
tachyon condensation and this involves NS sector states only. All objects
encountered in the WZW-theory in the previous sector will have an analog in
Berkovits' theory. We will define these analogs in such a way that the structure
of the WZW-theory is completely preserved.
\subsection{From the WZW-action to Berkovits' action}
\begin{itemize}
\item
We take for the analogs of the two exterior derivations
$d_z$ and $d_{\bar z}$, the BRST-operator $Q$ and the zero-mode $\eta_0$
respectively, see the first line  in table~\ref{WZWtoB}.
This choice is motivated by the fact that these two operators 
satisfy the properties
$$Q^2 = 0, \quad \eta_0^2=0\quad\mbox{and}\quad \left\{Q,\eta_0\right\} =0,$$
which are the analogs of
$$d_z^2=0, \quad d_{\bar z}^2 =0\quad\mbox{and}\quad
\left\{ d_z, d_{\bar z} \right\} =0.$$
The nilpotent BRST operator is given by
$$
Q= \oint dz j_B(z)
= \oint dz \Bigl\{  c \bigl( T_m + T_{\eta\xi} + T_\phi)
+ c \partial c b +\eta \,e^\phi
\, G_m - \eta\p \eta e^{2\phi} b \Bigr\}\, ,
$$
where
$$
T_{\eta\xi}=\p\xi\,\eta\quad\mbox{and}\quad T_\phi=-{1\over 2} \p\phi \p
\phi -\p^2\phi \, ,
$$
$T_m$ is the matter stress tensor and 
$G_m$ is the matter supercurrent. 
The operator $\eta_0$ is the zero-mode of one of the superghosts, see
section~\ref{superstring}.

\begin{table}[h]
\begin{center}
\begin{tabular}{|l|p{5cm}|p{5cm}|}
\hline
& Wess-Zumino-Witten& Berkovits' string field theory\\
\hline
\hline
\STRUT
1&the exterior derivatives $d_z$ and $d_{\bar z}$ 
& the BRST-operator $Q$ and the zero-mode $\eta_0$\\
\hline
\STRUT
2 & a $(p,q)$-form & a state in the CFT with ghost number $p$ and 
picture number $-q$\\
\hline
\STRUT
3 & $X(z,\bar z)$ & a string field $\Phi$ with ghost number 0 and 
picture number 0\\
\hline
\STRUT
4 &the gauge parameters  $\epsilon_L(\bar z)$ and $\epsilon_R( z)$ 
& the states $\state{\Xi_L}$  and 
$\state{\Xi_R}$ with ghost number 0 and picture number 0 satisfying 
$Q \state{\Xi_L}=0$ and $\eta_0\state{\Xi_R}=0$\\
\hline
\end{tabular}\caption{The elements of WZW theory and 
their analogs in Berkovits' open \ssft\ on a BPS $D$-brane. The CFT contains
only the GSO(+) sector.\label{WZWtoB}}
\end{center}
\end{table}
\item
Acting on a $(p,q)$-form with $d_z$ or $d_{\bar z}$ gives a $(p+1,q)$-form
or a $(p,q+1)$-form respectively. The BRST-operator $Q$ has ghost number 1 and
picture number\footnote{See table~\ref{t:convgp} for our conventions about 
picture and ghost
number assignments.} 0 and the zero mode $\eta_0$ has ghost number~0 and 
picture number~$-1$.
Therefore we define the
analog of a $(p,q)$-form to be a state in the CFT with ghost number~$p$ and
picture number~$-q$, see the second line in table~\ref{WZWtoB}. 
In particular, the
analog of the matrix $X(z,\bar z)$ is a state with ghost number~0 
and picture number~0.
\item
The CFT we use is the combined
conformal field theory of a $c=15$ superconformal matter system, times an
appropriate superghost sector\footnote{See
appendix~\ref{superstring} for a review of these conformal field theories}.
At this moment we are discussing the string field action for a 
BPS $D$-brane. Therefore the states in the CFT we are using will always be in
the GSO$(+)$ sector of the theory. 
\item
As was the case in Witten's open \sft\ action (see section~\ref{s:SFT}), the
analog of the combination of integration and taking the trace will be the \sft\
interaction as defined in definition~\ref{dubbra}
$$ 
\Corr{A_1 \cdots A_n} = 
\corr{f_1^{n} \circ A_1(0)\  \cdots\  f_n^{n} \circ A_n(0)} .$$ 
Since we have, in general, non-integer weight vertex operators, we
should be more careful in defining $f\circ A$ for such vertex
operators.
Noting that 
$$
f^{n\prime}_k(0) = {4 i\over n} e^{2 \pi i{k-1 \over n}} \equiv {4\over
n}
e^{2\pi i ({k-1\over n} + {1\over 4})}\, ,
$$
we adopt the following definition of $f^{n}_k\circ\varphi(0)$ for a
primary
vertex operator $\varphi(x)$ of conformal weight $h$:
\be\label{ct:branch}
f^{n}_k\circ\varphi(0) = \bigg|\Big({4\over n}\Big)^h\bigg| e^{2\pi i h
({k-1 \over n}+{1\over
4})}\varphi(f^{n}_k(0))\, .
\ee
Since all secondary vertex operators can be obtained as products of
derivatives of primary vertex operators, this uniquely defines
$f^{n}_k\circ A(0)$ for all vertex operators. 
\item
For the superghost system we have to take the large
Hilbert space\index{large Hilbert space}~\cite{FMS},
i.e.~the Hilbert space that includes
the $\xi_0$ mode. This is forced upon us for the following reason. In the WZW
theory we have
$$\int_M\omega =0 \quad\mbox{if } \omega \mbox{ is not a } (1,1)-\mbox{form}.$$
The analog of this in Berkovits' theory reads
$$\Corr{\Psi} = 0 \quad\mbox{ if } \Psi \mbox{ has ghost number }
  \neq 1  \mbox{ or picture number}\neq -1.
$$
If we include the $\xi_0$ zero mode, this property holds,  indeed we have
$$
\corr{\xi(z) c\p c\p^2 c(f) e^{-2\phi(y)}} = 2.
$$
The left hand side of this equation has  
ghost number~1 and picture number~$-1$ as we wanted to be the case.
\item
The two chiral functions $\epsilon_L(\bar z)$ and $\epsilon_R( z)$ which appear
in the gauge transformation\refpj{WZW:gt} of the WZW action are characterized by
$$d_z \epsilon_L =0 \quad\mbox{and}\quad d_{\bar z} \epsilon_R =0.$$
Therefore the analogs of these gauge parameters $\epsilon_L$ 
and $\epsilon_R$ are the states $\state{\Xi_L}$ and $\state{\Xi_R}$
which have ghost number 0, picture number 0 and satisfy\footnote{The cohomology of the zero mode $\eta_0$ and of the
BRST-operator $Q$ is trivial~\cite{9912121}. Indeed we have $\{ \eta_0,\xi_0 \} = 1 $ and 
$\{Q, (\xi Y)_0 \} = 1$. Here $(\xi Y)_0$ is the zero-mode of the vertex
operator $\xi Y(z)$, where $Y = - \p\xi c e^{-2 \phi}$ 
is an inverse picture changing operator~\cite{FMS:Y}.  
Therefore, there exist states such that 
$\state{\Xi_L} = Q \state{\Omega_L}$ and 
$\state{\Xi_R} = \eta_0 \state{\Omega_R}$.
}
$$Q \state{\Xi_L}=0\quad\mbox{and}\quad\eta_0\state{\Xi_R}=0.$$
The gauge transformation motivated by\refpj{WZW:gt} reads
$$\delta e^{\Phi} = \Xi_L e^{\Phi} + e^{\Phi} \Xi_R.$$
Again for BPS-branes these states will be restricted to lie in the
GSO$(+)$ sector.
\end{itemize}
\subsection{Berkovits' action for a BPS $D$-brane}
Using the translation provided in the previous section, the Neveu-Schwarz part 
of Berkovits'
superstring field theory action~\cite{0001084} reads
\kaderzonderrm{equation}{B:action}{
S={1\over 2g^2}\BigCorr{ (e^{-\Phi} Q e^{\Phi}) 
(e^{-\Phi}\eta_0 e^\Phi)
- \int_0^1 dt 
(e^{-t\Phi}\p_t e^{t\Phi})\{ (e^{-t\Phi}Q e^{t\Phi}),
(e^{-t\Phi}\eta_0 e^{t\Phi})\}} 
}
This is the action Berkovits proposes as the correct off-shell description of
open superstring theory. It is easy to see that it has the correct on-shell
degrees of freedom. However, it is not known if this action gives rise to the
correct scattering amplitudes. Only the four point amplitude has been shown to
agree with perturbative superstring theory~\cite{9912120}. 
This is in sharp contrast
with Witten's open string field theory of the previous chapter. In the latter
case, it has been proven that Witten's field theory generates the correct
$S$-matrix.

An analysis analogous to that in ref.~\cite{Sen:univ} shows that the string field
action\refpj{B:action} describes a $D9$-brane with mass (we use $\ap=1$)
$$M = {1 \over 2 \pi^2 g^2}\ .$$
The details of this calculation can be found in appendix C of ref.~\cite{BSZ}.
\paragraph{Gauge invariance}
This action is invariant under the gauge transformation~\cite{0001084}
\be \label{B:gatr}
\delta e^{\Phi} = \Xi_L e^{\Phi} + e^{\Phi}\Xi_R \, . 
\ee
Here the gauge transformation parameters $\Xi_L$ and $\Xi_R$ 
are GSO$(+)$ vertex operators with ghost number and picture number values 0. 
The properties needed to show that this is a gauge invariance are
\be\label{B:prop}
\left\{ \begin{array}[h]{ll}
1.& Q I = 0 \quad\mbox{and}\quad \eta_0 I = 0 \\
2a.&Q \mbox{ and } \eta_0 \mbox{ are graded derivations:}\\
&\displaystyle{\qquad Q(A \star B) = 
Q A  \star B+(-1)^{|A|}A \star Q B,} \\
&\displaystyle{\qquad \eta_0(A \star B) = 
\eta_0 A  \star B+(-1)^{|A|}A \star \eta_0 B,} \\
2b.&\mbox{The Stokes property holds:}\\
& \Corr{ Q \cdot A } = 0 \quad\mbox{and}\quad \Corr{ \eta_0 \cdot A } = 0 
\quad\mbox{for all operators } A\\
3.&\mbox{String field theory interactions are 
cyclically symmetric:}\\ 
& \Corr{A\ B} = (-1)^{|A||B|} \Corr{B\ A} .\\
\end{array}\right.  
\ee
\paragraph{About the proof of these properties}
\begin{itemize}
\item[1.]
It is well known that there is an identity $I$ in the star-algebra. We will not
need the exact form of this identity in the rest of the thesis, it can be
found in the literature~\cite{GrossJev,conserv,Ellwood:I}. The 
proof\footnote{One can deduce that $Q I$ = $\eta_0 I = 0$ from the fact that the
operators $Q$ and $\eta_0$ are derivations of the star algebra. However, there
are some subtleties. As explained in ref.~\cite{conserv}, 
the string field $I$ is not an identity on all string
fields $A$. Therefore a direct proof  is necessary. 
For other anomalies see ref.~\cite{assanom}.} that the identity is
annihilated by the BRST operator $Q$ and by $\eta_0$ can for example be found
in ref.~\cite{conserv}. 
\item[2.]\label{toprove2}
These proofs are completely worked out in sections~\ref{s:ders} and
\ref{s:cons}. However it is
necessary to elaborate at this point on the minus sign in the second term of the
derivation property.\\
The integer~$|A|$ is the analog of the degree of a differential form. As
discussed above (see table~\ref{WZWtoB}) the analog of the degree is the ghost
number minus the picture number of a state. Is is easy to see that for states in
the GSO$(+)$ sector
$$(-1)^{\mbox{\scriptsize{ghost number}}(A) -
\mbox{\scriptsize{picture number}}(A)} = 
 (-1)^{\mbox{\scriptsize{Grassmann number}}(A)}\ .$$
The minus sign in the second term of the derivation property then originates
form pulling the Grassmann odd operator $Q$ through the field $A$.
\item[3.]
This proof is very similar to the proof of the cyclicity in 
Witten's \sft, see
section~\ref{s:proofs}. It can be found in appendix~A of ref.~\cite{BSZ}.
\end{itemize}
At the linearized level, the gauge invariance can be fixed by
imposing the Feynman-Siegel gauge
\be\label{B:FSgauge}\index{Siegel gauge!in Berkovits' sft}
b_0 \state{\Phi}=0\quad\mbox{and}\quad\xi_0 \state{\Phi}=0\ .
\ee
A proof that this is a valid gauge choice can be found in appendix~\ref{s:FS}.
\paragraph{The equation of motion}
The equation of motion reads
\be\label{B:eom}
\eta_0\left( e^{-\Phi} Q e^{\Phi} \right)=0\ . 
\ee
\index{string field theory!Berkovits' sft for a BPS $D$-brane|)}
\section{Berkovits' action for a non-BPS $D$-brane}\label{s:nonBPS}
\index{string field theory!Berkovits' sft for a non-BPS $D$-brane|(}
The open string states living on a single
non-BPS D-brane are  divided into two classes, GSO$(+)$ states
and GSO$(-)$ states. We want to preserve the algebraic structure reviewed in the
previous sector, in particular we still want 
$$\qquad \wh Q(\wh A \star \wh B) = 
\wh Q \wh A  \star \wh B+(-1)^{|\wh A|}\wh A \star \wh Q \wh B,$$
where $\wh A$ and $\wh B$ are now states in a CFT containing both the GSO$(+)$
and GSO$(-)$ sector. As mentioned in the previous section, the sign in the
second term of this derivation property originates from pulling the Grassmann
odd operator $\wh Q$ through the field $\wh A$. Is is easy to verify however
that
$$(-1)^{\mbox{\scriptsize{Grassmann}}(\wh A)}= 
 (-1)^{F(\wh A) + |\wh A|}\ .$$
Here $F( \wh A)$ is the world sheet fermion number of $\wh A$ and $|\wh A|$ is
the ghost number of $\wh A$ minus the picture number of $\wh A$. 
Therefore, for states $\wh A$ in the GSO$(-)$ sector, we will have to make sure
that there is an additional sign when moving $\wh Q$ through $\wh A$. This is
done by tensoring with $2\times 2$ matrices 
carrying internal Chan-Paton (CP) indices.
These are added\footnote{Note that this definition shows that these matrices do
not really carry conventional CP indices;  had it been 
so, both
$Q$ and $\eta_0$ should have been tensored with the
identity matrix, as such operators should not change the sector
the strings live in~\cite{9705038}.} both to the vertex operators and
to $Q$ and $\eta_0$.

\kaderzonderrm{eqnarray}{}{ \nonumber \\[-6ex]
\wh Q &=& Q \otimes \sigma_3 
\quad\mbox{and}\quad \wh \eta_0 = \eta_0\otimes \sigma_3\ , \nonumber\\[1ex]
\wh A &=& A_{+} \otimes I + A_{-} \otimes\sigma_1
\quad\mbox{ if } |\wh A| \mbox{ is even}\ , \label{hats}\\[1ex]
\wh A &=&  A_{+} \otimes\sigma_3 
 + A_{-} \otimes i \sigma_2
\quad\mbox{ if } |\wh A| \mbox{ is odd}\ .\nonumber
}
Here $I$ is the $2 \times 2$ identity matrix and $\sigma_1,\sigma_2,\sigma_3$
are the three Pauli matrices. The Chan-Paton factors above are chosen in such a
way that 
\begin{itemize}
\item
Pulling $\wh Q$ through $\wh A$ will produce no minus sign if
$|\wh A|$ is even\footnote{Indeed, pulling 
$Q$ through $A_{-}$ will give a minus sign because $A_{-}$
is Grassmann odd, this minus sign is compensated by commutating $\sigma_3$ and
$\sigma_1$.
}.
On the other hand, pulling $\wh Q$ through $\wh A$ does produce a minus sign if
$|\wh A|$ is odd.
\item
The Chan-Paton factor of $\wh Q \wh A$ is consistent 
with the choice\refpj{hats}.
\item
The Chan-Paton factor of $\wh A \star \wh B$ is consistent with the 
choice\refpj{hats}.
\end{itemize}
Finally, we define
\be \label{e200}
\Corr{ \wh A_1\ldots \wh A_n } = \mbox{Tr}  \corr{f^{n}_1 \circ
\ha_1(0)\cdots
f^{(n)}_n\circ \ha_n(0)},
\ee
where the trace is over the internal CP matrices.
We shall adopt the convention that fields or operators with internal
CP factors
included are denoted by symbols with a
hat on them, and fields or operators without internal
CP factors included are
denoted by symbols without a hat, as in the previous sections. 

The string field  action for the non-BPS D-brane takes the  
same structural form as that in\refpj{B:action} and is given 
by~\cite{0001084}
\kaderzonderrm{equation}{B:ACTION}{
S={1\over 4g^2} \BigCorr{ (e^{-\hp} \hq e^{\hp})(e^{-\hp}\he e^\hp) -
\int_0^1 dt (e^{-t\hp}\p_t e^{t\hp})\{ (e^{-t\hp}\hq e^{t\hp}),
(e^{-t\hp}\he e^{t\hp})\}}  
}
Here we have divided the overall normalization by a factor
of two in order to compensate for the trace operation on the
internal matrices.
By the judicious choice of Chan-Paton factors, this action is invariant under 
the gauge transformation~\cite{0001084}
\be \label{egtrsa}
\delta e^{\hp} = \wh\Xi_L\ e^{\hp} + e^{\hp}\ \wh\Xi_R \, , 
\ee
where, as before the gauge transformation parameters $\wh\Xi_L$ and
$\wh\Xi_R$ have ghost number~0 and picture number~0. 
The proof of gauge
invariance is formally identical to the one given in appendix~\ref{s:FS}. 
The equation of motion is just\refpj{B:eom} with hats on fields and operators
$$
\wh\eta_0\left( e^{-\wh\Phi} \wh Q e^{\wh\Phi} \right)=0\ . 
$$
\index{string field theory!Berkovits' sft for a non-BPS $D$-brane|)}

Let us recapitulate what we have done so far. We have written down two actions,
one describing the BPS $D9$-brane and one the non-BPS
$D9$-brane. We have derived the equations of motion and
elaborated upon the gauge invariance. In the next sections we will use
Berkovits' action for the non-BPS $D9$-brane to calculate the tachyon potential.
We will use this potential to verify Sen's first conjecture approximately.
\section{The tachyon potential at level $(2,4)$}\label{s:level4}
In this section we will give the tachyon potential\footnote{The level~$(0,0)$
calculation was first done by Berkovits~\cite{0001084}, the level~$(3/2,3)$
calculation by Berkovits, Sen and Zwiebach~\cite{BSZ}. The level~$(2,4)$
calculation was performed by the author in collaboration with Joris
Raeymaekers~\cite{0003220}.} up to level $(2,4)$. This
notation means that we have fields up to level~2 and keep terms up to 
level~4 in the potential. The level of a field is just the conformal 
weight shifted by~$1/2$. In this way the tachyon is a level~0 field. 
The level of a monomial in the potential is defined to be the sum of
the levels of the fields entering into it.
The calculation of the tachyon potential consists of several parts:
\begin{enumerate}
\item
Write down the vertex operators in the string field up to level~2. The string
field is gauge fixed by imposing the Feynman-Siegel gauge, see\refpj{B:FSgauge}.
As was the case in Witten's open bosonic string field theory, see
section~\ref{s:twist}, the string field action has a $\IZ_2$ twist under which
string field components associated with a vertex operator of dimension~$h$ carry
charge $(-1)^{h+1}$ for even $2 h$, and $(-1)^{h+1/2}$ for odd $2 h$. 
We refer the reader to appendix~B of ref.~\cite{BSZ} for a proof of this twist 
symmetry. The tachyon vertex operator, having dimension $-1/2$, is even under
this twist transformation. Thus we can restrict the string field to be twist
even.
A list of contributing states up to level~2 is provided in 
table~\ref{t:Berstates}.
\item
Compute the finite \cts\ of these vertex operators. 
A list of these \cts\ and detailed information about
their derivation can be found in appendix~\ref{s:transf}.
\item
Compute the necessary CFT correlations. This is usually done by performing Wick
contractions. There are many of these correlators to be computed, but this 
can be done with the aid of a machine. For
this part we have written a Mathematica code in collaboration with Joris
Raeymaekers. 
\end{enumerate}
\begin{table}[h]
\begin{center}
\begin{tabular}{|l|p{4.5cm}|l|l|p{0.5cm}|}\hline
Level\STRUT
     &states
     & vertex operators &GSO& $\mathbb{Z}_2$\\
\hline 
\hline    
0\STRUT &  $\xi_0 c_1\state{-1}$&$ T = \xi c e^{-\phi}$& $-$&$+$\\ 
\hline
1/2\STRUT&$\xi_0\xi_{-1}c_0 c_1 \state{-2}$
  &$ \xi\p\xi \p c\  c e^{-2 \phi}$& +&$-$\\ 
\hline  
1\STRUT&$\xi_0 c_1 \phi_{-1}\state{-1}$
  &$ \xi c\  \p\phi e^{-\phi}$& $-$&$-$\\ 
\hline
3/2\STRUT &$ 2 c_1 c_{-1} \xi_0\xi_{-1} \state{-2}$
            &$A = c\partial^2 c\xi\partial\xi\ e^{-2\phi}$& & \\
    & $ \xi_0 \eta_{-1}  |0\rangle$&
    $E =  \xi\eta$ &+&+\\
   & $  \xi_0 c_1 G^m_{-3/2}  |-1\rangle$
    &$F = \xi c
    G^m\ e^{-\phi}$& & \\ 
\hline
2\STRUT &  $ \xi_0 c_1 \left[(\phi_{-1})^2-\phi_{-2}\right]|-1\rangle$
  &$ K = \xi c \ \partial^2\left(e^{-\phi}\right)$ & &\\
  &$ \xi_0 c_1 \phi_{-2} |-1\rangle$
  &$ L = \xi c\ \partial^2\phi\ e^{-\phi}$ & & \\
 & $\xi_0 c_1 L^m_{-2}  |-1\rangle$
 &$ M =\xi c T^m\ e^{-\phi}$& $-$&+\\
 &$2 \xi_0 c_{-1}   |-1\rangle$
 & $N = \xi \partial^2 c\ e^{-\phi}$& & \\
 &$\xi_0 \xi_{-1}\eta_{-1}c_1  |-1\rangle$&$ P = \xi\partial\xi\eta c\ 
e^{-\phi}$& & \\\hline
\end{tabular}
\caption{The states up to level two.
We
use the notation $|q\rangle$ for the state corresponding with the
operator $e^{q \phi}$ (see appendix~\protect\ref{superstring}).
The contributing states are the ones with $\IZ_2$-charge $+1$. States in the
GSO$(+)$ and GSO$(-)$ sector are to be tensored with $I$ and $\sigma_1$
respectively, see equation\refpj{hats}.
\label{t:Berstates}
}

\end{center}
\end{table} 
Taking all this together, the string field $\widehat\Phi$ up to level~2 reads
$$
\widehat\Phi = t \widehat T + a \widehat A +e \widehat E + f \widehat
F+ k \widehat K + l \widehat L +m \widehat M+n \widehat N + p \widehat P.$$
In the state formalism this is
\bqs
\state{\Phi}&=& A_{0,1}\ \xi_0 c_1\state{-1}\\ 
& & + A_{3/2,1}\ \xi_{-1}\xi_0 c_{-1}c_1\state{-2}
+A_{3/2,2}\  G^{m}_{-3/2}\xi_0 c_1\state{-1}
+A_{3/2,3}\  \eta_{-1}\xi_0 \state{0}\\
& & 
+ A_{2,1}\ \xi_0 c_{-1}\state{-1}+A_{2,2}\ L^{m}_{-2}\xi_0c_1\state{-1}\\
& & +A_{2,3}\ \eta_{-1}\xi_{-1}\xi_0c_1\state{-1}
+A_{2,4}\ \xi_0c_1\phi_{-2}\state{-1}
+A_{2,5}\ \xi_0c_1\phi_{-1}\phi_{-1}\state{-1}\ .
\eqs
The relations between the coefficients $A_{k,l}$ and $t,a,e, \ldots$ can easily
be read of from table~\ref{t:Berstates}.
We give the 
potential with 
coefficients evaluated\footnote{In ref.~\cite{0003220} we had listed the
potential as a function of the variables ${t,a,e,f,k, \ldots}$. 
We have changed this because we will compute higher levels 
in section~\ref{s:high}. 
There we will use a different calculation method 
which is naturally based on the state-formalism
instead of the vertex operator formalism. The
potential at level $(2,4)$ has thus been calculated 
using two different methods. Both methods give the same result. } 
numerically up to 6 significant digits (subscripts refer to the level).
\bqs
V(\widehat\Phi )&=& -S(\widehat\Phi)\\
&=& 2 \pi^2 M(V_0 + V_{3/2} + V_2 + V_3 + V_{7/2} + V_4)
\eqs
\bqs
\lefteqn{V(\widehat\Phi)_0=
-0.25A_{0,1}^2 + 0.5A_{0,1}^4}\\[4pt]
\lefteqn{V(\widehat\Phi)_{3/2}=
-0.5A_{ 3/2,1}A_{ 0,1}^2 + 
  0.25A_{ 3/2,3}A_{ 0,1}^2 + 
  0.518729A_{ 3/2,3}A_{ 0,1}^4}\\[4pt]
\lefteqn{V(\widehat\Phi)_{2}=
1.41667A_{2,1}A_{0,1}^3 - 
  3.75A_{2,2}A_{0,1}^3
  - 0.25A_{2,3}A_{0,1}^3}\\  
  &&+1.83333A_{2,4}A_{0,1}^3
  + 2.16667A_{2,5}A_{0,1}^3\\[4pt]
\lefteqn{V(\widehat\Phi)_{3}=
-1.A_{ 3/2,1}A_{ 3/2,3} - 
  2.48202A_{ 3/2,1}A_{ 3/2,3}A_{ 0,1}^2}\\ 
  &&-5.47589A_{ 3/2,2}A_{ 3/2,3}A_{ 0,1}^2 
 + 5.A_{ 3/2,2}^2 + 
  5.82107A_{ 0,1}^2A_{ 3/2,2}^2 \\
  & &- 0.66544A_{ 0,1}^2A_{ 3/2,3}^2 + 
  0.277778A_{ 0,1}^4A_{ 3/2,3}^2\\[4pt]
\lefteqn{V(\widehat\Phi)_{7/2}=
-0.407407A_{ 0,1}A_{ 3/2,1}A_{ 2,1} + 
  0.648148A_{ 0,1}A_{ 3/2,3}A_{ 2,1}}\\[4pt]&&+ 
  1.38889A_{ 0,1}A_{ 3/2,1}A_{ 2,2}  - 
  8.88889A_{ 0,1}A_{ 3/2,2}A_{ 2,2}\ - 
  0.694444A_{ 0,1}A_{ 3/2,3}A_{ 2,2}\\[4pt]&& + 
  0.777778A_{ 0,1}A_{ 3/2,1}A_{ 2,3} + 
  2.96296A_{ 0,1}A_{ 3/2,2}A_{ 2,3} + 
  0.944444A_{ 0,1}A_{ 3/2,3}A_{ 2,3}\\[4pt]&& - 
  3.55556A_{ 0,1}A_{ 3/2,1}A_{ 2,4}  - 
  11.8519A_{ 0,1}A_{ 3/2,2}A_{ 2,4} + 
  0.296296A_{ 0,1}A_{ 3/2,3}A_{ 2,4}\\[4pt]&& - 
  5.07407A_{ 0,1}A_{ 3/2,1}A_{ 2,5} - 
  11.8519A_{ 0,1}A_{ 3/2,2}A_{ 2,5} + 
  0.166667A_{ 0,1}A_{ 3/2,3}A_{ 2,5}\\[4pt]&& + 
  2.38682A_{ 3/2,3}A_{ 2,1}A_{ 0,1}^3  - 
  4.35732A_{ 3/2,3}A_{ 2,2}A_{ 0,1}^3+ 
  0.605194A_{ 3/2,3}A_{ 2,3}A_{ 0,1}^3\\[4pt]&& + 
  1.94348A_{ 3/2,3}A_{ 2,4}A_{ 0,1}^3+ 
  4.81648A_{ 3/2,3}A_{ 2,5}A_{ 0,1}^3\\[6pt]
\lefteqn{V(\widehat\Phi)_{4}=
-7.96875A_{2,1}A_{2,2}A_{0,1}^2 + 
  0.25A_{2,1}A_{2,3}A_{0,1}^2 }\\[4pt]&&+ 
  1.40625A_{2,2}A_{2,3}A_{0,1}^2 + 
  5.95833A_{2,1}A_{2,4}A_{0,1}^2- 
  10.3125A_{2,2}A_{2,4}A_{0,1}^2 \\[4pt]&&+ 
  0.875A_{2,3}A_{2,4}A_{0,1}^2+ 
  8.72917A_{2,1}A_{2,5}A_{0,1}^2 - 
  12.1875A_{2,2}A_{2,5}A_{0,1}^2 \\[4pt]&&+ 
  2.3125A_{2,3}A_{2,5}A_{0,1}^2+ 
  14.2083A_{2,4}A_{2,5}A_{0,1}^2 - 
  0.75A_{2,1}^2 \\[4pt]&&+ 
  1.45833A_{0,1}^2A_{2,1}^2+ 
  5.625A_{2,2}^2 + 
  14.7656A_{0,1}^2A_{2,2}^2 \\[4pt]&&- 
  0.75A_{2,3}^2 - 
  1.5A_{0,1}^2A_{2,3}^2 + 
  1.5A_{2,4}^2 \\[4pt]&&+
  6.70833A_{0,1}^2A_{2,4}^2 + 
  1.5A_{2,5}^2 + 
  17.8958A_{0,1}^2A_{2,5}^2\\
\eqs
After having calculated the potential, one can look for a stable extremum.
\subsection*{The stable vacuum at level $(0, 0)$}
At this level we keep only the field $T$ and find
$$V_0(\hp) = -0.25\ A_{0,1}^2 + 0.5\ A_{0,1}^4.$$
At the minimum 
$A_{0,1} = 0.5$ we have  $V = -0.616850 M.$
Hence, at level~0 we get $62\%$ of the conjectured value $V = -M$. 
\subsection*{The stable vacuum at level $(3/2, 3)$}
At this level we keep only the fields $T, A, E$ and $F$ and truncate the
interaction at level 3. At the extremum
$$ A_{0,1} = 0.58882 ,\ A_{3/2,1} = 0.112726,\ 
A_{3/2,2} = -0.0126028,\ A_{3/2,3}=-0.0931749, $$
we have $V = - 0.854458 M$. 
This is around 85\% of the conjectured value. 
\subsection*{The stable vacuum at level $(2, 4)$}
\begin{figure}
\begin{center}
\epsfbox{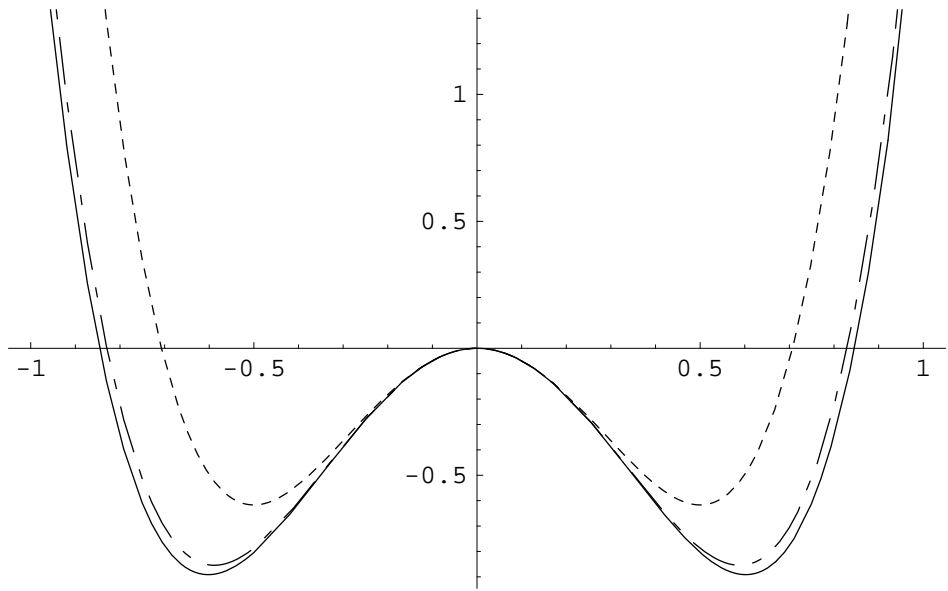}
\caption{In the potential computed up to
level~$4$ all the fields but $t$ appear only quadratically.  They can
be eliminated to give the effective potential $V(t)/M$ 
at level 0 (dotted line), 
level 3 (dashed line)  and level 4 (full line).}
\end{center}
\end{figure}
The potential has an extremum at which 
$$
\begin{array}{rclrclrcl}
A_{0,1}&=& 0.602101\, ,\\
A_{3/2,1}&=&0.104387\, ,&  A_{3/2,2} &=& -0.0138164\,,
& A_{3/2,3} &=& -0.0430366\,, \\
A_{2,1} &=& 0.0946226\,, & A_{2,2} &=&  0.0322127\,,
 & A_{2,3} &=& -0.021291\,,\\ 
A_{2,4} &=& -0.0348532\,, & A_{2,5} &=& -0.01019\,.
\end{array}
$$
At this extremum $V = -0.891287 \, M$.
Hence at the level $4$ we get $89 \%$ of the conjectured value $V= - M$.
\subsection*{Comparison with the results in ref.~\cite{0004015}}
The tachyon potential up to level~$(2,4)$ has also been calculated
in ref.~\cite{0004015}. However, there are some discrepancies between our results 
and the one in ref.~\cite{0004015} on which we will now comment. 
The potential in ref.~\cite{0004015}
is given as a function of 9~variables $t, a,e,f,v_1,v_2, \ldots,v_5$ which are
related to our variables $A_{k,l}$ through
$$
\begin{array}{lll}
t = A_{0,1} & a = {1 \over 2} A_{3/2,1} & e =-A_{3/2,3}\\
f = -A_{3/2,2}& v_1 = A_{2,2}& v_2 = {1 \over 2} A_{2,1}\\
v_3 = - A_{2,3}& v_4 = -2  A_{2,5}& v_5 = A_{2,4}-2 A_{2,5}.
\end{array}
$$ 
We will now make a detailed comparison of the results. Notice that
in ref.~\cite{0004015} the action~$S$ is written down whereas we have listed the
potential~$V$.
\begin{itemize}
\item{$S_0$:\ }
Both results agree.
\item{$S_{3/2}$:\ }
The three monomials in ref.~\cite{0004015} should all 
be multiplied with $(-1)$ to be
in correspondence with our result and the one in ref.~\cite{BSZ}.
\item{$S_2$:\ }
Both results agree.
\item{$S_{3}$:\ }
The 7 monomials in ref.~\cite{0004015} should all be multiplied 
with $(-1)$ to be
in correspondence with our result and the one in ref.~\cite{BSZ}.
\item{$S_{7/2}$:\ }
According to our calculation, 5 out of the 19 monomials of $S_{7/2}$ are
incorrect. Specifically, $44/27\ t a v_3$ 
should
be $14/9\ t a v_3$. There is also a term  $44/27\ t a v_2$ which is correct, so this
is perhaps a typing mistake.
More bizarre, the 5~terms having numerical coefficients, i.e.~the coefficients of
$t^3 e v_{\alpha}$, all have the wrong sign.
\item{$S_{4}$:\ } Out of the 21 monomials, 2 are wrong. $5/4\ t^2 v_3^2$ should
be $3/2\ t^2 v_3^2$ and $19/48\ t^2 v_2 v_3 $ should
be $1/2\ t^2 v_2 v_3$.
\end{itemize}
We think that our results are correct for two reasons. First of all, we have
done the calculation with the help of a computer program. Once the program is
correct, there is no room for mistakes. This is in contrast with the authors
of ref.~\cite{0004015} who did the calculation by hand. It is highly likely that the
program is correct because it reproduces correctly 
the potential at level~$({3 \over 2},3)$ which was calculated in ref.~\cite{BSZ}.
Secondly, as explained in sections~\ref{s:cons} and~\ref{s:high}, we also
calculated the potential with a completely different method. Hence, we have
obtained the level~$(2,4)$ potential with two different methods.
Both methods give the same result.
\subsection*{Higher levels}
We would like to establish agreement
beyond a shadow of doubt, as has been done for the bosonic
string~\cite{0002237}, by including more fields. Apart from this,
it is also important to have
accurate information about the tachyon condensate itself. Indeed, we would like
to study fluctuations around this stable vacuum to see if and how closed strings
are described in Berkovits' open string field theory.

The calculational scheme used in the previous section, however, is not suited to
do higher level computations. Indeed, finite \cts\ are complicated for
non-primary fields and their computation becomes very cumbersome at higher
levels. In combination with the 
exponential\footnote{It is easy to write down the partition function for
the number of components $N_n$ of the string field at every 
level~$n$ (see appendix~\ref{s:partition}):
\bqs
&& \sum_{n \ge 0} N_n q^n = 1 + q + 3\ q^{3/2} + 5\ q^2 + 8\ q^{5/2}
 + 13\ q^3 + 
    23\ q^{7/2} + 38\ q^4
      + 58\ q^{9/2}\\ &&+ 89\ q^5 + 
    141\ q^{11/2} + 216\ q^6 + 318\ q^{13/2} + 470\ q^7 + 
    699\ q^{15/2} + 1019\ q^8 + O(q^{9/2})\ .
\eqs}
 growth of the number of fields at higher 
levels, it
is clear that we should look for an alternative calculation method. 

Conservation laws provide an elegant and efficient alternative. In the
next section we will develop these identities for the superstring field theory 
following ref.~\cite{conserv} where this method was 
developed in the bosonic case.
\section{Conservation laws}\label{s:cons}
\index{conservation laws!theory|(}
\subsection[for primary fields]{Conservation laws for primary fields}
\begin{figure}[ht]
\begin{center}
\epsfxsize=10cm
\epsfysize=10cm
\begin{psfrags}
\psfrag{P1}[][]{$P_1$}
\psfrag{P2}[][]{$P_2$}
\psfrag{P3}[][]{$P_3$}
\psfrag{P4}[][]{$P_4$}
\psfrag{P5}[][]{$P_5$}
\psfrag{P6}[][]{$P_6$}
\psfrag{C1}[][]{$\mathcal{C}_1$}
\psfrag{C2}[][]{$\mathcal{C}_2$}
\psfrag{C3}[][]{$\mathcal{C}_3$}
\psfrag{C4}[][]{$\mathcal{C}_4$}
\psfrag{C5}[][]{$\mathcal{C}_5$}
\psfrag{C6}[][]{$\mathcal{C}_6$}
\psfrag{CG}[][]{$\mathcal{C}$}
\epsfbox{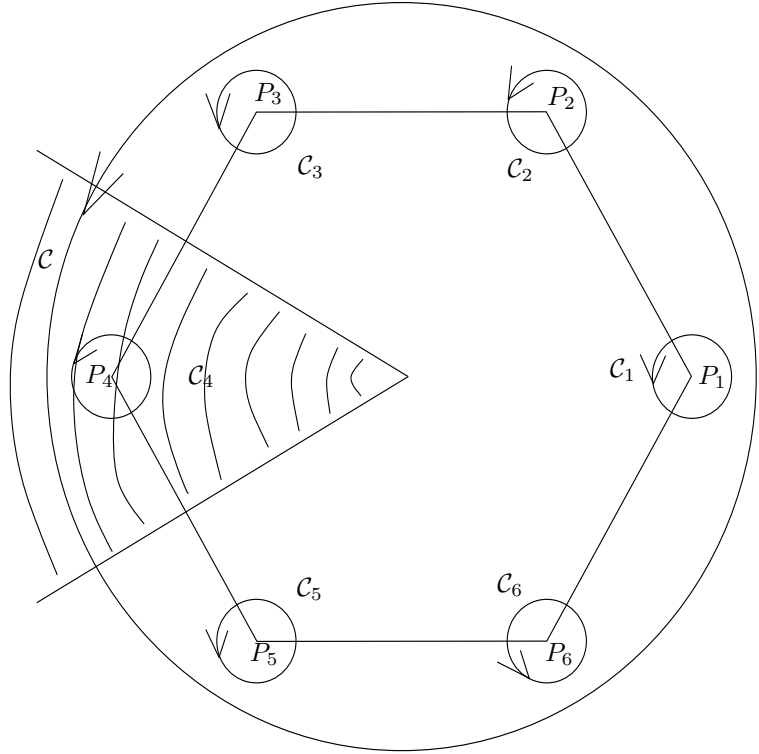}
\end{psfrags} 
\caption{Conservation laws for the 6~string interaction. The
contour $\mathcal{C}$ can be deformed into the sum of 6 contours  
$\mathcal{C}_i$ around the 6 punctures $P_i = f^6_i(0)$. As an illustration we
have also drawn the patch of the local coordinate~$z_4$.}
\label{f:cons}
\end{center}
\end{figure}

Suppose we have a conformal field $\mathcal{O}$ of weight $h$. If $\varphi(z)$
is a density of conformal weight $1-h$, then  $\varphi(z)\mathcal{O}(z) dz$
transforms as a
1-form, so it can be naturally integrated along 1-cycles. 
For this 1-form to
be regular at infinity we need
\be\label{cons:reg}
\lim_{z\to\infty} z^{2-2 h} \varphi(z) < \infty.
\ee 
Thanks to the holomorphicity of $\varphi(z)$, integration
contours in the $z$-plane can be continuously deformed as long as we do not
cross a singularity. Consider a contour $\mathcal{C}$ which encircles 
the $n$~punctures at $f_1^n(0), \ldots,f_n^n(0)$, see figure~\ref{f:cons}. 
For arbitrary vertex operators $\Phi_i$, the correlator
\be\label{cons:eq1}
\corr{\oint_{\mathcal{C}} \varphi(z)\pO(z)dz 
\ f_1^n\circ\Phi_1(0)\cdots f_n^n\circ\Phi_n(0)}
\ee
vanishes identically by shrinking the contour $\mathcal{C}$ to zero size around the point
at infinity. 
Since\refpj{cons:eq1} is zero for
arbitrary $\Phi_i$, we can write
$$
\etats{V_n}\oint_{\mathcal{C}}\varphi(z)\pO(z)dz =0.
$$
Deforming the contour $\mathcal{C}$ into the sum of $n$ contours 
$\mathcal{C}_i$ around the $n$~punctures, and referring the 1-form to the local coordinates $z_i = f^n_i(z)$, we
obtain the basic relation
\kaderzonderrm{equation}{cons:eq2}{
\etats{V_n}\sum_{i=1}^n\oint \varphi^{(i)}(z_i)\mathcal{O}(z_i)dz_i =0,
}
where $\varphi^{(i)}(z_i) = \varphi(f^n_i (z_i) ) \left(
f^{n\ \prime}_i(z_i)\right)^{1-h}$. 
Since $\varphi(z)$ will not necessarily have integer weight, it is necessary to
be more careful about its \ct. We again adopt the
convention\refpj{ct:branch}:
$$
f^{n}_k\circ\varphi(0) = \left({4\over n}\right)^{1-h}e^{2\pi i (1-h)
({k-1 \over n}+{1\over 4})}\ \varphi(f^{n}_k(0))\, .
$$
In this way $\varphi\pO(z)$ transforms as a 1-form:
\bqs
f\circ\ \varphi\pO (0) &=& \left({4\over n}\right) e^{2\pi i 
({k-1 \over n}+{1\over 4})}\ (\varphi \pO) (f^{n}_k(0))\\
&=& f^{n\prime}_k(0)\ (\varphi \pO)  (f^{n}_k(0)). 
\eqs
Let us take a conformal field $\pO$ of weight~1 as a special case.
Taking $\varphi(z)$ constant, we find $\etats{V_n}\pO_0=0$. In particular, we
have
$$\etats{V_n}Q = 0 \qquad\mbox{and}\qquad\etats{V_n}\eta_0\ .$$
This proves the Stokes property in\refpj{B:prop}.
As a generic example we can take 
$$\varphi(z) = (z-1)^{-k+h-1}\ .$$
This gives rise to a conservation law of the form
\bq
\lefteqn{\etats{V_n} \left( a^{(1)}_{-k}\mathcal{O}_{-k}^{(1)} + 
a_{-k+1}^{(1)}\mathcal{O}_{-k+1}^{(1)}+\cdots
\right)+}\nonumber\\
& &
\sum_{i=2}^n\etats{V_n} \left( a_{1-h}^{(i)}\mathcal{O}_{1-h}^{(i)} + 
a_{2-h}^{(i)}\mathcal{O}_{2-h}^{(i)}+\cdots \right)\ =0\label{cons:primaries}\, ,
\eq
where the coefficients $a^{(j)}_l$ are complex numbers.
We have to take $-k \le h-1$ to satisfy the regularity 
condition\refpj{cons:reg}. Some examples of conservation laws involving the
ghost field~$\eta$ and the super stress tensor~$G$ can be found in the
appendices~\ref{s:conseta} and~\ref{s:consG}.
\subsection[for the stress tensor]{Conservation laws for the stress tensor}
The stress energy tensor transforms as
\be\label{cons:trT}
f\circ T (z) = (f')^2 T(f) + {c \over 12} S(f,z),
\ee
where the Schwarzian derivative is defined by
$$ S(f,z) = {f''' \over f'} - {3 \over 2} \left({f'' \over f'}\right)^2.$$
Let $\varphi(z)$ be a holomorphic vector field, thus $\varphi(z)$ transforms as
$f\circ\varphi = (f')^{-1} \varphi(f)$. For $\varphi$ to be regular at infinity,
$\lim_{z\to\infty} z^{-2} \varphi(z)$ must be a constant (including zero). The
product $\varphi(z) T(z) dz$ transforms as a 1-form (except for a correction due
to the central term):
$$
f\circ T(z)\varphi dz = \left(\tilde T(f) - {c \over 12} S(f,z)\right)
\varphi(z) dz, $$
and can be integrated along 1-cycles. We consider a contour which encircles all
punctures and shrink it to zero around the point at infinity. At this point it is
important that under inversion the Schwarzian derivative vanishes and thus there
is no contribution from the central term in\refpj{cons:trT}. So we find
$$ \etats{V_n} \sum_{i=1}^n \oint dz_i\ \varphi^{(i)} 
\left[T(z_i) - {c \over 12} S(f_i (z_i),z_i) \right]=0.$$
Some examples of conservation laws for the stress tensor are given in
appendix~\ref{s:consT}.
\subsection[for the current $\p\phi$]
{Conservation laws for the current $\p\phi$}
The field $\p\phi$ has weight~1 and its \ct\ is 
$f\circ\p\phi = f' \p\phi(f) - f''/f'$. Let $\varphi(z)$ be a scalar.
The product $\varphi(z)\p\phi(z) dz$ transforms as a 1-form, except for a
correction due to the anomalous term:
$$f\circ \p\phi(z) = \left( \p\phi(f) + {f''\over f'}\right) \varphi(z) dz.$$
This time $\etats{V_n} \oint_\mathcal{C} dz \p\phi(z) \varphi(z)$ is not zero,
since to shrink the contour around infinity we need to perform the
transformation $z\to -1/z$, and the current is not covariant under this. One
finds instead
$$
\etats{V_n} \oint_{\mathcal{C}} dz\ \p\phi(z)\varphi(z) dz 
-2 \oint _{z=0}{\varphi(-1/z) \over z} dz = 0 $$
Here the second integral is evaluated along a contour that goes around the
origin counter clock wise. We can now deform the contour 
$\mathcal{C}$ into the
sum of $n$ contours $\mathcal{C}_i$ around the $n$ punctures and pass to the
local coordinates. The conservation law reads
$$
\etats{V_n}\left( \sum_{i=1}^n \oint\left[\p\phi(z_i) + {f''_i(z_i) \over
f'_i(z_i)}\right] \varphi(z_i) dz_i  
-2 \oint _{z=0}{\varphi(-1/z) \over z} dz \right)= 0\ . $$
We refer to ref.~\cite{conserv} for some concrete examples.
\index{conservation laws!theory|)}
\subsection[and string field interactions]
{Calculation of string field interactions using conservation laws}
We will now give some illustrations of the use of these conservation laws for
the calculation of \sft\ interactions. More information of the notation used can
be found in appendix~\ref{CFT}.
\paragraph{Example 1. \ $\Corr{L_{-2},L_{-3},1,1}$}\ \\
Making use of equation\refpj{C:eq2} we can trade the oscillator $L_{-3}$ for a
sum of annihilation operators
\bqs
\Corr{L_{-2},L_{-3},1,1} &=& -{7 \over 8} \Corr{L_{-1}L_{-2},1,1,1}+
{13 \over8}\Corr{L_{0}L_{-2},1,1,1}\\
&& -{11 \over8}\Corr{L_{1}L_{-2},1,1,1}+{5 \over16}\Corr{L_{2}L_{-2},1,1,1}
\eqs
Commuting these annihilation operators to the right, this is equal to
$$-{7 \over 8} \Corr{L_{-3},1,1,1}+
{13 \over4}\Corr{L_{-2},1,1,1}+{5 c\over32}\Corr{1,1,1,1}$$
The first term is zero as can be seen by using\refpj{C:eq2} to remove $L_{-3}$.
For the second term, we use\refpj{C:eq1} to find
$$\Corr{L_{-2},1,1,1}= -{c \over 8}\Corr{1,1,1,1}= -{c \over 8}$$
Therefore the final result is
$$-{13 \over 4}{c \over 8} + { 5 c \over 32} = -{c \over 4}$$
\paragraph{Example 2. \ $\Corr{L_{-2},L_{-2},1,1}$}\ \\
In a similar way as in the previous example one can calculate that this
interaction equals
$${c^2 \over 64} + {c \over 8}$$
\paragraph{Example 3\protect\footnote{If one wants to
calculate the result of this example with vertex operators, one has to
calculate the finite conformal transformation $f \circ : \p T T:$. This is
already a bit tedious.}. \ $\Corr{L_{-3}L_{-2},L_{-2},1,1}$}
Using the conservation law\refpj{C:eq2} we find
\bqs
\corr{L_{-3}L_{-2}, \ L_{-2},  \ 1,  \ 1} &=& -{7 \over 8}
  \corr{L_{-2}, \ L_{-1}L_{-2},  \ 1,  \ 1}
  -{13 \over 8}\corr{L_{-2}, \ L_0 L_{-2},  \ 1,  \ 1}\\
& &   -{11 \over 8}\corr{L_{-2}, \ L_1 L_{-2},  \ 1,  \ 1}
  -{5 \over 16}\corr{L_{-2}, \ L_2 L_{-2},  \ 1,  \ 1}\\
   &=& -{7 \over 8}
  \corr{L_{-2}, \ L_{-3},  \ 1,  \ 1}
  -{13 \over 4}\corr{L_{-2}, L_{-2},  \ 1,  \ 1}\\
& &  
  -{5 c\over 32}\corr{L_{-2}, 1,  \ 1,  \ 1}\\
\eqs
These terms have been calculated in the previous examples and we find the final
result $ -3\ c /16  - c^2/ 32 $.
\section{Derivations of the star algebra}\label{s:ders}
In a rather similar way as in the previous section we can prove that for all
primary fields $\pO$ of integer weight~$h$ the combination
$\pO_n + (-1)^{n+h-1} \pO_{-n}$ is conserved on the vertex, that is
\be\label{der:eq1}
\etats{V_N} \sum_{j=1}^N\left( \pO_n^{(j)} + (-1)^{n+h-1} \pO_{-n}^{(j)}\right)
=0.
\ee
\begin{proof}
These equations are special cases of the conservation laws derived in section
\ref{s:cons} and can be obtained by adding the $N$ cyclic versions of
these laws. A direct and more elegant derivation is as follows. Consider the
density defined by
$$\varphi^{(i)}(z_i) = z_i^{n+h-1} + (-1)^{n+h-1} z_i^{-n+h-1}$$ around each of
the $N$ punctures. This density has by definition weight~$1-h$. It
is defined globally on the Riemann surface, indeed on the boundaries of the
different coordinate patches we have -- as explained around
equation\refpj{gluing1} -- $z_1z_2 = -1$ and cyclically. Therefore under $z_1
\to -1/z_2$ we have
\bqs  
\lefteqn{z_1^{n+h-1} + (-1)^{n+h-1} z_1^{-n+h-1}}\\
&\to& \left(1 / z_2^2 \right)^{1-h}
\left\{ (-1)^{n+h-1}\ z_2^{-n-h+1} + (-1)^{n+h-1}
(-1)^{n-h+1}z_2^{n-h+1}\right\}\\
&=& (-1)^{n+h+1} z_2^{-n+h-1} + (-1)^{2 n} z_2^{n+h-1}
\eqs
Thus $\varphi^{(i)}$ is clearly defined globally if $n \in \IZ$.
\end{proof}

Equation\refpj{der:eq1} implies the following important derivative property.
\kader{prop}{prop:der}{Suppose $D =  \pO_n + (-1)^{n+h-1} \pO_{-n}$ where $\pO$
is a primary field of integer weight $h$, then $D$ is a derivation of the
star-algebra
$$D\left( \state{A} \star \state{B} \right) = 
\left(D \state{A}\right) \star \state{B} +(-1)^{|D||A|} 
\state{A} \star D\state{B}.
$$
Here $|\Phi|$ denotes the Grassmann number of the field $\Phi$.
}
\begin{proof}
It is sufficient to prove that the bpz inner product with all states $\state{C}$
gives
$$\etats{C}D\left( \state{A} \star \state{B} \right) = 
\etats{C}\left(D\state{A}\right) \star \state{B} +(-1)^{|D||A|} 
\etats{C}\state{A} \star D\state{B}.$$
Using property~\ref{prop:V2} on the left hand side and the definition of the 
star product\refpj{starprod} on the right hand side,
this equation is equivalent with
$$\etats{V_2}\state{C}\  D\left( \state{A} \star \state{B} \right) = 
\etats{V_3}\state{C} \left(D \state{A}\right) \state{B} +
(-1)^{|D||A|} 
\etats{V_3}\state{C} \state{A} \ D\state{B}$$
According to the definition of the bpz-conjugate\refpj{bpzosc} 
we have $\mbox{bpz} (D) = - D$. After using the property~\ref{prop:bpz} and
again the definition of the star product, we have
$$0=(-1)^{|D||C|}\etats{V_3}\left(D\state{C}\right)\state{A} \state{B} + 
\etats{V_3}\state{C} \left(D \state{A}\right) \state{B} +
(-1)^{|D||A|} 
\etats{V_3}\state{C} \state{A} \ D\state{B},$$
but this the exactly the content of equation\refpj{der:eq1}.
\end{proof}
Special cases of these derivations are
\begin{itemize}
\item
$Q$ and $\eta_0$ are derivatives of the star
algebra as we needed to prove on page~\pageref{toprove2}.
\item
$K_n =
L^{\mbox{\scriptsize{tot}}}_n - (-1)^n L^{\mbox{\scriptsize{tot}}}_{-n}$ 
are well
known to generate the reparametrisation symmetries of the 
vertex~\cite{art:supW}.
\item\label{der:alpha}
The derivation $\alpha^\mu_n + (-1)^n \alpha^\mu_{-n}$ will be important in the
toy model of tachyon condensation we develop in chapter~\ref{c:TM}, see
page~\pageref{TM:der}.
\end{itemize} 
\section{The tachyon potential at level $(4,8)$}\label{s:high}
In section~\ref{s:level4} we calculated the potential up to level~$(2,4)$. The
difference in potential energy 
between the unstable and stable vacuum was in this approximation
$89,1\%$. In this section we continue the level truncation up to level~$(4,8)$.
For this calculation we have written a Mathematica code for the evaluation of
\sfi\ interactions using conservation laws.
This program was tested thoroughly.
\begin{itemize}
\item
We tested the derivation properties treated in section~\ref{s:ders}. More
concretely we tested\refpj{der:eq1} for several hundreds of \sft\ interactions
with among others the operator $\pO_n + (-1)^{n+h-1} \pO_{-n}$ being $Q$ and
$\eta_0$.
\item
We checked on several hundreds of states that our implementation of the BRST
operator satisfies 
\begin{itemize}
\item
the correct commutation relations with the other operators
\item
nilpotency
\end{itemize}
\item
We checked the cyclicity property.
\item
The results up to level~$(2,4)$ agree  with the calculations done using vertex
operators and finite \cts, see section~\ref{s:level4}.
\end{itemize}
The potentials at level~$(2,5)$ and level~$(2,6)$ can be found in 
appendix~\ref{s:V25} and~\ref{s:V26}. At level~$(7/2,7)$ we must include the
23~fields having level~$7/2$ and at level~$(4,8)$ we must include an
additional 38~fields. A list of these fields can be found in the appendices~\ref{s:f72}
and~\ref{s:f4}. At these levels, the potential becomes too lengthly to put on
paper. The number of monomials in the potential can be found in
appendix~\ref{s:terms}. The expectation values of the tachyon string field at
the successive levels can be found in appendix~\ref{s:VEV}. Plugging these
expectation values into the action, one can calculate the depth of the tachyon
potential, see table~\ref{t:conj}.
\begin{table}[h]
\begin{center}
\begin{tabular}{*{8}{|c}|}
\hline
level \STRUT& $(0,0)$&$(3/2,3)$& $(2,4)$&$(2,5)$&$(2,6)$& $(7/2,7)$&$(4,8)$ \\
\hline
$V/M$ \STRUT & 62\% & 85\% & 89\%  & 90.8 \% & 91.7\% & 93.8\%&94.4\%\\
\hline
\end{tabular}
\end{center}
\label{t:conj}
\end{table}
In the level truncation scheme, the depth of the potential seems to converge to
the mass of the $D9$-brane. We consider this as a strong test of the correctness
of Berkovits' action. This action seems to be the correct off-shell description
of open superstring theory.
\section{Conclusions and topics for further research}
Let us briefly comment on topics not discussed in this chapter and possible
lines of further research. 
\begin{enumerate}
\item
It is believed that a $D$-brane with a space-dependent
tachyon profile, in the form of a localized ``kink'' of energy, describes a
$D$-brane of lower dimension~\cite{9902105}. The existence of kink solutions 
describing lower-dimensional D-branes has been studied with the level 
truncation method~\cite{tokinks}. Their tensions have been found to 
agree with the known
tensions of the lower-dimensional branes.
\item
Berkovits' superstring field theory is not background independent. The
background has to be on-shell. This means that
the world sheet theory of the open strings
moving in this background has to be conformally invariant. A background
independent open string field theory has been proposed~\cite{0011033,0010108}. 
In this theory, Sen's first and second conjecture have been proved
exactly~\cite{0010108}.
\item
Besides Berkovits' proposal, other open superstring field theories have been
proposed. Firstly, Witten proposed a cubic Chern-Simons-like open superstring
field theory~\cite{art:supW}. However, infinities arise in the calculation of
scattering amplitudes~\cite{Wendt} and in the proof of gauge invariance. The
tachyon potential in this theory does not lead to the conjectured
behavior~\cite{0004112}. Secondly, a modified cubic superstring field theory
was proposed~\cite{Russen}. 
Neither in this theory does the tachyon potential show the correct 
behavior, see ref.~\cite{0011117,0107197}, where the potential is calculated up to
level~$(2,6)$\footnote{The different 
versions of the e-print~\cite{0011117} contain
different numerical results.} 
and ref.~\cite{Joris} for a level~$(5/2,5)$ calculation.
\item
It would be nice to develop  a more efficient calculation scheme to discuss
tachyon condensation in Berkovits' superstring field theory. 
A natural idea~\cite{ephi} to 
accomplish this, is to do calculations using the \sfi\ $e^{\Phi}$ instead of the
\sfi\ $\Phi$. Then one does not have to expand Berkovits' action and calculate
all the higher order interactions. In the level truncation method used so far,
these higher order interactions are the origin of extremely time consuming
combinatorics, see table~\ref{t:termen}.
\item
The most important unsolved problem in this area of research is to find 
a closed form expression of the string field which represents the stable 
vacuum. It seems likely that one needs this closed form to discuss the
following question. \emph{Is Berkovits' theory a field theory of 
closed strings when expanded 
around the stable vacuum?} If it turns out that we get a workable theory of
closed strings in this way, it is possible to look for closed string
backgrounds with minimal energy. Then it would be possible to solve a major
unsolved problem in string theory:
\begin{center}
{\Large $\mathcal{W}${\it hat is the correct vacuum of string theory?}}
\end{center}
\end{enumerate}

\chapter{A Toy Model}\label{c:TM}
If we want to learn for example
how closed strings arise in the true vacuum of open \sft, it seems plausible
that we need a closed expression for the true vacuum. However this seems to be
very complicated for two reasons. The first reason is that there are an infinite
number of fields acquiring expectation values in the minimum and the second one
is the complicated form of the interaction in open \sft. In this 
chapter\footnote{This chapter is an expanded version of some of the topics
discussed in~\protect\cite{art:TM}.} we try to
learn something about the exact form of the minimum by looking for the
exact form of the minimum in a ``baby version'' of 
\sft\footnote{Other toy models of tachyon condensation were considered
in~\cite{Minahan,padic,GP}.}. In the baby version of
\sft\ we forget about the ghosts and instead of taking the infinite set of
oscillators   
\bqs
 \alpha^\mu_n&&\mbox{ where } \mu: 1,\ldots,26\mbox{ and } n: 1,\ldots,+\infty\\
     && \mbox{ with } [\alpha^\mu_m,\alpha^\nu_n] = m\ \eta^{\mu\nu}\delta_{m+n},
\eqs
we take only one operator $a$ with $[a,\crea]$ = 1. In this way the baby version
is very easy, deriving all the equations in this chapter involves only
undergraduate quantum mechanics. 
\paragraph{}
In section~\ref{TM:action} we will give the action of the toy model. 
In its most general form, the model depends  on some parameters that 
enter in the definition of a  star product and are the analog of 
the Neumann coefficients in bosonic \sft. These parameters
are further constrained if we insist that the toy model star product
satisfies some of the properties that are present in the full \sft.
More specifically, the \sft\ star product satisfies the following properties:
\begin{itemize}
\item
The three-string interaction term is cyclically symmetric, see
equation\refpj{W:prop}.
\item
The star product is associative, see property~\ref{starass}.
\item
Operators of the form $a - a^\dagger$ act as derivations of the star-algebra,
see page~\pageref{der:alpha}.  
\end{itemize}
We impose cyclicity of the interaction term in our toy 
model in
section~\ref{TM:cyc}. We deduce the equations of 
motion in section~\ref{TM:eom}. 
In section~\ref{TM:eom} we define 
the star product for the toy model. 
We discuss the restrictions following from imposing associativity of
the star product in section~\ref{TM:ass}. It turns out
that we are left with 3
different possibilities, hereafter called case I, II and
III. As is the case for the bosonic \sft\ we can also look if there is a
derivation $D = a - a^\dagger$ of the star-algebra. This further restricts the
cases I, II and III to case Id, IId and again Id respectively. This is 
explained
in section~\ref{TM:der}, where we also discuss the existence of an identity
of the star-algebra.

After having set the stage we can start looking for exact solutions. In 
section~\ref{TM:CaseI} we give an exact form for the minimum in case I, in 
section~\ref{TM:othersols} we mention the other exact solutions we have found.
In section~\ref{TM:CaseIId}, we discuss the case IId which perhaps bears
the most resemblance to the full string field theory problem.
In this case, it is possible  to recast the equation of motion in
the form of an ordinary second order nonlinear differential equation.
This equation is not of the Painlev\'{e} type and we have not been able to
find an exact solution. 
Here too, it is possible to get very accurate information about the
stable vacuum using the level truncation method.
We conclude in
section~\ref{TM:end} with some suggestions for further research. 
\section{The action}\label{TM:action}	
The toy model we consider has the following action:
\be\label{Toyaction}
S(\psi) = \frac{1}{2} \etats{\psi}(L_0-1) \state{\psi} + \frac{1}{3} \etats{V}
\state{\psi} \state{\psi}\state{\psi}
\ee 
where $L_0$ is the usual kinetic operator $L_0= a^\dagger a$ 
and $[a,a^\dagger] = 1$. Let us denote the Hilbert space which is built in the
usual way by $\Hil$. 
The ``string field'' $\state{\psi}$ is simply a state in this
Hilbert space $\Hil$ and can thus be expanded as 
$$
\state{\psi} = \psi_0 \vac + \psi_1 a^\dagger\vac + \psi_2 (a^\dagger)^2\vac +
\cdots,
$$
where the coefficients $\psi_0, \psi_1, \cdots$ 
are complex numbers. To illustrate the analogy
between this toy model and the bosonic \sft, the complex numbers
$\psi_i$ in the toy model correspond to the space-time fields in the bosonic
\sft\ in the Siegel-gauge\refpj{W:Siegel}. The term $-1$ in the kinetic part of the action 
should be thought of as the zero point energy in the bosonic string. 
In this way, the state $\vac$ has negative energy. 

The interaction term is by definition the following:
\be\label{Toyint} 
\etats{V}
\state{\psi} \state{\psi}\state{\psi}=\  
_{123}\leftvac \exp ( \frac{1}{2} \sum_{i,j=1}^3 N_{ij} a_i a_j)\
\state{\psi}_1\state{\psi}_2\state{\psi}_3.
\ee
The numbers $N_{ij}$ mimic the Neumann coefficients treated in
section~\ref{s:Neumann}. The Neumann coefficients in Witten's \sft\ carry
additional indices, they look like $N_{ij,kl}\ \eta^{\mu\nu}$ where $k,l = 1,
\ldots,\infty$ label the different modes of the string and $\mu,\nu =
1,\ldots,26$
are space-time indices.

We see that we have made 3 copies of the Hilbert space $\Hil$ for notational
device. The extra
subscript on a state denotes the copy the state is in:
\bqs
&&\mbox{if } \state{\psi} = \sum_m \psi_m a^{\dagger m} \vac\ \in \Hil, \\
&&\mbox{then } 
\state{\psi}_i = \sum_m \psi_m a^{\dagger m}_i \vac_i \in \Hil_i\quad\mbox{  for }
i=1,2,3.
\eqs
By definition we have the following commutation relations in
$\Hil_1\otimes\Hil_2\otimes\Hil_3$:
\be\label{Toycomm}
[a_i,a_j^\dagger] = \delta_{ij}.
\ee
Hence the interaction term\refpj{Toyint} of the toy model is the inner product
between the state $\state{V} = \exp ( \frac{1}{2} \sum N_{ij}
a_i^\dagger a_j^\dagger)\vac_{123} \in\Hil_1\otimes\Hil_2\otimes\Hil_3$ and 
$\state{\psi}_1 \otimes\state{\psi}_2 \otimes \state{\psi}_3$.

As an example let us calculate the action for 
$\state{\psi} = t \vac + u\ a^\dagger \vac,$ i.e. the level 1 part of the action\refpj{Toyaction}.
The kinetic part is obviously
$$-{1 \over 2} t^2 $$
and the interaction is
\bqs
&&{1 \over 3}\ _{123}\leftvac \left(1+ {1 \over 2} N_{ij}a_ia_j\right) 
\left(t + u\ a_1^\dagger\right) \left(t + u\ a_2^\dagger\right) 
\left(t + u\ a_3^\dagger\right)
\vac_{123} = \\
&&{1 \over 3} t^3 + {1 \over 3}\left( N_{12}+N_{13}+N_{23}\right)t u^2
\eqs
\section{Cyclicity}\label{TM:cyc}
As is the case for the full \sft\ we would like the interaction to be 
\emph{cyclic}\index{cyclicity!in toy model}:
\be\label{Toycyc} 
\etats{V} \state{A}\state{B}\state{C} = \etats{V} \state{B}\state{C}\state{A}.
\ee
Let us see what restrictions this gives for the matrix $N$. Imposing the
cyclicity\refpj{Toycyc} leads to $N_{11} = N_{22} = N_{33}$,  
$N_{12} = N_{23} = N_{31}$ and $N_{13} = N_{21} = N_{32}$. Hence $N$ will be of
the following form:
$$ N = \left(\begin{array}{ccc}
N_{11} &N_{12} &N_{13}\\
N_{13}& N_{11} & N_{12}\\
N_{12}& N_{13}  & N_{11}\\
\end{array} \right)
.$$
Because the oscillators $a_1,a_2$ and $a_3$ commute among each other, the matrix
$N$ can be chosen to be symmetric without losing generality. Hence we have 
fixed  the matrix $N$ to be of the form
$$ N_{ij} = \left(\begin{array}{ccc}
2 \lambda &\mu &\mu\\
\mu&2 \lambda & \mu\\
\mu& \mu  &2 \lambda\\
\end{array} \right)
.$$
From this form it is clear that imposing cyclicity in our toy-model forces 
the star product to be commutative as well.
\section{The equation of motion}\label{TM:eom}
If we impose the condition that the interaction is cyclically symmetric, 
the equation of motion reads
\be\label{Toyeom}
(a^\dagger a -1) \state{\psi} +\state{\psi} \ast \state{\psi} = 0. 
\ee
Here we have introduced the star product, it is defined by
\bq\label{Toystar}\index{star product!in toy model}
&&\state{\psi} \ast \state{\eta} =\\
&&_{23}\leftvac \exp( \frac{1}{2} \sum_{i,j = 2}^3 N_{ij} a_i a_j +
\sum_{i=2}^3 a_1^\dagger N_{1i} a_i + 
\frac{1}{2} a_1^\dagger N_{11} a_1^\dagger )\
\vac_1 \state{\psi}_2\state{\eta}_3\nonumber
\eq
Let us give some examples of the star product:
$$\vac \ast \vac = e^{\lambda a^{\dagger 2}}\vac.$$
The star product of two coherent states gives a squeezed state
\bqs
&&e^{l_1 \crea} \vac \ast e^{l_2 \crea} \vac =\\
&& \exp\left( \lambda (l_1^2 + l_2^2) + \mu\ l_1 l_2 \right)
\exp \left( \lambda a^{\dagger 2} + \mu ( l_1 + l_2) \crea\right) \vac
.\eqs
By taking derivatives one can calculate lots of star products e.g.
\bqs
\vac\ast\crea\vac &=& \mu\crea e^{\lambda a^{\dagger 2}}\vac\\
\crea\vac\ast\crea\vac &=& (\mu+\mu^2 a^{\dagger 2}) e^{\lambda a^{\dagger 2}}\vac\\
\eqs

We can also give the equation of motion as a differential equation. Let us use a
short hand notation for the \sfi\ $\psi$: 
$$
\state{\psi} = \sum_{n=0}^{\infty} \psi_n (\crea)^n \vac
\equiv \psi(\crea)\vac.
$$
If we use $ \partial_i = \partial/\partial x_i$, the equation of motion reads
\bqs
&&\left( x \frac{\partial}{\partial x} -1 \right) \psi(x)+\\
&&\exp\left( \frac{1}{2} \sum_{i,j = 2}^3 N_{ij} 
\partial_i\partial_j+
x \sum_{i=2}^3 N_{1i} \partial_i+ 
\frac{1}{2} N_{11} x^2 \right)
\left.\psi(x_2) \psi(x_3)\right|_{x_2 = x_3 = 0} = 0. 
\eqs
Here we have used that
$$
\leftvac a\ F(\crea) \vac = \left.\frac{\partial}{\partial \crea} 
F(\crea)\right|_{\crea = 0}. 
$$
This equation is thus a non-linear differential equation of infinite order.
The fact that it has infinite order makes it impossible to write down a finite
recursion relation in $x$.
\section{Associativity}\label{TM:ass}
In the full string field theory  the star product is \emph{associative}.
I will now check the associativity in our model on a basis of coherent states. The
star-product of two coherent states is easy to calculate:
$$
e^{l_1 \crea} \vac \ast e^{l_2 \crea} \vac =
A \exp \left( \lambda a^{\dagger 2} + \mu ( l_1 + l_2) \crea\right) \vac
$$
with 
$ A = \exp\left( \lambda (l_1^2 + l_2^2) + \mu\ l_1 l_2 \right)$. Then using
property\refpj{corr2} we  find
\bq\label{Toy3star}
&&\left(e^{l_1 \crea} \vac \ast e^{l_2 \crea} \vac\right)
\ast e^{l_3 \crea}\vac =A B \frac{1}{\sqrt{1 - 4 \lambda^2}}\times\\
&&
\exp\left\{\frac{\lambda \mu^2 (l_1+l_2)^2 + \mu^2 (\crea + l_3) (l_1+l_2) 
+ \lambda (\crea + l_3)^2 \mu^2}{1-4 \lambda^2}\right\} \vac \nonumber
\eq  
with
$B = \exp \left( \lambda a^{\dagger 2} + \lambda l_3^2 + \mu \crea l_3\right)$.
Imposing cyclicity among $l_1,l_2,l_3$ we find that the star-product is
associative only in the following three cases, hereafter called case I, II and
III:
\begin{enumerate}
\item[I.]
$\mu = 0 $, then $N = \left(\begin{array}{ccc}
2 \lambda &0 &0\\
0&2 \lambda & 0\\
0& 0 &2 \lambda\\
\end{array} \right) $
\item[II.] 
$ 2 \lambda = \mu -1$, then $N = \left(\begin{array}{ccc}
\mu-1 &\mu &\mu\\
\mu&\mu-1 & \mu\\
\mu&\mu  &\mu-1\\
\end{array} \right) $
\item[III.] 
$ \lambda = 1/2$, then $N = \left(\begin{array}{ccc}
1 &\mu &\mu\\
\mu&1 & \mu\\
\mu&\mu  &1\\
\end{array} \right) $
\end{enumerate} 
However, due to the factor $1 / \sqrt{1 - 4 \lambda^2}$ in
equation\refpj{Toy3star} the star product of 3 coherent states diverges in the
last case. Therefore we should look for another proof of associativity in 
this case. We will not do this, we just discard this case. 
\section[A derivation]{We want to have a derivation}\label{TM:der}
Let us now look if $D = a - a^\dagger$ is a derivation of the $\ast$-algebra:
$$
D(A \ast B) =DA \ast B + A \ast DB
\mbox{\quad where $A$ and $B$ are two string fields}.
$$
This is analogous to $\alpha^\mu_1-\alpha^\mu_{-1}$ being a derivative 
in the full \sft\footnote{See page~\pageref{der:alpha}.}. 
It is easy to see that for $D$ to be a derivation we need
\be\label{Toyder1}
\sum_i ( a_i - \crea_i)\state{V}=0.
\ee
Let us calculate the left hand side of\refpj{Toyder1}:
\bqs
\sum_i ( a_i - \crea_i)\state{V} &=& 
\sum_i ( \parder{\crea_i} - \crea_i)\state{V}\\
&=& \sum_i ( N_{ij}\crea_j- \crea_i)\state{V}
\eqs
This is zero if and only if $(\ 1\ 1\ 1\ ) \cdot (N-1) = 0$.
\begin{itemize}
\item{case I}\\
We need $3 ( 2 \lambda -1) = 0 $ so $\lambda = 1/2$. Hence $D$ is a derivation
if and only if $N=1$. We call this trivial case hereafter case Id.
\item{case II}\\
We need $ 2 \mu + \mu -2=0$, hence $\mu = 2/3$. In this case we have
\be\label{NIId}
N = \left(\begin{array}{ccc}
-1/3& 2/3&2/3\\
2/3 &-1/3& 2/3\\
2/3 & 2/3& -1/3\\
\end{array} \right)
\ee
Here after we call this subcase IId.
\item{case III}\\
We need $2 \mu =0$ so $\mu=0$. This reduces to case Id.
\end{itemize}

For the case Id we will show in
section\refpj{TM:CaseI} that there is no non-perturbative vacuum. Therefore we
consider the value\refpj{NIId} as the most important special case that we
would like to solve exactly in our toy-model.
\paragraph{}
Of course, when one is given an algebra, one likes to look for special elements.
There exists an identity string field $I$, i.e.~a
string field obeying $I \ast A = A$ for all string fields $A$, only in case II.
In this case we have for the identity $I$
$$
\state{I} = \frac{\sqrt{2 \mu -1}}{\mu} 
\exp \frac{1-\mu}{4 \mu-2}a^{\dagger 2}\vac
.$$
Using the correlators derived in appendix~\ref{TM:corr}, the reader can
easily check that 
$$ \state{I} \ast e^{l \crea} \vac =  e^{l \crea} \vac $$
for all coherent states $e^{l \crea} \vac $, thus proving that $I$ is the
identity. Proving that there is no identity if $N$ does not belong to case II
is most easily done by first arguing that the identity should be a Gaussian in
the creation operator $\crea$ and thereafter showing that one can not find a
Gaussian which acts as the identity on all coherent states. In case IId 
the identity string field reduces to
\be\label{Toyspid}
\state{I} = \frac{2}{\sqrt{3}}\exp \frac{1}{2} a^{\dagger 2}\vac
\ee
\section{Twist invariance}\label{TM:twist}\index{twist invariance!in toy model}
The bosonic \sft\ has a twist invariance, see section~\ref{s:twist}. We have a
twist invariance in the toy model as well. Indeed,  
the action\refpj{Toyaction} is invariant under the
following $\mathbb{Z}_2$-twist $S$:
\bqs
a,a^\dagger &\to& -a,-a^\dagger\\
\state{\psi} = \psi(\crea) \vac &\to& \psi(-\crea)\vac.
\eqs
If we write $\mathcal{H} = \mathcal{H}_{\mbox{\scriptsize{even}}}\oplus
\mathcal{H}_{\mbox{\scriptsize{odd}}}$, 
where $\mathcal{H}_{\mbox{\scriptsize{even}}}$ are the states even in
$a^\dagger$ and $\mathcal{H}_{\mbox{\scriptsize{odd}}}$ are the states odd 
in $a^\dagger$,
the action roughly looks like
$$S = \mbox{even}^2 + \mbox{odd}^2 +  
\mbox{even}^3 + \mbox{even}\cdot\mbox{odd}^2.$$
Consequently the equations of motion look like:
$$
\left\{
\begin{array}{lcl}
\displaystyle{\parder{\mbox{even}}}&:&  \displaystyle{\mbox{even}+ \mbox{even}^2 
+ \mbox{odd}^2=0}\\[2ex] 
\displaystyle{\parder{\mbox{odd}}}&:&  \displaystyle{\mbox{odd}+ 
\mbox{even}\cdot \mbox{odd}=0}\\ \end{array}\right.
$$
This means in particular that it is consistent with the equations of motion 
to put the odd components to zero when looking for solutions of the equations of
motion. In the next few sections we will write down some solutions in particular
cases.

\section{Exact solution in case I}\label{TM:CaseI}
We can give an exact formula for the minimum in case I, in this case we have 
$$N = \left(\begin{array}{ccc}
2 \lambda &0 &0\\
0&2 \lambda & 0\\
0& 0 &2 \lambda\\
\end{array} \right) \equiv \left(\begin{array}{ccc}
- 2 l^2 &0 &0\\
0&-2 l^2 & 0\\
0& 0 &-2 l^2\\
\end{array} \right),
$$
where $\lambda = -l^2 $. We need to solve the following equation:
$$
\left(x \parder{x} -1\right) \psi(x) + 
\left.\exp-l^2 \left(\partial_2^2 + \partial_3^2 + x^2\right)\psi(x_2) \psi(x_3)
\right|_{x_2=x_3=0} = 0 
$$
This is
 $$
\left(x \parder{x} -1\right) \psi(x) + e^{ - l ^2 x^2} a^2 = 0, 
$$
where $a$ is just a number 
$$a = \left.e^{- l^2 \partial^2_x} \psi(x) \right|_{x=0}. 
$$
A solution of this differential equation is
$$ \psi(x) = \frac{1}{1-4 l^4} \phi( l x), $$ 
where $\phi(x)$ is the function
\bqs
\phi(x)&=& \exp(-x^2) + \sqrt{\pi} x\ \erf ( x)\\
&=& - \sum_{m=0}^{+\infty}\frac{(- x^2)^m}{m!\ ( 2 m -1)}
\eqs
thus
\be\label{Toyexact1}
\state{\mbox{true vac}} = - \frac{1}{1-4 \lambda^2}\sum_{m=0}^{+\infty}
\frac{(\lambda\ a^{\dagger 2})^m}{m!\ ( 2 m -1)}\vac
\ee
The energy difference between the false and true vacuum is $ E =
-\frac{1}{6} \left( 1 - 4 \lambda^2\right)^{-3/2}$. It is easy to see that there
is only a true vacuum for $|\lambda| < 1/2$. Also the level-truncation does not
seem to converge in the case $|\lambda| \geq 1/2$, see 
section~\ref{TM:leveltrunc}. Notice that for the special
case Id there does not seem to be a true vacuum.
\section{Other exact solutions}\label{TM:othersols}
We can also find the exact minimum in case II when $\mu = 1$, i.e.~when 
$$N = \left(\begin{array}{ccc}
0&1 &1\\
1&0 &1\\
1&1 &0\\
\end{array} \right). $$
We need to solve the following equation:
$$
\left(x \parder{x} -1\right) \psi(x) + 
\left.\exp\left(\partial_2 \partial_3  + x( \partial_2 + \partial_3)\right)\psi(x_2) \psi(x_3)
\right|_{x_2=x_3=0} = 0 
.$$
A solution of this equation is $\psi(x) = 1$. 
$$
\state{\mbox{false vac}} = 0 \vac \qquad
\state{\mbox{true vac}} = 1 \vac
$$
More generally we can also solve $$N = \left(\begin{array}{ccc}
0&\mu &\mu\\
\mu&0 &\mu\\
\mu&\mu &0\\
\end{array} \right),$$
for general $\mu$, again the solution is  $\psi(x) = 1$. However this case is
not associative if $\mu\neq 1$.
\section{Towards the exact solution in case IId?}\label{TM:CaseIId}
\subsection*{The star product in momentum space}
In section~\ref{TM:ass} we have deduced that $D = a-\crea$ is a
derivation of the star algebra. If we write the creation and 
annihilation operators in terms of the momentum 
and coordinate operators:
$$
\left\{
\begin{array}{ll}
\crea=& \frac{1}{\sqrt{2}}\ (p + i x),\\
a=& \frac{1}{\sqrt{2}}\ (p -i x),\\
\end{array}\right.
$$
we see that $D$ is proportional to $\p/\p p$. 
Therefore it is tempting to think that the star product will reduce
to an ordinary product in momentum space, this is indeed true. If we write the
states in momentum representation:
$$\state{\psi} = \int dp\  \psi(p) \state{p}_p,$$
where the states $\state{p}_p$ are the eigenstates of the momentum operator
$\hat p$, normalized in such a way that $\langle p_1 \state{p_2} =
\delta(p_1-p_2)$ -- we use the extra subscript to denote which representation we
are using -- , we find 
\be\label{Toyordprod}
\state{\psi} \ast \state{\eta} = \int dp\ \pi^{1/4} \sqrt{\frac{3}{2}}\ \psi(p)
\eta(p) \state{p}_p.
\ee
\paragraph{First proof}
We will give an overview of the main steps in the derivation of this result.\\
We can easily see that the following equation holds:
\be\label{Toybew1}
\etats{V} \state{\psi_1} \state{\psi_2}\state{\psi_3}  
= \int d^3k\ _a\leftvac\exp \frac{1}{2} a_i N_{ij} a_j \state{\vec k}_a \ 
\psi_1(k_1) \psi_2(k_2)  \psi_3(k_3),
\ee
where the subscript denotes that we are working in the $a,\crea$ representation.
The state  $\state{\vec k}_a$ is the eigenstate of the momentum operator written
in the $a,\crea$ representation, in one dimension this is
\be\label{Toybew2}
\state{k}_a = \frac{1}{\pi^{1/4}} \exp(-\frac{1}{2} a^{\dagger 2}+ 
  \sqrt{2}\ k\ \crea -
\frac{k^2}{2}) \vac_a
.\ee
Indeed we find
\bqs
\hat k \state{k}_a &=& {1 \over \sqrt{2}} ( a + \crea ) \state{k}_a\\
&=& ( - \crea + \sqrt{2}\ k + \crea)\ {1 \over \sqrt{2}} \state{k}_a\\
&=&  k \state{k}_a,
\eqs
thus proving that $\state{k}_a$ is an eigenstate of the operator $\hat k$. It
is left to prove that this state has correct normalization:
$$_a \langle k \state{k'}_a =\delta(k-k').$$
This is most easily done by calculating the $k$-representation of the ground
state $\vac_a$
$$_a\langle 0 \state{k}_a = {1 \over \pi^{1/4}} e^{- k^2 /2}$$
which is indeed the well-known Gaussian.

Next let us diagonalize the matrix $N$ with an orthogonal matrix $O$, perform a
change of variables and do a linear transformation on the oscillators:
$$\left\{
\begin{array}{l}
O^T N O = \left(\begin{array}{ccc} 
              1&&\\ &-1&\\ && -1
	      \end{array}\right) \equiv D
\mbox{\qquad with }O = \left(\begin{array}{ccc} 
              1/\sqrt{3} & \star &\star \\
	      1/\sqrt{3} & \star &\star \\ 
	      1/\sqrt{3} & \star &\star \\ 
	      \end{array}\right)\\
a_i = O_{ij} b_j,\mbox{\qquad again } [b_i,b_j] = \delta_{ij}\mbox{ holds}\\
k_i = O_{ij} k'_j.
\end{array}\right.
$$	  
Then the equation\refpj{Toybew1} becomes 
$$
\int\frac{d^3 k'}{\pi^{3/4}}\  _b\leftvac \exp \frac{1}{2} b \cdot D \cdot b 
- \frac{1}{2}\ b^{\dagger 2} + \sqrt{2}\ k'\
	      b^\dagger - \frac{1}{2} k'^2\vac_b\
	 \psi_1(k_1) \psi_2(k_2)  \psi_3(k_3).     $$
The correlators involving the oscillators $b_2$ and $b_3$ give rise to 
delta functions $\delta(k'_2)$ and $\delta(k'_3)$ upon using 
equation\refpj{Toybew2}, for the correlator involving the oscillator
$b_1$ one should use equation\refpj{corr2}. Taking all this together we find
\be\label{Toybew3}
\etats{V} \state{\psi_1} \state{\psi_2}\state{\psi_3} = \pi^{1/4}
\sqrt{\frac{3}{2}} \int dp\  \psi_1(p) \psi_2(p)  \psi_3(p),
\ee 
from which\refpj{Toyordprod} follows. 
\paragraph{Second proof}
We can also give an alternative proof working with coherent states. The
interacting term with 3 coherent states is easy to calculate: if $\state{\psi_i}
= \exp\ ( l_i \crea )\vac$, then
$$ \etats{V} \state{\psi_1}\state{\psi_2} \state{\psi_3} = e^{\ l^T N l}.$$
Let us verify if we get the same result in momentum space. Using\refpj{Toybew2}
we find that  a coherent state is given by a Gaussian in momentum space:
$$ e^{ l \crea} \vac_a =  
\frac{1}{\pi^{1/4}} \exp(-\frac{1}{2}\ l^2+ \sqrt{2}\ l\ k -\frac{k^2}{2}) 
.$$
Equation\refpj{Toybew3} then holds by Gaussian integration.

As a check on our result we will verify that the state $\state{I}$ given
by\refpj{Toyspid} is the identity in momentum space. In momentum space we have
$$
\state{I} = \sqrt{\frac{2}{3}}\ \frac{1}{\pi^{1/4}}\  1 \mbox{\qquad as a function in
momentum space}, $$
therefore we have for arbitrary states $\psi$
$$ \state{I} \ast \state{\psi} = \sqrt{\frac{2}{3}}\ \frac{1}{\pi^{1/4}}\  1 \cdot
	\pi^{1/4}\sqrt{\frac{3}{2}}\  \psi(k) = \psi(k), $$
as it should be.       	    
\subsection*{The equation of motion in momentum space}
The equation of motion we want to solve now becomes in momentum space
\bqs
&&(a^\dagger a -1) \state{\psi} +\state{\psi} \ast \state{\psi} \\
&& = {1 \over 2} \left( - {\p^2 \over \p p^2} + p^2 -3\right)\psi + \pi^{1/4}
\sqrt{{3 \over 2}}\psi(p)^2 =0.  
\eqs
If we drop some constants, the differential equation we are left with reads
$$ {\p^2 \over \p p^2} \psi(p) = (p^2 -3)\psi(p) + \psi(p)^2.$$
So we see that instead of the infinite order differential equation we started
with, we have now a second order non-linear differential equation. However we
have not found an exact solution of this equation in the literature. 
\section[Level truncation]{Approximating the true vacuum by level truncation}
\label{TM:leveltrunc}\index{level truncation!in toy model}
Let us now make a brief study of the level truncation method in this toy model.
The level truncation method proceeds as follows. 
For the level $n$ approximation one takes terms in the ``string field'' up to
$a^{\dagger n}\vac$, thereafter one calculates the action and minimizes it.

It is easy to program the calculation of the potential in for example 
Mathematica. If the reader likes to play around with these calculations, we
refer him to appendix~\ref{TM:programs} where we give some of these programs.
\subsection*{Case I}
As an example for case I we take $\lambda = 1/4$. Of course, in this case 
$\mu=0$. Let us give successive approximations to the minimum up to level~10.
Hence, we truncate the state upto level~10
$$
\state{\psi}= \psi_0\vac+ \psi_2 a^{\dagger 2}\vac + \cdots+ 
\psi_{10} a^{\dagger
10}\vac\ .
$$
The potential is then given by
$$
S(\state{\psi})= 
\sum_{n=0}^5 {1\over2} (2 n-1)( 2 n)!\ \psi_{2n}^2 +
\frac{1}{3} \left(\sum_{n=0}^5 \frac{ (2 n)!}{4^n n!}\psi_{2 n}\right)^3 \ .
$$
Using this potential we find at the different levels the following results:
\begin{description}
\item{at level $0$:}  $ \state{\psi} = 1.\vac$\\ with $S(\psi) =
-0.166667$, this is $65\%$ of the exact answer.
\item{at level $2$:}  $\state{\psi} = (1.30612\ -0.326531\ a^{\dagger 2})\vac$
\\
with $S(\psi) = -0.248785$, this is $97\%$ of the exact answer.
\item{at level $4$:}  $\state{\psi} = (1.32976\  -0.332441\ a^{\dagger 2}
  -0.0138 517\ a^{\dagger 4})\vac$ \\
with $S(\psi) = -0.255570$, $99.6\%$.
\item{at level $6$:}  $\state{\psi} = (1.33276\  -0.333190\ a^{\dagger 2}
  -0.0138829\ a^{\dagger 4}-0.000694147\  a^{\dagger 6})\vac$\\ 
with $S(\psi) = -0.256435$,  $99.936\%$.
\item{at level $8$:}  $
\state{\psi} = (1.33323\  -0.333308\ a^{\dagger 2}
  -0.0138878\ a^{\dagger 4}-0.000694392\  a^{\dagger 6}$\\
  \hspace*{2.5 cm}$-0.0000309996\ a^{\dagger 8})\vac $\\
with $S(\psi) =-0.256571 $, $99.989\%$.
\item{at level $10$:}  $
\state{\psi} = (1.33331\  -0.333329\ a^{\dagger 2}
  -0.0138887\ a^{\dagger 4}-0.000694434\  a^{\dagger 6}$\\
  \hspace*{2.5 cm}$\-0.0000310015 a^{\dagger 8}-1.20562\cdot 10^{-6}a^{\dagger 10})\vac $\\
with $S(\psi) = -0.256595$, $99.998\%$.
\end{description}
We can see that the level truncation scheme converges pretty well to the exact
form of the true vacuum\refpj{Toyexact1}
\bqs
\state{\mbox{true vac}} &=& ( \frac{4}{3} - \frac{1}{3} a^{\dagger 2}
- \frac{1}{72} a^{\dagger 4}\\
& & - \frac{1}{1440} a^{\dagger 6}- \frac{1}{32256} a^{\dagger 8}
- \frac{1}{829440} a^{\dagger 10})\vac + \cdots\\
&=& (1.33333- 0.333333 a^{\dagger 2}-0.0138889 \cre{4} - 0.000694444 \cre{6}\\
 & & -3.10020\cdot 10^{-5} \cre{8}-1.20563\cdot10^{-6} \cre{10})\vac+ \cdots 
\eqs
with $$S(\mbox{true vac}) = -\frac{1}{6} (3/4)^{-3/2} = -0.256600.$$ 
We can also use the equation of motion to eliminate the higher level fields. We
then get as successive approximations to the 
effective potential the following picture:
\begin{figure}
\begin{center}
\epsfbox{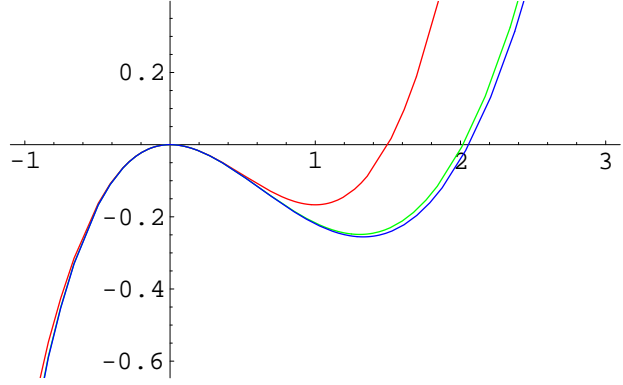}
\end{center}
\label{f1}
\caption{The effective actions at level 0, level 2 and level 4.}
\end{figure}
\subsection*{Case IId}
Even though we are not able to find a closed form solution in this case, we
can get good approximate results with the level truncation method.
In this case we give the potential only for fields up to level 4. It reads:
\bqs
&&S(\state{\psi})=\frac{-{\psi_0}^2}{2} + 
  \frac{{\psi_0}^3}{3} - 
  \frac{{\psi_0}^2\,\psi_2}{3} + 
  {\psi_2}^2 + 
  \psi_0\,{\psi_2}^2 + 
  \frac{13\,{\psi_2}^3}{27} + 
  \frac{{\psi_0}^2\,\psi_4}{3}\\&& - 
  \frac{34\,\psi_0\,\psi_2\,
     \psi_4}{9} + 
  \frac{41\,{\psi_2}^2\,\psi_4}
   {27} + 36\,{\psi_4}^2 + 
  \frac{227\,\psi_0\,{\psi_4}^2}
   {27} + \frac{319\,\psi_2\,
     {\psi_4}^2}{27} + 
  \frac{1249\,{\psi_4}^3}{81}
\eqs
Again we can minimize this action and we find
\begin{description}
\item{at level $0$:}  $ \state{\psi} = 1.\vac$\\ with $S(\psi) =
-0.166667$.
\item{at level $2$:}  $\state{\psi} = (1.05083\ +0.0870701\ a^{\dagger 2})\vac$
\\
with $S(\psi) = -0.181514$
\item{at level $4$:}  $\state{\psi} = (1.0508\  +0.0867394\ a^{\dagger 2}
  -0.000383389\ a^{\dagger 4})\vac$ \\
with $S(\psi) = -0.181521$
\item{at level $6$:}  $\state{\psi} = (1.05082\  +0.0867768\ a^{\dagger 2}
  -0.000408059\ a^{\dagger 4}-0.0000352206\  a^{\dagger 6})\vac$\\ 
with $S(\psi) = -0.181523$
\item{at level $8$:}  $
\state{\psi} = (1.05082\  +0.0867771\ a^{\dagger 2}
  -0.000412528\ a^{\dagger 4}-0.0000341415\  a^{\dagger 6}$\\
  \hspace*{2.5 cm}$+1.788 \cdot 10^{-6}\ a^{\dagger 8})\vac $\\
with $S(\psi) =-0.181524 $
\item{at level $10$:}  $
\state{\psi} = (1.05082\  +0.0867771\ a^{\dagger 2}
  -0.000412537\ a^{\dagger 4}-0.0000339848\  a^{\dagger 6}$\\
  \hspace*{2.5 cm}$+1.76475 \cdot 10^{-6} a^{\dagger 8}
  -4.54233\cdot 10^{-8}a^{\dagger 10})\vac $\\
with $S(\psi) = -0.181524$
\end{description}
%
\section{Conclusions and topics for further research}\label{TM:end}
We simplified Witten's open \sft\ by dropping all the ghosts and keeping only
one matter oscillator. The model we constructed closely resembles the full \sft\
on the following points:
\begin{itemize}
\item
There is a false vacuum and a stable vacuum.
\item
The interaction is given in terms of ``Neumann coefficients'' and can be written
by using an associative star product.
\item
There is a notion of level truncation which seems to converge to the correct
answers.
\end{itemize}
For some special values of one of the parameters of the
model, we were able to obtain the exact solution for the stable vacuum state and
the value of the potential at the minimum. 

For other values of the parameters we did not succeed to
construct the exact minimum of the tachyon potential. 
This does not mean that it is impossible to solve 
Witten's \sft\ exactly. In the full \sft\ there is a lot
more symmetry around: for example Witten's \sft\ has a huge gauge invariance and
one could try to solve the equation of motion by making a pure-gauge like 
ansatz~\cite{Schnabl:B}, see page~\pageref{solSch}.

Therefore maybe a natural thing to
do is to set up a toy model that includes some of the ghost oscillators in such
a way that there is also a gauge invariance. 
Another research topic can be to set up a toy model of Berkovits' \sft. It also
should not be too difficult to try to mathematically prove the convergence 
of the level truncation in these toy models. Maybe it teaches us something
about why the level truncation method converges in the string field theories.

\appendix
\renewcommand{\chaptermark}[1]{\markboth{\appendixname\ \thechapter.\ #1}{}}
\chapter{Conformal Field Theory}\label{CFT}
In this appendix, we review the elements of conformal field theory needed 
in this thesis. In the first section we will treat the bosonic string in a flat
background, in the second section we will treat the superstring. We will 
follow refs.~\cite{Polchinski,LeClair} closely.
\section{The bosonic string}\label{s:bosstring}
\index{conformal field theory! of bosonic string|(}
We will study the bosonic string in orthonormal gauge, and use a Euclidean
metric on the world-sheet. After gauge-fixing, the action for modes propagating
on the world-sheet has the form:
\be\label{Polyakov}
\frac{1}{2 \pi } \int d^2 z\left( \frac{1}{\ap}\partial X^\mu \bar\partial X_\mu
+ b \bar\partial c + \bar b \partial \bar c\right)
,\ee
where $X^\mu$ is the space-time coordinate and $c$, $b$ are the
reparametrisation ghost and antighost\index{ghosts!bc@$bc$}. The action\refpj{Polyakov} is an example
of a conformally-invariant quantum field theory. The action is invariant under
general conformal transformations $z \to f(z)$, while the fundamental fields
transform as conformal tensors. The transformation law of a tensor (or in the
language of conformal field theory, a \textit{primary conformal
field}\index{primary}) is
determined by its scaling dimension. We will write the transformation
law\index{conformal transformation} of a
primary field $\mathcal{O}(z)$ of dimension $d$\index{conformal weight} as
\be\label{conftrans}
f \circ \mathcal{O}(z) = (f'(z))^d \mathcal{O}(f(z))
\ee
The fields $\partial X^\mu$, $c$ and $b$ transform as primary fields of dimension
1, $-1$ and 2 respectively. Some examples of the calculations of 
transformation laws for non-primary fields are given in 
appendix~\ref{s:transf}. 

In radial quantization, charges are defined as integrals around circles:
\be
Q = \oint dz\ j(z).
\ee
If, as is often the case, the charge density is an analytic function, the
contour of integration may be freely deformed. Since the Hilbert space
interpretation of a correlation function sets the operators in (radial)
time-order, equal-time commutators of charges with operators $\mathcal{O}(w)$ may be
written as differences of correlations functions with the contour displaced
slightly to either side of the point $w$. In other words,
\be\label{charges}
\langle \cdots \left[ Q, \mathcal{O}(w)\right] \cdots \rangle = 
\langle \cdots  \oint dz j(z) \mathcal{O}(w)\cdots \rangle,
\ee
where the contour encircles the point $w$. Equal-time commutators may then be
related directly to singularities of the operator-product expansion of $j(z)$
and $\mathcal{O}(w)$. A particular important set of charges are the Virasoro operators,
the Fourier components of the energy-momentum tensor $T$:
\be
L_n = \oint dz z^{n+1} T(z).
\ee
In any conformally-invariant theory, $T$ is an analytic function of $z$; in the
bosonic string, it has the explicit form
\bqs
T &=& T^{\mbox{\scriptsize{matter}}}+ T^{\mbox{\scriptsize{ghost}}}\\ 
 &=& -\frac{1}{\ap} \partial X^\mu \bar\partial X_\mu +
(\partial b ) c -2 \partial(bc).
\eqs
The operator product of $T$ with a primary field has the general structure
\be\label{OPEprim}
T(z) \mathcal{O}(w) \sim \frac{d}{(z-w)^2} \mathcal{O}(w) + \frac{1}{z-w} \partial\mathcal{O}(w),  
\ee
this relation is equivalent, by the use of\refpj{charges}, to the commutator
\be
\left[ L_n, \mathcal{O}(w)\right] = d n w^n \mathcal{O} (w) + w^{n+1} \partial \mathcal{O}(w)
.\ee
This is the infinitesimal form of\refpj{conftrans}, for the particular
variation
\be
L_n \Leftrightarrow  w \to w+ \epsilon w^{n+1}
.\ee
Notice that $L_0$ generates an infinitesimal dilatation, this operator is
precisely the Hamiltonian of radial quantization introduced above.
It is convenient to define the Fourier decomposition of an arbitrary primary
field by
\be\label{Fconv}
\mathcal{O}(z) = \sum_{n = -\infty}^{\infty} \mathcal{O}_n z^{-n-d} \qquad
\mathcal{O}_n = \oint z^{n+d -1} \mathcal{O}(z).
\ee
The notation is arranged so that\refpj{OPEprim} leads to
\be
\left[ L_0, \mathcal{O}_n\right] = -n \mathcal{O}_n.
\ee
Hence the $\mathcal{O}_n$ are ladder operators for $L_0$. 
Following the convention for the Fourier decompositions of the
modes\refpj{Fconv}, 
we define for the bosonic string
\bq\label{modeexpansion}
\alpha_m^{\mu} &=& i \left(\frac{2}{\ap}\right)^{1/2} \oint dz z^m \partial
X^\mu(z)   \nonumber \\
b _m &=& \oint dz z^{m+1} b(z)\\
c _m &=& \oint dz z^{m-2} c(z) \nonumber.
\eq
From this decomposition, one can calculate the mode expansion of the Virasoro
generators
\bqs
L_m^{\mbox{\scriptsize{matter}}}&=&
\frac{1}{2}\sum_{n}\oscNO{\alpha^{\mu}_n\ \alpha_{m-n,\mu}}\\
L_m^{\mbox{\scriptsize{ghost}}}&=&\sum_{n}(2m-n) \oscNO{b_n\ c_{m-n}}-\delta_{m,0}\
,
\eqs
here, $\oscNO{\cdots}$ denotes creation-annihilation normal ordening.

It is a well-known fact that the physical spectrum is given by the 
cohomology of
the BRST-charge. In string theory, the BRST-charge is the zero-mode of the
BRST-current
\bq
Q&=& \oint dz\left(c T^{\mbox{\scriptsize{matter}}}+ 
{1 \over 2} T^{\mbox{\scriptsize{ghost}}}\right)\nonumber \\
&=& \sum_{n}\oscNO{c_{-n} L_n^{\mbox{\scriptsize{matter}}}}+ {1 \over 2}
\sum_{n}\oscNO{c_{-n} L_n^{\mbox{\scriptsize{ghost}}}} -{1 \over
2}c_0\label{BRST:bos}
\eq

The action\refpj{Polyakov} leads to a Gaussian path integral. Hence, all
correlators can be computed by doing Wick-contractions
\be\label{basiscorr}
\wick{1}{\partial<1 X^\mu(z) \partial>1 X^\nu(w)\rangle}  = 
-\frac{\ap}{2}\ \eta^{\mu\nu}\frac{1}{(z-w)^2} \qquad
\wick{1}{<1 b(z) >1c(w)}= \frac{1}{z-w}
\ee
Using\refpj{basiscorr} together with\refpj{charges} and
\refpj{modeexpansion}, we find the following commutation relations
\be
\left[ \alpha_m^{\mu},\alpha_n^{\nu}\right] = m \delta_{m+n}  \eta^{\mu\nu},
\qquad 
\left\{ b_m, c_n \right\} = \delta_{m+n}.
\ee

In radial quantization, the point $z=0$ ( or, more generally, the center of the
concentric circles ) represents $t = -\infty$. In quantum field theory, one
usually defines the vacuum as the state which develops from $t = - \infty$, the
generalization to radial quantization is to define the
$SL(2,\mathbb{C})$-invariant vacuum 
$\vac$\index{SL@$SL(2,\IC)$!invariant vacuum}  to be the
state which develops from the point $z = 0$ when we put no operator
there\footnote{In this we differ from Polchinski~\cite{Polchinski}. He uses the
notation  $\state{1}$ for the  $SL(2,\mathbb{C})$-invariant vacuum.}. More
concretely,
\be
\langle\cdots | \mathcal{O}_n\vac 
= \langle \cdots  \oint z^{n+d -1} \mathcal{O}(z) \rangle,
\ee
with the contour enclosing no other operators. From this it follows that
$$
\mathcal{O}_n \vac = 0 \quad\mbox{if }\quad n \geq 1-d.
$$
For the bosonic string, this tells us that
\bq
\alpha_n^{\mu} \vac &=& 0 \mbox{ for } n \geq 0\nonumber\\
c_n \vac &=& 0 \mbox{ for } n \geq 2\nonumber\\
b_n \vac &=& 0 \mbox{ for } n \geq -1\label{annihilators}\\
L_n \vac &=& 0 \mbox{ for } n \geq -1\nonumber
\eq

The adjoint of the last relation in\refpj{annihilators} gives
\be
\leftvac L_n = 0 \mbox{ for } n \leq +1 
\ee
Thus, the three generators $L_{-1}, L_0, L_1$ annihilate both of $\vac$ and
$\leftvac$ and thus are symmetries of all \co\ field theory matrix elements.
These three charges, plus the corresponding charges built of anti-analytic
fields, generate the $SL (2,\IC)$ subgroup\index{SL@$SL(2,\IC)$} of \cts
\be
z \to \frac{a z +b}{c z + d}, \quad a d -b c =1.
\ee
We will refer to $\vac$ henceforth as the $SL (2,\mathbb{C})$
vacuum.

Since $b$ has the same dimension as $T$, the operators $b_{-1}, b_0, b_1$ also
annihilate both $\vac$ and $\leftvac$. On the other hand, the operators 
$c_{-1}, c_0, c_1$ annihilate \textit{neither} of these states. We may interpret
this by saying that the three  $SL (2,\mathbb{C})$ \trs\ of the \co\ plane are
zero modes of the field $c(z)$, and that  these must be saturated in order to
obtain a nonzero matrix element. If this interpretation were correct, we would
expect:
\be\label{zeromodes}
\langle c(z_1) c(z_2) c(z_3) \rangle = \mbox{det} | \mathcal{Z}_i(z_j)|
\ee
where the $\mathcal{Z}_i(z)$ are the three zero modes: $\mathcal{Z}_i(z) =
(1,z,z^2)$ for $ i = -1,0,1$. Indeed, if we choose a normalization by writing 
\be\label{bc:norm}
\leftvac c_{-1} c_0 c_1 \vac = 1
\ee 
and use this together with the Fourier expansion\refpj{modeexpansion}, 
we find\refpj{zeromodes}. In general, \cft\ matrix elements will be non-vanishing
only if they contain 3 more $c$ operators than $b$ operators. 
\index{conformal field theory! of bosonic string|)}
\section{The superstring}\label{superstring}
\index{conformal field theory! of superstring|(}
After gauge-fixing the action for the matter modes has the form:
\be
\frac{1}{4 \pi } \int d^2 z\left( \frac{2}{\ap}\partial X^\mu \bar\partial X_\mu
+ \psi^\mu \bar\p\psi_\mu + \tilde\psi^\mu \p\tilde\psi_\mu\right) 
,\ee
and the ghost part reads 
\be
\frac{1}{2\pi} \int d^2 z\left(b \bar\partial c + \beta\bar\p\gamma +\mbox{left}
\leftrightarrow \mbox{right}\right)
.\ee
Here the fields $\psi^\mu$ are space-time vectors but world-sheet fermions with
conformal weight $1/2$. The fields $\beta$ and $\gamma$ are the 
superreparametrisation ghost and anti-ghost\index{ghosts!betagamma@$\beta\gamma$}, having weight $3/2$ and $-1/2$ 
respectively. From the definition\refpj{annihilators} of the $SL (2,\mathbb{C})$ 
invariant vacuum, it follows that for the superstring the following
annihilation relations hold in the NS-sector:
\bqs
\psi_r^{\mu} \vac &=& 0 \mbox{ for } r \geq 1/2\\
\beta_r \vac &=& 0 \mbox{ for } r \geq -1/2\\
\gamma_r \vac &=& 0 \mbox{ for } r \geq 3/2\\
\eqs

In the above analysis we assumed that the vacuum $\vac$ was  
$SL (2,\mathbb{C})$ invariant. However the spectrum is unbounded above and
below, so the choice of vacuum state is somewhat arbitrary~\cite{FMS}.
This point is
familiar in the case of Fermi statistics, where a Fermi sea-level must be
specified; likewise, for Bose statistics we must state the energy level below
which all the levels are filled. We call this the Bose sea-level. Therefore let
us 
define different vacua $\state{q}$ by
\bq\label{defvacq1}
&&\beta_r\state{q} = 0 \qquad r >-q-3/2,\\
&&\gamma_r\state{q} = 0 \qquad r \ge q+3/2.
\eq
So our conventions are still consistent, the $SL (2,\mathbb{C})$-invariant 
vacuum $\vac$ is the one given by $q=0$. The
$q$-vacua generate inequivalent representations of the $\beta$-$\gamma$
algebra -- a finite number of field operators cannot fill the state. 
If we bosonize the ghost system as $\beta = e^{-\phi}\p\xi$ and 
$\gamma = \eta e^{\phi}$~\cite{FMS}
\index{ghosts!etaxi@$\eta\xi$}\index{ghosts!phi@$\phi$}, we
can see that the coherent states $e^{q \phi}$ interpolate between the various
Bose sea-levels:
\be\label{defvacq2}
\state{q} = e^{q \phi}(0)\vac.
\ee
We have to show that the annihilation relations\refpj{defvacq1} hold to prove
that $e^{q \phi}(0) \vac$ can be identified with $\state{q}$. This is easy, indeed
\bqs
\beta_r \state{q} &=& \oint dz\  z^{r+1/2} \beta(z) e^{q \phi}(0)\vac\\
&=&\oint dz\ z^{r+1/2}\ e^{-\phi}\p\xi(z)\  e^{q \phi}(0)\vac\\
&=&-\oint dz\ z^{r+1/2}\ \p\xi(z)\ z^q\ :e^{-\phi}(z) e^{q \phi}(0):\vac\\
&=& 0 \qquad\mbox{if } r> -q-3/2
\eqs
and similarly
 \bqs
\gamma_r \state{q} &=& \oint dz\ z^{r-3/2}\ \eta e^{\phi}(z)\  e^{q \phi}(0)\vac\\
&=&\oint dz\ z^{r-3/2}\ \eta(z)\ z^{-q}\ :e^{\phi}(z) e^{q \phi}(0):\vac\\
&=& 0 \qquad\mbox{if } r> 1/2+q
\eqs
It is also easy to verify that the following annihilation relations for the
linear dilaton hold
\bqs
\phi_n\state{q} &=& 0\quad\mbox{if}\quad n \ge 1,\\
\phi_0\state{q} &=& -q \state{q}.
\eqs
The state $\state{q}$ is not the $SL(2,\IC)$-invariant vacuum and consequently
we find for the modes of the stress-energy tensor
$$T^{\phi} = -{1 \over 2} (\p\phi)^2 - \p^2\phi,$$
\bqs
L^\phi_n\state{q} &=& 0\quad\mbox{if}\quad n \ge 1,\\
L^\phi_0\state{q} &=& (-{1\over 2} q^2 -q) \state{q},\\
L^\phi_{-1}\state{q} &\not=& 0.
\eqs
We define the ghost number current $j_g$ and the picture number current $j_p$ 
as follows:
\bqs\index{picture number}\index{ghost number}
j_g&=& -bc-\beta\gamma \\
&=& -bc - \p\phi\qquad\mbox{(after bosonisation)}\\
j_p &=& -\eta\xi-\p\phi 
\eqs
Therefore the ghost number and the picture assignments are:
\begin{center}\index{picture}\index{ghost number}\label{t:convgp}
\begin{tabular}{|l|c|c|}\hline
\mbox{field}     & \mbox {ghost number }& \mbox {picture number} \\
\hline     
$b, \beta$& $-1$ & 0\\
\hline
$c, \gamma$ \STRUT &$1$ & 0\\
\hline
$e^{ q \phi}$ \STRUT &$q$ & $q$\\
\hline
$\eta$ \STRUT &$0$ & $-1$\\
\hline
$\xi$ \STRUT &$0$ & 1\\
\hline
\end{tabular}
\end{center}
\index{conformal field theory! of superstring|)}
\section{The bpz inner product}\label{s:bpz}
The Fock space of the \cft\ can be naturally equipped with a nondegenerate
bilinear inner product that was first introduced in~\cite{BPZ} and goes under
the name of \emph{bpz inner product}\index{bpz inner product}. 
\kader{definition}{def:bpz}{The bpz inner product of two states 
$\state{A}$ and $\state{B}$ is the following
correlation function on the sphere:
$$
\Bigl(\state{A},\state{B}\Bigr)_{\mbox{\scriptsize{\bpz}}} \equiv \langle A \state{B}
= \corr{I \circ A (0)\ B(0)} 
$$
}
Here $I(z) = -1/z$. The physical interpretation of this inner product is as
follows. Each state in the Hilbert space of the conformal field theory may be 
created from $\vac$ by a corresponding vertex operator
\be
\state{A}= A(0) \vac,
\ee
this is the so-called state -- operator 
mapping\footnote{See ref.~\cite{Polchinski} 
for more information about this mapping.}.  It is natural to
define the left vacuum as the state which develops by evolving backward from $t
= \infty$, that is, from $z = \infty$. The
dual of the state $\state{A}$ is then created by  $A(z)$ as the state created by
the vertex operator formed by acting on $A(z)$ by the inversion $I$. This 
inversion $I$ is a $SL(2,\IC)$-transformation which takes the origin to infinity while taking
the unit circle to itself.

We can also define the so-called \emph{bpz conjugation}\index{bpz conjugation}.
\kader{definition}{bpzosc}{The bpz conjugation of the modes of a quasi-primary field
$\pO$ of weight~$h$ is 
$$\bpz(\mathcal{O}_n) =\mathcal{O}_{-n} (-1)^{n+h}$$
}
Notice that this definition in particular holds for the Fourier 
components of the stress-energy tensor, indeed, for 
$SL(2,\mathbb{C})$-transformations the stress-energy tensor
behaves as a true primary field. 
The above definition is motivated by the following property.
\kader{prop}{prop:bpz}{
$$\Bigl( \state{A}, \pO_n \state{B} \Bigr)_{\mbox{\scriptsize{\bpz}}} 
= (-1)^{|\pO||A|}\ 
\Bigl( \bpz{(\pO_n)} \state{A}, \state{B} \Bigr)_{\mbox{\scriptsize{\bpz}}}$$
Here $|\Phi|$ is the Grassmann number of the operator $\Phi$.
}
\begin{proof}
In this proof we will make use of notations and ideas developed in
sections~\ref{s:gluing} and~\ref{s:cons}. Property~\ref{prop:bpz} is equivalent
to the following conservation law
$$\etats{V_2} \left( \pO^{(1)}_{-n} + (-1)^{n+h-1} \pO^{(2)}_n \right) =0\ .$$
This equation is rather easy to derive. Indeed, let us take a primary field
$\varphi(z) = z^{n+h-1}$ of weight $1-h$. By deforming contours, see
section~\ref{s:cons}, we find
\be\label{bpz:eq1}
\etats{V_2} \left(\oint dz\ (I \circ \varphi(z))\pO(z) 
+\oint dz\ \varphi(z) \pO(z)\right) =0\ .  
\ee 
Here we have
$$
I\circ \varphi(z) = z^{-2(1-h)} \varphi(-1 /z) 
= (-1)^{n+h-1} z^{h-n-1}\ .
$$
Hence\refpj{bpz:eq1} leads to
$$\etats{V_2} \left( (-1)^{n+h-1}\pO^{(1)}_{-n} + 
\pO^{(2)}_n \right) =0\ .$$
\end{proof}
\section{Basic formulas and conventions}\label{conventions}
The following is a list of the formulas of Polchinski's 
book~\cite{Polchinski} we need
most.
\begin{itemize}
\item\index{ghosts!bc@$bc$}
$bc$-system, this is a CFT with central charge $-26$.\\
The ghosts $b$ and $c$ are primaries of conformal weight 2 and $-1$ 
respectively. The stress tensor is
$$T^{bc} = (\p b)c - 2 \p(bc),$$
and the basic OPE reads
$$b(z) c(w) \sim {1 \over z-w}\qquad\mbox{so}\qquad \{b_m,c_n\} = \delta_{m+n} $$
\item\index{ghosts!betagamma@$\beta\gamma$}
$\beta\gamma$-system, this is a CFT with central charge $+11$.\\
$\beta$ has weight 3/2, $\gamma$ has weight $-1/2$.
The stress tensor is 
$$T^{\beta\gamma} = (\p \beta)\gamma - {3\over 2}\ \p(\beta\gamma)$$
The OPE is 
$$\beta(z) \gamma(w) \sim -{1 \over z-w}\qquad\mbox{so}\qquad [\beta_r,\gamma_s] = -\delta_{r+s} $$
\item
bosonised $\beta\gamma$-system\footnote{
In this we differ form Polchinski. We adopt the convention that all fermions
anticommute with all fermions. The field $e^{q \phi}$ is a fermion for odd $q$
and a boson for even $q$. For odd $q$ it will anticommute with fields $\eta$ and
$\xi$, therefore the order in\refpj{bosspoken} is important for the $\eta\xi$
and $\phi\phi$  OPE's to reproduce the $\beta\gamma$ OPE.
}\\
\be\label{bosspoken}
\beta = e^{-\phi}\p\xi  \qquad\mbox{and}\qquad \gamma = \eta e^{\phi}
\ee
After bosonisation we have $T^{\beta\gamma} = T^\phi + T^{\eta\xi}$.
\begin{itemize}    
\item\index{ghosts!phi@$\phi$}
linear dilaton, the central charge is $+13$.\\
$\p\phi$ is a non-primary field with weight 1:
$$T^{\phi}(z)\p\phi(w) = {-2\over (z-w)^3} + {\p\phi(w) \over (z-w)^2} + 
{\p^2\phi(w) \over z-w}, $$
$$\mbox{where the stress tensor is } T^{\phi} = -{1 \over 2} (\p\phi)^2 - \p^2\phi.$$ So we have
$$[L^{\phi}_m,\phi_n] = -n \phi_{m+n} - m(m+1)\delta_{m+n}.$$

The vertex operator $e^{l\phi}$ has weight $-{1 \over 2} l^2 -l$.
\bqs
\phi(z)\phi(w) &\sim& -\log(z-w)\\
\p\phi(z)\p\phi(w) &\sim& -{1 \over (z-w)^2}\qquad\mbox{so}
\qquad [\phi_m,\phi_n] = -m\delta_{m+n}\\
\p\phi(z) e^{l \phi}(w) &\sim& -{l \over z-w}e^{l\phi}(w)\\
e^{q\phi}(z) e^{l\phi}(w) &\sim& (z-w)^{-q l} :e^{q\phi}(z) e^{l\phi}(w):\\
\eqs
\item\index{ghosts!etaxi@$\eta\xi$}
$\eta\xi$-system, the central charge is $-2$.\\
$\eta$ has weight 1, $\xi$ has weight $0$, and the stress tensor is 
$$T^{\eta\xi} = -\eta\p\xi.$$
$$\eta(z) \xi(w) \sim {1 \over z-w}
\qquad\mbox{so}\qquad \{\eta_m,\xi_n\} = \delta_{m+n}$$
\end{itemize}
\end{itemize}

\chapter{Some Technical Results}\label{c:TEX}
\section{Fixing the gauge invariance}\label{s:FS}
\index{Siegel gauge!in Berkovits' sft|(}
The linearized gauge transformation\refpj{B:gatr} reads 
$$
\delta \state{\Phi} =\state{\Xi_L} +\state{\Xi_R} 
\qquad\mbox{with}\quad Q\state{\Xi_L}=0 \mbox{ and }\eta_0\state{\Xi_R}=0\ .   
$$
We will prove that at the linearized level, the gauge invariance can be fixed by
imposing the Feynman-Siegel gauge\refpj{B:FSgauge}
$$
b_0 \state{\Phi}=0\quad\mbox{and}\quad\xi_0 \state{\Phi}=0\ .
$$
\begin{proof} The proof consists of two parts.
\begin{itemize}
\item
We
first show that the Siegel slice intersects each gauge orbit at least once. 
Suppose we have a state $\state{\Phi}$ with weight~$h \neq 0$. 
If
we define 
\bqs
\state{\tilde\Phi} &=&\state{\Phi}-{1 \over h} Q b_0\state{\Phi} - 
\eta_0 \xi_0\left( \state{\Phi}-{1 \over h} Q b_0\state{\Phi}\right) \\
&=& \xi_0 \eta_0 \left( \state{\Phi}-{1 \over h} Q b_0\state{\Phi}\right)\ ,
\eqs
then $\state{\tilde\Phi}$ belongs to the same gauge orbit and satisfies the
gauge condition\refpj{B:FSgauge}. 
\item
Next we prove that the Feynman-Siegel
gauge\refpj{B:FSgauge} fixes the gauge invariance completely, i.e.~there are no
pure gauge directions satisfying the Feynman-Siegel gauge condition. Suppose we
have a state $\state{\Xi}=\state{\Xi_L} +\state{\Xi_R}$ which satisfies the
Feynman-Siegel gauge and which is pure gauge
\bq Q \state{\Xi_L}&=&0 \qquad\mbox{and}\label{B:FSeq1}\\
\eta_0 \state{\Xi_R}&=&0\label{B:FSeq2} \ .
\eq
Then we have
\bqs
Q \eta_0 \state{\Xi} &=&Q \eta_0 \state{\Xi_L} + Q \eta_0 \state{\Xi_R}   
\stackrel{\ref{B:FSeq2}}{=}- \eta_0 Q\state{\Xi_L} \\
&\stackrel{\ref{B:FSeq1}}{=} &0 \\
b_0 \eta_0 \state{\Xi} &=& -\eta_0 b_0 \state{\Xi} 
\stackrel{\ref{B:FSgauge}}{=} 0
\eqs
From these two equations follows 
$$ \{ Q, b_0\} \eta_0 \state{\Xi} =0, \quad\mbox{hence}\quad
L_0^{\mbox{\scriptsize{tot}}} \eta_0 \state{\Xi} =0\ .$$
Because the conformal weight of $\state{\Xi}$ is different from zero, it follows
that
$$\eta_0 \state{\Xi}=0\ .$$
From this fact and the Feynman-Siegel gauge $\xi_0\state{\Xi}=0$ follows
trivially
$$\state{\Xi} = \{ \xi_0,\eta_0\} \state{\Xi} =0\ . $$
\end{itemize}
\end{proof}
\index{Siegel gauge!in Berkovits' sft|)}
\section{The conformal transformations of the fields}\label{s:transf}
\index{conformal transformation! of some non-primary fields|(}
We now list the conformal transformations of the fields used in the calculation
of the tachyon potential in section~\ref{s:level4}, see table~\ref{t:Berstates}. To shorten the notation we denote $f = f(z)$.
\begin{eqnarray*}
f \circ T(z) &=& (f'(z))^{-1/2} T(f)
\\
f \circ A(z) &=& f'(z) A(f) -
      \frac{f''(z)}{f'(z)}c\partial c\ \xi\partial\xi\ e^{-2 \phi}(f)
\\
f \circ E(z) &=& f'(z) E(f) -\frac{f''(z)}{2 f'(z)}
\\
f \circ F(z) &=& f'(z) F(f)
\\
f \circ K(z) &=& (f'(z))^{3/2} K(f) + 
2 \frac{f''(z)}{f'(z)} (f'(z))^{1/2}\xi c
\partial\left( e^{-\phi}\right)(f)+
\\*& &   
+ \left[ \frac{1}{2}\frac{f'''}{f'}-\frac{1}{4}
\left(\frac{f''}{f'}\right)^2 \right] (f'(z))^{-1/2} \xi c
e^{-\phi}(f)
\\
f \circ L(z) &=& (f'(z))^{3/2} L(f) +  \frac{f''(z)}{f'(z)}(f'(z))^{1/2} 
 \xi c \partial\phi\ e^{-\phi}(f)+
\\*& & 
+ \left[ \frac{3}{4}\left(\frac{f''}{f'}\right)^2-
\frac{2}{3}\frac{f'''}{f'}\right] (f'(z))^{-1/2} 
\xi c e^{-\phi}(f)
\\
f \circ M(z) &=& (f'(z))^{3/2} M(f) + \frac{15}{12} \left[\frac{f'''}{
f'} - \frac{3}{2}\left( \frac{f''}{f'}\right)^2\right](f'(z))^{-1/2}
\xi c e^{-\phi}(f)
\\
f \circ N(z) &=& (f'(z))^{3/2} N(f) -
 \frac{f''(z)}{f'(z)}(f'(z))^{1/2} \xi \partial c\ e^{-\phi}(f)+
\\*& & 
+ \left[2 \left(\frac{f''}{f'}\right)^2- \frac{f'''}{f'}\right]
(f'(z))^{-1/2} \xi c\ e^{-\phi}(f)
\\
f \circ P(z) &=& (f'(z))^{3/2} P(f) +
\frac{1}{2}\frac{f''(z)}{f'(z)}(f'(z))^{1/2} \partial\xi\ c
e^{-\phi}(f)+
\\*& & 
+ \left[ \frac{1}{4} \left(\frac{f''}{f'}\right)^2-
\frac{1}{6}\frac{f'''}{f'}\right] (f'(z))^{-1/2} \xi c e^{-\phi}(f)
\end{eqnarray*} 
We will now give two examples of the calculation of finite \cts.
\subsection*{The  \ct\ of $L = \xi c \p^2 \phi\ e^{-\phi}$}
It is clear that 
$f\circ L = f\circ \xi \ f \circ c \ f\circ :\p^2\phi e^{-\phi}:$.
To calculate the \ct\ of the $\phi$-piece we will follow the following strategy:
\begin{enumerate}
\item
We will deduce the \ct\ of $\p\phi$ by using the fact that $\p\phi$ is the
bosonisation of $:\beta\gamma:\ =  \p\phi$.
\item
It is trivial to obtain the \ct\ of $\p^2\phi$ from the \ct\ of $\p\phi$. 
\item
The \ct\ of all normal ordered vertex operators will be obtained by a
point-splitting regularization.
\end{enumerate}
While calculating OPE's we use the conventions of Polchinski's book, see section
\ref{conventions}\\
\emph{1. The \ct\ of $\p\phi$}\\
At the end of the calculation the small number $\ep$  is taken to zero.
\bqs
\lefteqn{f \circ :\beta\gamma:(z) =
 f \circ \left\{ \beta(z+ \ep) \gamma(z) +
\frac{1}{\ep} \right\}}\\
&=& f'(z+\ep)^{3/2} f'(z)^{-1/2} \beta(f(z+ \ep)) \gamma(f(z)) + \frac{1}{\ep}\\
&=& f'(z) : \beta\gamma:(f(z)) - f'(z+ \ep)^{3/2} f'(z)^{-1/2}
\frac{1}{f(z+\ep)-f(z)} + \frac{1}{\ep}\\
&=& f'(z) : \beta\gamma:(f(z)) - {f''(z) \over f'(z)}
\eqs
\emph{2. The \ct\ of $\p^2\phi$}\\
\bqs
f\circ \p^2 \phi &=& \p \left[ f \circ \p\phi \right] = 
\p \left[ f'\p\phi(f)- {f'' \over f'}\right]\\
&=& f'' \p\phi(f) + f'^2 \p^2\phi - { f' f''' - f'' f'' \over f'^2}\\
&=& f'^2 \p^2 \phi (f) + f''\p\phi(f)+ 
\left[ \left( {f'' \over f'}\right)^2 - { f'''
\over f'}\right]\\
\eqs 
\emph{3. The \ct\ of $\p^2\phi e^{-\phi}$}\\
To define the normal ordered products we need the following OPE's:
\bqs
\p\phi(z) e^{l\phi}(w) &=& -{ l \over z-w} e^{ l \phi}(w) + : \p\phi(z)
e^{l\phi}:(w)\\
\p^2\phi(z) e^{l\phi}(w) &=& { l \over (z-w)^2} e^{ l \phi}(w) + : \p^2\phi(z)
e^{l\phi}:(w)\\
\eqs
So we have
\bqs
: \p\phi e^{l\phi}:(z)&=& \lim_{\ep\to 0}\quad \p\phi(z) e^{l\phi}(z+\ep)-
{l \over \ep} e^{l\phi}(z+\ep)\\
: \p^2\phi e^{l\phi}:(z)&=& \lim_{\ep\to 0}\quad \p^2\phi(z) e^{l\phi}(z+\ep)-
{l \over \ep^2} e^{l\phi}(z+\ep)\\
\eqs
The vertex operator $e^{l \phi}$ is a primary field of weight $-{1 \over 2} l^2
-l$. Taking all the previous results together we have (suppressing the limit)
\bqs \lefteqn{f\circ :\p^2\phi e^{-\phi}:(z) = 
f\circ\left\{\p^2\phi (z) e^{-\phi}(z+\ep) + {1 \over \ep^2}
e^{-\phi}(z+\ep)\right\}}\\
&=& \left( f'(z)^2 \p^2\phi(f(z)) + f''(z)\p\phi(f(z))+ 
\left[ \left( {f'' \over f'}\right)^2 - { f'''\over f'}\right]\right)\\
&&\qquad\qquad\cdot
f'(z+\ep)^{1/2} e^{-\phi}(f(z+\ep))\\
&&+{1 \over \ep^2} f'(z+\ep)^{1/2} e^{-\phi}(f(z+\ep))\\
&=& f'(z)^{5/2}:\p^2\phi e^{-\phi}:(f(z)) 
- { f'(z)^2 f'(z+\ep)^{1/2} \over \left[ f(z+\ep) - f(z)\right]^2 }
e^{-\phi}(f(z+\ep))\\
&& + f''(z) f'(z)^{1/2}:\p\phi e^{-\phi}:(f(z)) 
- { f''(z) f'(z+\ep)^{1/2} \over f(z+\ep) - f(z) } e^{-\phi}(f(z+\ep))\\
&& +\left[ \left( {f'' \over f'}\right)^2 - { f'''\over f'}\right] f'(z)^{1/2} 
e^{-\phi}(f)+{1 \over \ep^2} f'(z+\ep)^{1/2} e^{-\phi} (f(z+\ep))
\eqs
Taking the limit $\ep\to 0 $ one finds
\bqs 
\lefteqn{f\circ :\p^2\phi e^{-\phi}:(z) = f'(z)^{5/2}:\p^2\phi e^{-\phi}:(f)}\\
& & +
 {f''\over f'} f'(z)^{3/2}:\p\phi e^{-\phi}:(f)
 + \left[{3 \over 4}  \left( {f'' \over f'}\right)^2 
       -{2 \over 3} { f'''\over f'}\right] f'(z)^{1/2} e^{-\phi}
\eqs
To finish the calculation one needs $f\circ\xi = \xi(f)$ and $f\circ c =
(f')^{-1} c(f)$.
\subsection*{The  \ct\ of $L = \xi\p\xi\ \eta c e^{-\phi}$}
We will also do this \ct\ explicitly because it will give us a little more
insight in the normal ordering procedure. 

To define the normal ordered product $:\xi\p\xi\ \eta:$ we first have to compute
the OPE between  $\xi\p\xi$ and $\eta$
\bqs
\xi\p\xi(z)\ \eta(w) &=& -{1\over z-w} \p\xi(z) -{1\over (z-w)^2 }\xi(z) 
+ :\xi\p\xi(z)\ \eta(w):\\
&=&  -{1\over z-w} \p\xi(w) -\p^2 \xi(w)\\
&& -{1\over (z-w)^2} \xi(w ) -{1\over (z-w)} \p \xi(w )-{1\over 2} \p^2 \xi(w)\\
&& + :\xi\p\xi(z)\ \eta(w):+ O(z-w) \\
\eqs
So we find 
$$ :\xi\p\xi\ \eta(z): = \lim_{\ep\to 0} \left\{ \xi\p\xi(z+\ep)\eta(z) 
+ {1\over \ep^2} \xi(z) +  {2\over \ep} \p\xi(z) + {3\over 2} \p^2
\xi(z)\right\}.$$
We emphasize at this point that to define a normal ordered product we have thus
to subtract a finite piece as well.

Because $\xi$ is a primary field of weight 0 and $\eta$ is a primary field of
weight 1, we find
\bqs
\lefteqn{f\circ :\xi\p\xi\ \eta(z):= 
f'(z+\ep) f'(z)\  \xi\p\xi(f(z+\ep))\eta(f(z))}\\
&& + {1\over \ep^2} \xi(f(z)) +  {2\over \ep} f'(z)\ \p\xi(f(z)) 
+ {3\over 2} f''(z)\ \p\xi(f(z))+ {3\over 2} f'(z)^2\ \p^2\xi(f(z))\\
&=& f'(z)^2\  :\xi\p\xi\ \eta:(f(z))\\
&&- f'(z+\ep) f'(z) \left\{ {\xi(f(z))\over [ f(z+\ep) -f(z)]^2}
+2 {\p\xi(f(z))\over f(z+\ep) -f(z)}+{3\over  2} \p^2\xi(f(z))\right\}\\ 
&& + {1\over \ep^2} \xi(f) +  {2\over \ep} f'\ \p\xi(f) 
+ {3\over 2} f''\ \p\xi(f)+ {3\over 2} (f')^2\ \p^2\xi(f)\\
&=&f'(z)^2\  :\xi\p\xi\ \eta:(f)+ \left[{1\over 4} \left({f''\over f'}\right)^2
-{1\over 6} {f'''\over f'}\right] \xi(f) + {1\over2} f'' \p\xi(f)
\eqs 
Thus confirming the \ct\ on the field $P$. As an extra check one can calculate 
$f\circ :\xi\p\xi\ \eta:$ by doing the point splitting between the fields 
$:\xi\eta:$ and $\p\xi$. Needless to say, this gives the same result. That would
not be the case if one did not subtract a finite piece to define the normal ordered
product.
\index{conformal transformation! of some non-primary fields|)}
\section{Partition function for Berkovits' string field
theory}\label{s:partition}
It is easy to write the partition 
function\index{partition function!in Berkovits' super
sft} if we let all oscillators act on the
vacuum $\state{\Omega} = c_1 \state{-1}$, where $\state{-1}$ is the vacuum with
picture $-1$, see section~\ref{superstring}. The vacuum $\state{\Omega}$ is
equivalently defined by
\bqs
b_n \state{\Omega}&=& 0 \qquad\mbox{if    } n \ge 0,\\
c_n \state{\Omega}&=& 0 \qquad\mbox{if    } n \ge 1,\\
\gamma_r,\beta_r \state{\Omega}&=& 0 \qquad\mbox{if    } r \ge 1/2,\\
L_n \state{\Omega}&=& 0 \qquad\mbox{if    } n \ge -1,\\
G_r \state{\Omega}&=& 0 \qquad\mbox{if    } r \ge -1/2.\\
\eqs
Notice that $c_0$ does not annihilate this vacuum. However, due to the
Siegel-gauge, fields with non-zero conformal weight will not contain the $c_0$
oscillator. 
Defining $N_{n,k}$ to be the number of fields after gauge fixing 
at level $n$ with ghost number $k$, we have
\bqs
\sum_{n\ge 0,\ k \in\mathbb{Z}} N_{n,k}\ q^n g^k &=&
(b\ '\mbox{s}) (c\ '\mbox{s})(\beta\ '\mbox{s}) (\gamma\ '\mbox{s})
(L^{\mbox{m}}\ '\mbox{s}) (G^{\mbox{m}}\ '\mbox{s})\\
&=&\prod_{n=1}^{\infty}(1+ q^n g^{-1}) \prod_{n=1}^{\infty}(1+ q^n g)
\prod_{n=1/2}^{\infty}{1\over 1-q^n g^{-1}}\\
&& \prod_{n=1/2}^{\infty}{1\over 1-q^n g}
 \prod_{n=2}^{\infty}{1\over 1-q^n } \prod_{n=3/2}^{\infty}
(1+q^n) \\
\eqs 
We expand these products and keep only the ghost number~0 terms. Hence we find
for the number of space-time fields in the string field at every level:
\bqs
 \sum_{n \ge 0} N_{n,1} q^n&=& 1 + q + 3\ q^{3/2} + 5\ q^2 + 8\ q^{5/2}
 + 13\ q^3 + 23\ q^{7/2}\\ 
 && + 38\ q^4+ 58\ q^{9/2} + 89\ q^5 + 
    141\ q^{11/2} + 216\ q^6 + 318\ q^{13/2} \\ &&+ 470\ q^7 + 
    699\ q^{15/2} + 1019\ q^8 + O(q^{9/2})
\eqs
As a check we can see that there is 1 field at level~0, this is the tachyon.
There are 3 fields at level~$3/2$, these are the 3 fields used by Berkovits, Sen
and Zwiebach~\cite{BSZ}. There are 5 fields at level~2, this also agrees with
the table~\ref{t:Berstates}.
\section{Examples of conservation laws}\label{s:vbcons}
\index{conservation laws! examples|(}
\subsection[for the field $\eta$]{Conservation laws for the field $\eta$}\label{s:conseta}
\begin{itemize}
\item{Some examples for $n=3$.}\\
If we take $$\varphi(z) = {4 i \over 3} \left({1\over z-1}+ {1\over2}\right),$$
we obtain the conservation law
\bqs
\lefteqn{0=\etats{V_3}\left(\eta_{-1} + \frac{5}{27}\ \eta_1 - 
  \frac{32}{729}\ \eta_3 + 
  \frac{416}{19683}\ \eta_5 + \cdots\right)^{(1)}+ }\\
&&\etats{V_3}\left(
  \frac{2}{3\,{\sqrt{3}}}\ \eta_0 - 
  \frac{16}{27}\ \eta_1 + 
  \frac{32}{81\,{\sqrt{3}}}\ \eta_2 + 
  \frac{16}{729}\ \eta_3 - 
  \frac{64}{729\,{\sqrt{3}}}\ \eta_4 
  + \cdots \right)^{(2)}+\\
&&\etats{V_3}\left(-\frac{2}{3\,{\sqrt{3}}}\ \eta_0 - 
  \frac{16}{27}\ \eta_1 - 
  \frac{32}{81\,{\sqrt{3}}}\ \eta_2 + 
  \frac{16}{729}\ \eta_3 + 
  \frac{64}{729\,{\sqrt{3}}}\ \eta_4 
  +\cdots\right)^{(3)} 
\eqs
If we take  $$\varphi(z) = -{16 \over 9} \left({1\over (z-1)^2}+ {1\over z-1}\right),$$
we obtain 
\bqs
\lefteqn{0=\etats{V_3}\left(\ \eta_{-2} + \frac{22}{27}\ \eta_0 - 
  \frac{13}{243}\ \eta_2 + 
  \frac{512}{19683}\ \eta_4+\cdots\right)^{(1)}+ }\\
 &&\etats{V_3}\left( 
  \frac{16}{27}\ \eta_0 - 
  \frac{64}{81\,{\sqrt{3}}}\ \eta_1 + 
  \frac{128}{243}\ \eta_2 - 
  \frac{320}{729\,{\sqrt{3}}}\ \eta_3 - 
  \frac{256}{19683}\ \eta_4 
  +\cdots\right)^{(2)}+ \\
 &&\etats{V_3}\left(  \frac{16}{27}\ \eta_0 + 
  \frac{64}{81\,{\sqrt{3}}}\ \eta_1 + 
  \frac{128}{243}\ \eta_2 + 
  \frac{320}{729\,{\sqrt{3}}}\ \eta_3 - 
  \frac{256}{19683}\ \eta_4 
  +\cdots\right)^{(3)}  
 \eqs
\item{Some Examples for $n=4$.}\\
For $$\varphi(z) = i \left({1\over (z-1)} + \frac{1}{2}\right)$$ we have
 \bqs
0&=&\etats{V_4}\left(\ \eta_{-1} + \frac{1}{4}\ \eta_1 - 
  \frac{1}{16}\ \eta_3 + 
  \frac{1}{32}\ \eta_5+\cdots\right)^{(1)}+ \\
&& \etats{V_4}\left(\frac{1}{2}\ \eta_0 - \frac{1}{2}\ \eta_1 + 
  \frac{1}{4}\ \eta_2 - 
  \frac{1}{16}\ \eta_4+\cdots\right)^{(2)}+ \\
&&  \etats{V_4}\left(-\frac{1}{4}\ \eta_1 + 
  \frac{1}{16}\ \eta_3 - 
  \frac{1}{32}\ \eta_5+\cdots\right)^{(3)}+ \\
&&  \etats{V_4}\left(-\frac{1}{2}\ \eta_0 - \frac{1}{2}\ \eta_1 - 
  \frac{1}{4}\ \eta_2 + 
  \frac{1}{16}\ \eta_4+\cdots\right)^{(4)} \\
  \eqs
\end{itemize}
\subsection[for the super stress tensor $G$]{Conservation laws for the super stress tensor $G$}\label{s:consG}
In this case we will have a kind of
anti-cyclicity among the conservation laws. As an illustration we have for the
following functions:
\bqs
\varphi(z) &=&\left( -\frac{4}{3}  + 
    \frac{4\,i }{3} \right) \,
  {\sqrt{\frac{2}{3}}}\,
  \left( 1 + \frac{1}{z-1} \right)\\
\varphi(z) &=& \left( -\frac{2}{9}  + 
     \frac{2\,i }{9} \right) \,
   \sqrt{2}\,\left( 3\,i  + 
     \sqrt{3} \right)  + 
  \frac{\left( \frac{4}{3} - 
       \frac{4\,i }{3} \right) \,
     {\sqrt{\frac{2}{3}}}}{z-\omega}\\ 
\varphi(z)&=& \frac{\left( \frac{8}{3} - 
       \frac{8\,i }{3} \right) \,
     {\sqrt{2}}}{3\,i  + {\sqrt{3}}} -
   \frac{\left( \frac{4}{3} - 
       \frac{4\,i }{3} \right) \,
     {\sqrt{\frac{2}{3}}}}{z-\omega^2}    
\eqs  
these conservation laws respectively:
\bqs
\lefteqn{0 = \etats{V_3}\left(G_{-3/2} + \frac{49}{54}\ G_{1/2} - 
  \frac{181}{5832}\ G_{5/2}+\cdots  \right)^{(1)}+ }\\
  &&\etats{V_3}\left( \frac{4}{3\,{\sqrt{3}}}\ G_{-1/2} - 
  \frac{8}{27}\ G_{1/2} + 
  \frac{94}{81\,{\sqrt{3}}}\ G_{3/2} - 
  \frac{148}{729}\ G_{5/2}+\cdots\right)^{(2)}+\\
   &&\etats{V_3}\left( \frac{4}{3\,{\sqrt{3}}}\ G_{-1/2} + 
  \frac{8}{27}\ G_{1/2} + 
  \frac{94}{81\,{\sqrt{3}}}\ G_{3/2} + 
  \frac{148}{729}\ G_{5/2}+\cdots\right)^{(3)}\\  
\eqs
\bqs
\lefteqn{0= \etats{V_3}\left( -\frac{4}{3\,{\sqrt{3}}}\ G_{-1/2} - 
  \frac{8}{27}\ G_{1/2} - 
  \frac{94}{81\,{\sqrt{3}}}\ G_{3/2} - 
  \frac{148}{729}\ G_{5/2}+\cdots\right)^{(1)}+}\\ 
  && \etats{V_3}\left(G_{-3/2} + \frac{49}{54}\ G_{1/2} - 
  \frac{181}{5832}\ G_{5/2}+\cdots  \right)^{(2)}+ \\
  && \etats{V_3}\left( \frac{4}{3\,{\sqrt{3}}}\ G_{-1/2} - 
  \frac{8}{27}\ G_{1/2} + 
  \frac{94}{81\,{\sqrt{3}}}\ G_{3/2} - 
  \frac{148}{729}\ G_{5/2}+\cdots\right)^{(3)}\\
\eqs
\bqs
\lefteqn{0 = \etats{V_3}\left(- \frac{4}{3\,{\sqrt{3}}}\ G_{-1/2} + 
  \frac{8}{27}\ G_{1/2} - 
  \frac{94}{81\,{\sqrt{3}}}\ G_{3/2} + 
  \frac{148}{729}\ G_{5/2}+\cdots\right)^{(1)}+}\\
  && \etats{V_3}\left( -\frac{4}{3\,{\sqrt{3}}}\ G_{-1/2} - 
  \frac{8}{27}\ G_{1/2} - 
  \frac{94}{81\,{\sqrt{3}}}\ G_{3/2} - 
  \frac{148}{729}\ G_{5/2}+\cdots\right)^{(2)}+\\
   && \etats{V_3}\left(G_{-3/2} + \frac{49}{54}\ G_{1/2} - 
  \frac{181}{5832}\ G_{5/2}+\cdots  \right)^{(3)} \\ 
\eqs
\subsection[for the stress tensor $T$]{Conservation laws for the stress tensor $T$}\label{s:consT}
\emph{Some examples for the four-vertex}\\
Using the conformal field
$$\varphi(z) =  -{1 \over z - 1} - { 3 \over 2}  - 
{7 \over 4} (z - 1) - {5 \over 8}  ( z - 1)^2,$$
we find the following conservation law
\bq
0 &=& \etats{V_4}\left( {c \over 8} + L_{-2} - 
  \frac{7}{16}L_2 + 
  \frac{1}{8}L_4 + \cdots  \right)^{(1)}+\nonumber\\
  && \etats{V_4}\left( -\frac{3}{4}L_{-1}  - 
  \frac{1}{2}L_0 + \frac{1}{8}L_1 - 
  \frac{1}{2}L_2 + 
  \frac{11}{32}L_3 - 
  \frac{7}{64}L_5 + \cdots\right)^{(2)}+\nonumber\\
  && \etats{V_4}\left( L_0 + 
  \frac{7}{16}L_2 - 
  \frac{1}{8}L_4 + \cdots  \right)^{(3)}+\label{C:eq1}\\
   && \etats{V_4}\left( \frac{3}{4}L_{-1}  - 
  \frac{1}{2}L_0 - \frac{1}{8}L_1 - 
  \frac{1}{2}L_2 - 
  \frac{11}{32}L_3 +
  \frac{7}{64}L_5 + \cdots\right)^{(4)}\nonumber
 \eq  
If we take
$$\varphi(z) = {-i \over (z - 1)^2} - {2\ i  \over z - 1}
 - {11\ i\over 4}    - {7\ i\over 4}  (z - 1) -  {17\ i\over 16} ( z - 1)^2\ ,$$ 
 we have
 \bq
0 &=& \etats{V_4}\left(L_{-3} + \frac{15}{64} L_3  +
\cdots\right)^{(1)}+\nonumber\\
   & & \etats{V_4}\left(\frac{7}{8}L_{-1} + 
  \frac{13}{8}L_0 + 
  \frac{11}{8}L_1 + 
  \frac{5}{16}L_2 + 
  \frac{1}{2}L_3 - 
  \frac{29}{64}L_4 +\cdots  \right)^{(2)}+\nonumber\\
   & & \etats{V_4}\left(\frac{11}{4}L_{-1} + 
  \frac{7}{4}L_1 - 
  \frac{15}{64}L_3 + 
  \frac{31}{256}L_5 + \cdots \right)^{(3)}+\label{C:eq2}\\
   & & \etats{V_4}\left(\frac{7}{8}L_{-1} - 
  \frac{13}{8}L_0 + 
  \frac{11}{8}L_1 - 
  \frac{5}{16}L_2 + 
  \frac{1}{2}L_3 + 
  \frac{29}{64}L_4 +\cdots  \right)^{(4)}\nonumber
  \eq
\index{conservation laws! examples} 
\section{Higher levels in super string field theory}\label{s:meerV}
\subsection{The tachyon potential at level $(2,5)$}\label{s:V25}
\bqs
& & V(\widehat\Phi)_{9/2}=\\
& & 2.28089A_{ 3/2,1}A_{ 3/2,2}A_{ 3/2,3} + 
  1.25449A_{ 3/2,3}A_{ 3/2,1}^2 - 
  2.28089A_{ 3/2,1}A_{ 3/2,2}^2\\[4pt]&& - 
  2.28089A_{ 3/2,3}A_{ 3/2,2}^2 + 
  4.51847A_{ 3/2,3}A_{ 0,1}^2A_{ 3/2,2}^2 - 
  4.56178A_{ 3/2,2}^3 \\[4pt]&&+ 
  0.228089A_{ 3/2,1}A_{ 3/2,3}^2 - 
  0.545347A_{ 3/2,1}A_{ 0,1}^2\,
   A_{ 3/2,3}^2\\[4pt]&&
    + 1.6348A_{ 3/2,2}A_{ 0,1}^2\,
   A_{ 3/2,3}^2 + 0.214785A_{ 0,1}^2\,
   A_{ 3/2,3}^3 + 0.132886A_{ 0,1}^4\,
   A_{ 3/2,3}^3\\[3ex]
& & V(\widehat\Phi)_{5}=\\
& & -3.84446A_{ 0,1}A_{ 3/2,1}A_{ 3/2,3}A_{ 2,1} - 
  6.7407A_{ 0,1}A_{ 3/2,2}A_{ 3/2,3}A_{ 2,1} \\[4pt]&&+ 
  9.30758A_{ 0,1}A_{ 3/2,1}A_{ 3/2,3}A_{ 2,2} + 
  24.5749A_{ 0,1}A_{ 3/2,2}A_{ 3/2,3}A_{ 2,2}\\[4pt]&& - 
  0.5A_{ 0,1}A_{ 3/2,1}A_{ 3/2,3}A_{ 2,3} - 
  1.47927A_{ 0,1}A_{ 3/2,2}A_{ 3/2,3}A_{ 2,3}\\[4pt]&& - 
  1.84987A_{ 0,1}A_{ 3/2,1}A_{ 3/2,3}A_{ 2,4} + 
  0.580583A_{ 0,1}A_{ 3/2,2}A_{ 3/2,3}A_{ 2,4}\\[4pt]&& - 
  5.91216A_{ 0,1}A_{ 3/2,1}A_{ 3/2,3}A_{ 2,5}- 
  0.0444174A_{ 0,1}A_{ 3/2,2}A_{ 3/2,3}A_{ 2,5}\\[4pt]&& + 
  8.24651A_{ 0,1}A_{ 2,1}A_{ 3/2,2}^2 - 
  20.2757A_{ 0,1}A_{ 2,2}A_{ 3/2,2}^2\\[4pt]&& - 
  1.45527A_{ 0,1}A_{ 2,3}A_{ 3/2,2}^2 + 
  10.672A_{ 0,1}A_{ 2,4}A_{ 3/2,2}^2 \\[4pt]&&+ 
  12.6123A_{ 0,1}A_{ 2,5}A_{ 3/2,2}^2- 
  1.05709A_{ 0,1}A_{ 2,1}A_{ 3/2,3}^2 \\[4pt]&&+ 
  2.4954A_{ 0,1}A_{ 2,2}A_{ 3/2,3}^2 - 
  0.41728A_{ 0,1}A_{ 2,3}A_{ 3/2,3}^2\\[4pt]&& - 
  0.977327A_{ 0,1}A_{ 2,4}A_{ 3/2,3}^2 - 
  0.686273A_{ 0,1}A_{ 2,5}A_{ 3/2,3}^2 + 
  1.02007A_{ 2,1}A_{ 0,1}^3A_{ 3/2,3}^2\\[4pt]&& - 
  2.46914A_{ 2,2}A_{ 0,1}^3A_{ 3/2,3}^2 + 
  0.0399177A_{ 2,3}A_{ 0,1}^3A_{ 3/2,3}^2 + 
  0.706811A_{ 2,4}A_{ 0,1}^3A_{ 3/2,3}^2\\[4pt]&& + 
  1.4877A_{ 2,5}A_{ 0,1}^3A_{ 3/2,3}^2
\eqs
\subsection{The tachyon potential at level $(2,6)$}\label{s:V26}
\bqs
& & V(\widehat\Phi)_{11/2}=\\
& & 0.565844A_{ 3/2,1}A_{ 2,1}A_{ 2,2} - 
  3.6214A_{ 3/2,2}A_{ 2,1}A_{ 2,2} - 
  0.900206A_{ 3/2,3}A_{ 2,1}A_{ 2,2}\\[4pt]&& + 
  0.931413A_{ 3/2,1}A_{ 2,1}A_{ 2,3} + 
  1.20713A_{ 3/2,2}A_{ 2,1}A_{ 2,3} + 
  0.258573A_{ 3/2,3}A_{ 2,1}A_{ 2,3}\\[4pt]&& - 
  1.08025A_{ 3/2,1}A_{ 2,2}A_{ 2,3} - 
  5.10288A_{ 3/2,2}A_{ 2,2}A_{ 2,3} - 
  1.31173A_{ 3/2,3}A_{ 2,2}A_{ 2,3}\\[4pt]&& - 
  1.62414A_{ 3/2,1}A_{ 2,1}A_{ 2,4} - 
  4.82853A_{ 3/2,2}A_{ 2,1}A_{ 2,4} + 
  0.559671A_{ 3/2,3}A_{ 2,1}A_{ 2,4}\\[4pt]&& + 
  4.93827A_{ 3/2,1}A_{ 2,2}A_{ 2,4} + 
  16.4609A_{ 3/2,2}A_{ 2,2}A_{ 2,4} - 
  0.411523A_{ 3/2,3}A_{ 2,2}A_{ 2,4}\\[4pt]&& - 
  0.395062A_{ 3/2,1}A_{ 2,3}A_{ 2,4} + 
  0.438957A_{ 3/2,2}A_{ 2,3}A_{ 2,4} + 
  0.559671A_{ 3/2,3}A_{ 2,3}A_{ 2,4}\\[4pt]&& - 
  2.06722A_{ 3/2,1}A_{ 2,1}A_{ 2,5} - 
  4.82853A_{ 3/2,2}A_{ 2,1}A_{ 2,5} + 
  0.0404664A_{ 3/2,3}A_{ 2,1}A_{ 2,5}\\[4pt]&& + 
  7.04733A_{ 3/2,1}A_{ 2,2}A_{ 2,5} + 
  18.7654A_{ 3/2,2}A_{ 2,2}A_{ 2,5} - 
  0.231481A_{ 3/2,3}A_{ 2,2}A_{ 2,5}\\[4pt]&& - 
  0.618656A_{ 3/2,1}A_{ 2,3}A_{ 2,5} - 
  0.329218A_{ 3/2,2}A_{ 2,3}A_{ 2,5} + 
  0.314815A_{ 3/2,3}A_{ 2,3}A_{ 2,5}\\[4pt]&& + 
  1.27298A_{ 3/2,1}A_{ 2,4}A_{ 2,5} + 
  3.0727A_{ 3/2,2}A_{ 2,4}A_{ 2,5} + 
  0.274348A_{ 3/2,3}A_{ 2,4}A_{ 2,5}\\[4pt]&& - 
  15.037A_{ 3/2,3}A_{ 2,1}A_{ 2,2}A_{ 0,1}^2 + 
  1.24146A_{ 3/2,3}A_{ 2,1}A_{ 2,3}A_{ 0,1}^2\\[4pt]&& - 
  3.81272A_{ 3/2,3}A_{ 2,2}A_{ 2,3}A_{ 0,1}^2 + 
  5.15319A_{ 3/2,3}A_{ 2,1}A_{ 2,4}A_{ 0,1}^2\\[4pt]&& - 
  12.2439A_{ 3/2,3}A_{ 2,2}A_{ 2,4}A_{ 0,1}^2 + 
  0.673208A_{ 3/2,3}A_{ 2,3}A_{ 2,4}A_{ 0,1}^2\\[4pt]&& + 
  11.0543A_{ 3/2,3}A_{ 2,1}A_{ 2,5}A_{ 0,1}^2 - 
  30.3438A_{ 3/2,3}A_{ 2,2}A_{ 2,5}A_{ 0,1}^2\\[4pt]&& + 
  1.29973A_{ 3/2,3}A_{ 2,3}A_{ 2,5}A_{ 0,1}^2 + 
  4.83567A_{ 3/2,3}A_{ 2,4}A_{ 2,5}A_{ 0,1}^2 - 
  0.0390947A_{ 3/2,1}A_{ 2,1}^2 \\[4pt]&&+ 
  0.332305A_{ 3/2,3}A_{ 2,1}^2 + 
  3.68912A_{ 3/2,3}A_{ 0,1}^2A_{ 2,1}^2 - 
  2.28138A_{ 3/2,1}A_{ 2,2}^2 \\[4pt]&&+ 
  9.71193A_{ 3/2,2}A_{ 2,2}^2 - 
  1.49306A_{ 3/2,3}A_{ 2,2}^2 + 
  16.5939A_{ 3/2,3}A_{ 0,1}^2A_{ 2,2}^2\\[4pt]&& + 
  1.27778A_{ 3/2,1}A_{ 2,3}^2+ 
  2.96296A_{ 3/2,2}A_{ 2,3}^2 + 
  0.167695A_{ 3/2,3}A_{ 2,3}^2\\[4pt]&& - 
  0.831355A_{ 3/2,3}A_{ 0,1}^2A_{ 2,3}^2 - 
  1.0535A_{ 3/2,1}A_{ 2,4}^2 - 
  3.51166A_{ 3/2,2}A_{ 2,4}^2 \\[4pt]&&- 
  0.526749A_{ 3/2,3}A_{ 2,4}^2 + 
  3.09498A_{ 3/2,3}A_{ 0,1}^2A_{ 2,4}^2 + 
  4.86077A_{ 3/2,1}A_{ 2,5}^2 \\[4pt]&&+ 
  6.58436A_{ 3/2,2}A_{ 2,5}^2- 
  0.586763A_{ 3/2,3}A_{ 2,5}^2 + 
  5.18785A_{ 3/2,3}A_{ 0,1}^2A_{ 2,5}^2\\[3ex]
& & V(\widehat\Phi)_{6}=\\
& & -0.9375A_{ 0,1}A_{ 2,1}A_{ 2,2}A_{ 2,3} - 
  22.3438A_{ 0,1}A_{ 2,1}A_{ 2,2}A_{ 2,4} + 
  1.0625A_{ 0,1}A_{ 2,1}A_{ 2,3}A_{ 2,4}\\[4pt]&& - 
  3.28125A_{ 0,1}A_{ 2,2}A_{ 2,3}A_{ 2,4} - 
  32.7344A_{ 0,1}A_{ 2,1}A_{ 2,2}A_{ 2,5} + 
  1.75A_{ 0,1}A_{ 2,1}A_{ 2,3}A_{ 2,5}\\[4pt]&& - 
  8.67188A_{ 0,1}A_{ 2,2}A_{ 2,3}A_{ 2,5} + 
  20.3958A_{ 0,1}A_{ 2,1}A_{ 2,4}A_{ 2,5} - 
  53.2813A_{ 0,1}A_{ 2,2}A_{ 2,4}A_{ 2,5}\\[4pt]&& + 
  0.6875A_{ 0,1}A_{ 2,3}A_{ 2,4}A_{ 2,5} - 
  8.48958A_{ 3/2,1}A_{ 3/2,3}A_{ 3/2,2}^2 - 
  20.3125A_{ 3/2,3}A_{ 3/2,2}^3\\[4pt]&& + 
  12.6563A_{ 3/2,2}^4 + 
  1.97917A_{ 3/2,1}A_{ 3/2,2}A_{ 3/2,3}^2 + 
  1.16667A_{ 3/2,1}^2A_{ 3/2,3}^2\\[4pt]&& - 
  1.19792A_{ 3/2,2}^2A_{ 3/2,3}^2 + 
  2.75333A_{ 0,1}^2A_{ 3/2,2}^2\,
   A_{ 3/2,3}^2\\[4pt]&&
    + 0.03125A_{ 3/2,1}\,
   A_{ 3/2,3}^3 - 0.636974A_{ 3/2,1}\,
   A_{ 0,1}^2A_{ 3/2,3}^3\\[4pt]&& - 
  0.890861A_{ 3/2,2}A_{ 0,1}^2\,
   A_{ 3/2,3}^3 - 0.106644A_{ 0,1}^2\,
   A_{ 3/2,3}^4\\[4pt]&& + 0.0463388A_{ 0,1}^4\,
   A_{ 3/2,3}^4 - 5.46875A_{ 0,1}A_{ 2,2}\,
   A_{ 2,1}^2\\[4pt]&& + 0.59375A_{ 0,1}A_{ 2,3}\,
   A_{ 2,1}^2 + 5.27083A_{ 0,1}A_{ 2,4}\,
   A_{ 2,1}^2\\[4pt]&& + 8.35417A_{ 0,1}A_{ 2,5}\,
   A_{ 2,1}^2 + 0.645833A_{ 0,1}A_{ 2,1}^3\\[4pt]&& + 
  20.918A_{ 0,1}A_{ 2,1}A_{ 2,2}^2 - 
  3.69141A_{ 0,1}A_{ 2,3}A_{ 2,2}^2 + 
  27.0703A_{ 0,1}A_{ 2,4}A_{ 2,2}^2\\[4pt]&& + 
  31.9922A_{ 0,1}A_{ 2,5}A_{ 2,2}^2 - 
  27.1289A_{ 0,1}A_{ 2,2}^3 - 
  2.51042A_{ 0,1}A_{ 2,1}A_{ 2,3}^2\\[4pt]&& + 
  5.625A_{ 0,1}A_{ 2,2}A_{ 2,3}^2 - 
  3.52083A_{ 0,1}A_{ 2,4}A_{ 2,3}^2 - 
  4.79167A_{ 0,1}A_{ 2,5}A_{ 2,3}^2\\[4pt]&& + 
  0.25A_{ 0,1}A_{ 2,3}^3 + 
  10.8021A_{ 0,1}A_{ 2,1}A_{ 2,4}^2 - 
  25.1563A_{ 0,1}A_{ 2,2}A_{ 2,4}^2\\[4pt]&& + 
  0.09375A_{ 0,1}A_{ 2,3}A_{ 2,4}^2 + 
  15.0625A_{ 0,1}A_{ 2,5}A_{ 2,4}^2 + 
  7.5625A_{ 0,1}A_{ 2,4}^3\\[4pt]&& + 
  20.6927A_{ 0,1}A_{ 2,1}A_{ 2,5}^2 - 
  67.1094A_{ 0,1}A_{ 2,2}A_{ 2,5}^2 - 
  3.07813A_{ 0,1}A_{ 2,3}A_{ 2,5}^2\\[4pt]&& + 
  16.1563A_{ 0,1}A_{ 2,4}A_{ 2,5}^2 - 
  3.21875A_{ 0,1}A_{ 2,5}^3\\
\eqs
\subsection{The fields at level~$7/2$}\label{s:f72}
At level~$7/2$ there are 23~fields, see the partition function in
appendix~\ref{s:partition}.
These fields are all in the GSO$(+)$ sector and therefore are to be 
tensored with the unit CP factor, see equation\refpj{hats}.
The fields are:
\bqs
\lefteqn{\state{\Phi_{7/2}}=
A_{7/2,1}\, \xi_{-1} \xi_{0}c_{-3}c_{1} \state{-2} + 
A_{7/2,2}\,\xi_{-2} \xi_{0}c_{-2}c_{1} \state{-2} +}\\
& & 
A_{7/2,3}\, \xi_{-3} \xi_{0}c_{-1}c_{1}\state{-2}+ 
A_{7/2,4}\,L_{-2} \xi_{-1} \xi_{0}c_{-1}c_{1}\state{-2} +\\
& &  
A_{7/2,5}\,\xi_{-1} \xi_{0}c_{-1}c_{1} \phi_{-2}\state{-2}+
A_{7/2,6}\,\xi_{-1} \xi_{0}c_{-2}c_{1} \phi_{-1}\state{-2} +\\
& &  
A_{7/2,7}\,\xi_{-2} \xi_{0}c_{-1}c_{1} \phi_{-1}\state{-2}+ 
A_{7/2,8}\,\xi_{-1} \xi_{0}c_{-1}c_{1} \phi_{-1}\phi_{-1} \state{-2}+\\
& & 
A_{7/2,9}\,G_{-3/2} \xi_{0}c_{-1} \state{-1}+ 
A_{7/2,10}\, G_{-7/2} \xi_{0}c_{1}\state{-1}+\\
& & 
A_{7/2,11}\,L_{-2} G_{-3/2} \xi_{0}c_{1} \state{-1}+ 
A_{7/2,12}\, G_{-3/2} \eta_{-1} \xi_{-1}\xi_{0}c_{1} \state{-1}+\\
& &  
A_{7/2,13}\, G_{-3/2} \xi_{0}c_{1}\phi_{-2} \state{-1}+
A_{7/2,14}\,G_{-5/2} \xi_{0}c_{1} \phi_{-1}\state{-1}+\\
& & 
A_{7/2,15}\,G_{-3/2} \xi_{0}c_{1} \phi_{-1} \phi_{-1}\state{-1}+ 
A_{7/2,16}\,\eta_{-3} \xi_{0} \state{0}+\\
& & 
A_{7/2,17}\,L_{-2} \eta_{-1} \xi_{0}\state{0}+ 
A_{7/2,18}\,\eta_{-1} \xi_{0}b_{-3}c_{1} \state{0}+\\
& & 
A_{7/2,19}\, \eta_{-2} \xi_{0}b_{-2}c_{1} \state{0}+ 
A_{7/2,20}\,\eta_{-1} \xi_{0} \phi_{-2} \state{0}+\\
& &   
A_{7/2,21}\, \eta_{-2} \xi_{0} \phi_{-1}\state{0}+ 
A_{7/2,22}\,\eta_{-1} \xi_{0} b_{-2}c_{1} \phi_{-1}\state{0}+\\
& & 
A_{7/2,23}\,\eta_{-1} \xi_{0} \phi_{-1} \phi_{-1}\state{0} 
\eqs
\subsection{The fields at level~$4$}\label{s:f4}
At level~$4$ there are 38~fields, see the partition function in
appendix~\ref{s:partition}.
These fields are all in the GSO$(-)$ sector and therefore are to be 
tensored with $\sigma_1$, see equation\refpj{hats}.
The fields are:
\bqs
\lefteqn{\state{\Phi_{4}}=  
A_{4,1}\, \xi_{-2} \xi_{-1} \xi_{0}c_{-2} c_{-1} c_{1} \state{-3}+ 
A_{4,2}\, G_{-3/2} \xi_{-1} \xi_{0}c_{-2} c_{1} \state{-2}+}\\
& & 
A_{4,3}\, G_{-3/2} \xi_{-2} \xi_{0}c_{-1} c_{1} \state{-2}+
A_{4,4}\, G_{-5/2} \xi_{-1} \xi_{0}c_{-1} c_{1} \state{-2} +\\
& & 
A_{4,5}\, G_{-3/2} \xi_{-1} \xi_{0}c_{-1} c_{1} \phi_{-1} \state{-2} +
A_{4,6}\, \xi_{0} c_{-3} \state{-1}+ \\
& & 
A_{4,7}\, L_{-2} \xi_{0} c_{-1}\state{-1}+ 
A_{4,8}\, \eta_{-1} \xi_{-1}\xi_{0} c_{-1} \state{-1} +\\
& &  
A_{4,9}\, G_{-5/2} G_{-3/2} \xi_{0} c_{1}\state{-1}+ 
A_{4,10}\, L_{-4} \xi_{0}c_{1} \state{-1}+ \\
& &  
A_{4,11}\,L_{-2} L_{-2} \xi_{0} c_{1} \state{-1} +
A_{4,12}\, \xi_{-3} \eta_{-1} \xi_{0}c_{1} \state{-1}+\\
& &   
A_{4,13}\,\eta_{-2} \xi_{-2} \xi_{0} c_{1}\state{-1}+
A_{4,14}\, \eta_{-3} \xi_{-1}\xi_{0} c_{1} \state{-1} +\\
& & 
A_{4,15}\, L_{-2} \eta_{-1} \xi_{-1}\xi_{0} c_{1} \state{-1} +
A_{4,16}\, \xi_{0} b_{-2} c_{-2} c_{1}\state{-1}+\\
& & 
A_{4,17}\, \xi_{0} b_{-3} c_{-1} c_{1} \state{-1}+  
A_{4,18}\, \xi_{0} c_{1} \phi_{-4}\state{-1}+\\
& & 
A_{4,19}\, \xi_{0} c_{-1}\phi_{-2} \state{-1}+  
A_{4,20}\,L_{-2} \xi_{0} c_{1} \phi_{-2} \state{-1}+\\
& &  
A_{4,21}\, \eta_{-1} \xi_{-1} \xi_{0}c_{1} \phi_{-2} \state{-1} +
A_{4,22}\, \xi_{0} c_{1} \phi_{-2}\phi_{-2} \state{-1}+\\
& &   
A_{4,23}\,\xi_{0} c_{-2} \phi_{-1} \state{-1} + 
A_{4,24}\, L_{-3} \xi_{0} c_{1} \phi_{-1}\state{-1}+\\  
& & 
A_{4,25}\, \xi_{-2} \eta_{-1}\xi_{0} c_{1} \phi_{-1} \state{-1}+ 
A_{4,26}\, \eta_{-2} \xi_{-1} \xi_{0}c_{1} \phi_{-1} \state{-1}+\\
& &  
A_{4,27}\, \xi_{0} b_{-2} c_{-1} c_{1}\phi_{-1} \state{-1}+ 
A_{4,28}\,\xi_{0} c_{1} \phi_{-3} \phi_{-1}\state{-1}+\\
& &   
A_{4,29}\, \xi_{0} c_{-1}\phi_{-1} \phi_{-1} \state{-1} +
A_{4,30}\, L_{-2} \xi_{0} c_{1} \phi_{-1}\phi_{-1} \state{-1}+\\  
& & 
A_{4,31}\,\eta_{-1} \xi_{-1} \xi_{0} c_{1}\phi_{-1} \phi_{-1} \state{-1}+ 
A_{4,32}\, \xi_{0} c_{1} \phi_{-2}\phi_{-1} \phi_{-1} \state{-1} +\\
& & 
A_{4,33}\, \xi_{0} c_{1} \phi_{-1}\phi_{-1} \phi_{-1} \phi_{-1} \state{-1}+ 
A_{4,34}\, G_{-3/2} \eta_{-2} \xi_{0}\state{0}+\\
& &   
A_{4,35}\, G_{-5/2} \eta_{-1}\xi_{0} \state{0}+  
A_{4,36}\,G_{-3/2} \eta_{-1} \xi_{0}b_{-2} c_{1}\state{0}+\\  
& & 
A_{4,37}\, G_{-3/2} \eta_{-1}\xi_{0} \phi_{-1} \state{0} +
A_{4,38}\, \eta_{-2} \eta_{-1} \xi_{0}b_{-2} \state{1} 
\eqs
\subsection{The number of terms in the potential}\label{s:terms}
The number of interactions between fields satisfying 
\be\label{termnot0}
\left\{
\begin{array}{l}
\mbox{total  }\p\phi -\mbox{charge} = -4,-3,\mbox{ or } -2\\
\mbox{total world sheet number is even.}
\end{array}
\right.
\ee
Hence, these are the terms in the potential which are \emph{not manifestly zero}
for reasons of not satisfying some zero-mode constraint. 
Of course, it is possible
that some of these terms actually do vanish. For example, the coefficient of
$c_{3/2,2} c_{0,1}^2$ in $V_{(3/2,3)}$ is zero, though I see no easy a
priory reason for this.\\[1ex]
For the quadratic terms $\Phi_1\Phi_2$, because of the Feynman-Siegel 
gauge, (\ref{termnot0}) is further restricted to 
$$
\left\{
\begin{array}{l}
\mbox{total  }\p\phi -\mbox{charge } = -2\\
\mbox{total world sheet number is even}\\
\mbox{the weight of }\Phi_1 = \mbox{weight of }\Phi_2
\end{array}
\right.
$$
However, a lot of these quadratic
terms actually do vanish because the bpz-inner product is rather diagonal.
\begin{table}[h]
\begin{center}
\begin{tabular}{*{11}{|c}|}
\hline
degree
& 2 & 3& 4&5&6&7&8&9&10&11\\
\hline
\hline
$(0,0)$ \STRUT &1&0&1&0&0&0&0&0&0&0\\
\hline
$(\displaystyle{{3 \over 2}},3)$ \STRUT&3&3& 5& 1& 1&0&0&0&0&0\\[1ex]
\hline
$(2,4)$ \STRUT & 18& 18 & 25 & 6 & 1&0&0&0&0&0 \\ 
\hline
$(2,5)$ \STRUT &18&24& 45& 10& 6& 1&0&0&0&0\\
\hline
$(2,6)$ \STRUT &18& 69& 87& 25& 10 &1& 1&0&0&0\\
\hline
$(\displaystyle{{7 \over 2}},7)$\STRUT& 110 & 299 
&607 &147 &109 &14& 1& 0& 0&0\\[1ex]
\hline
$(4,8)$ \STRUT& 533 & 2036 & 2203 & 850 & 287 & 32 & 16 & 1&0&0\\
\hline
$(4,9)$ \STRUT& 533 & 2608 & 6164 & 1639 & 1191 & 163 & 33 & 2 & 1& 0\\
\hline
$(4,10)$ \STRUT&533 & 8317 & 11088 & 5010 & 2267 & 341 & 169 & 16 & 2& 0\\
\hline
$(\displaystyle{{11 \over 2}},11)$\STRUT&3778&24839 & 67302 & 19474  & 15227& 2118
& 487& 33& 16& 1\\[1ex]
\hline
\end{tabular} 
\end{center}
\caption{The number of terms in the level truncated 
potentials~$V$. We have listed the terms according to 
their degree.\protect\label{t:termen}}   
\end{table}
\subsection{Expectation value of the tachyon string field}\label{s:VEV}
\begin{center}
\begin{tabular}{|*{5}{c|}}
\hline
level \STRUT & $A_{0,1}$& $A_{3/2,1}$& $A_{3/2,2}$&  $A_{3/2,3}$\\ 
\hline
$(0,0)$ \STRUT& $0.5$& --&   --&  --\\
\hline 
$(3/2,3)$ \STRUT& 0.58882 &  0.112726& $-0.0126028$  & $-0.0931749$ \\
\hline
$(2,4)$ \STRUT&  0.602101& 0.104387 & $-0.0138164$ &   $-0.0430366$\\
\hline
$(2,5)$ \STRUT&  0.607944 &0.121297& $-0.0119856$& $-0.0589907$\\
\hline
$(2,6)$ \STRUT&0.613115&0.121907& $-0.0136997$& $-0.0494127$\\
\hline
$(7/2,7)$ \STRUT& 0.616571&0.121312& $-0.014544$&$ -0.0526311$\\
\hline
$(4,8)$ \STRUT & 0.615054& 0.125598&$ -0.012129$&$-0.0529716$\\
\hline
\end{tabular}
\end{center}

\begin{center}
\begin{tabular}{|*{6}{c|}}
\hline
level \STRUT & $A_{2,1}$&  
$A_{2,2}$& $A_{2,3}$&  $A_{2,4}$&  $A_{2,5}$\\ 
\hline
$(2,4)$ \STRUT&    0.0946226&  0.0322127& $-0.021291$&$-0.0348532$& $-0.01019$\\ 
\hline
$(2,5)$ \STRUT& 0.103101&0.0349833& $-0.0199327$& $-0.0308613$&$ -0.00934597$\\
\hline
$(2,6)$ \STRUT& 0.111049& 0.0379651& $-0.0161857$& $-0.0367398$&$ -0.0109718$\\
\hline
$(7/2,7)$ \STRUT& 0.116991& 0.0364333&$ -0.0179973$&$-0.0414937$&$-0.0111306$\\
\hline
$(4,8)$ \STRUT&0.120421&0.0367186&$-0.0162025$&$-0.0381402$&$-0.00548255$\\
\hline
\end{tabular}
\end{center}
Because of length, we do not list the expectation values of the 23~fields at
level~$7/2$ and the 38~fields at level~4.
\section{Programs for the toy model}\label{TM:programs}
The following function calculates expectation values:
$$
\mbox{Interaction}[i,j,k]=\
_{123}\etats{V}\ a_1^{\dagger\ i }\ a_2^{\dagger\ j }\  a_3^{\dagger\ k}\vac_{123},
$$     
the expectation values are calculated recursively by doing Wick-contractions.
\begin{verbatim}
Clear[Interaction]
Interaction[i_,j_,k_] :=  Interaction[i,j,k] =
Which[i < 0 || j < 0 || k < 0, 0,
  i >= 1, 
  (i-1) N11 Interaction[i-2,j,k] + j N12 Interaction[i-1,j-1,k] 
         + k N12 Interaction[i-1,j,k-1],
  i == 0 && j >= 1, 
  (j-1) N11 Interaction[i,j-2,k] + k N12 Interaction[i,j-1,k-1],
  i == 0 && j == 0 && k > 1, 
  (k-1) N11 Interaction[i,j,k-2] ,
  i == 0 && j == 0 && k == 1,0, 
  i == 0 && j == 0 && k == 0,1]  
\end{verbatim}\ \\[4mm]
The following function calculates the action for a finite number of components:
$$
\mbox{Action}[\{ \psi_0, \psi_2,\cdots,\psi_{2 k}\}] = S(\psi)
\mbox{ with } \state{\psi} = \sum _{i=0}^{k} \psi_{2i}a^{\dagger 2 i } \vac
$$
\begin{verbatim}
Action[field_] := Module[{i,j,k,aantal},
    (* Be careful, in Mathematica, 
    the first entry in a list has index 1 *)
 
aantal = Length[field] - 1;
1/2 Sum[(2 i-1) (2 i)! field[[i+1]]^2, {i,0,aantal}]
+ 1/3 Sum[ field[[i+1]]^3 Interaction[2 i,2 i,2 i], {i,0,aantal}]
+ 3/3 Sum[ field[[i+1]] field[[j+1]]^2 Interaction[2 i,2 j,2 j], 
             {i,0,aantal}, {j,0,i-1}]
+ 3/3 Sum[ field[[i+1]] field[[j+1]]^2 Interaction[2 i,2 j,2 j], 
             {i,0,aantal}, {j,i+1,aantal}] 
+ 6/3 Sum[ field[[i+1]] field[[j+1]] field[[k+1]] 
           Interaction[2 i,2 j,2 k],
             {i,0,aantal}, {j,i+1,aantal}, {k,j+1,aantal}]
	     ]
\end{verbatim}
\section{Some correlators}\label{TM:corr}
In this small appendix we gather some easy tools to calculate correlators. 
First of all we have 
$$
a\ F(\crea) \vac =\parder{\crea} F(\crea) \vac.$$ 
This is trivial to see if you think about the fact that $a$ and $\parder{\crea}$
have the same behavior:
$$ [a,\crea] = 1 \mbox{ and } a\vac = 0 ,$$
versus
$$ [\parder{\crea},\crea] = 1 \mbox{ and } \parder{\crea}\vac = 0.$$
From this follows the well know property of coherent states:
$$
F(a) e^{l \crea}\vac = F(l)e^ {l \crea}\vac
, 
$$
so
\be\label{corr1}
\leftvac F(a) e^{l \crea} \vac = F(l)
\ee
By writing arbitrary states as linear combinations of coherent states (this is
essentially a Fourier transformation) we can easily derive a lot of 
correlators, for example
\be\label{corr2}
 \leftvac e^{k a^2 + \rho a} e^{l a^{\dagger 2} + \sigma \crea} \vac=
\frac{1}{\sqrt{1 - 4 k l }} 
\exp\left( \frac{l \rho^2 + \sigma \rho + k \sigma^2}{1 - 4 k l }\right)
\ee
\begin{proof}The Fourier transformation of the in-state is again Gaussian:
$$
e^{l a^{\dagger 2} + \sigma \crea} = (4 \pi l )^{-1/2} \int_{-\infty}^{+\infty}
\!\!\! dy\ \exp\left(-\frac{1}{4} l^{-1} y^2 + \crea y + \sigma \crea\right),
$$ 
and using\refpj{corr1} one obtains
$$
\leftvac F(a)\ e^{l a^{\dagger 2} + \sigma \crea}\vac=
(4 \pi l )^{-1/2} \int_{-\infty}^{+\infty}
\!\!\! dy\ e^{-\frac{1}{4} l^{-1} y^2} F(y+\sigma)
.$$
 Take $F(a) = e^{k a^2 + \rho a }$ and calculate the Gaussian integral.
 \end{proof}

\selectlanguage{dutch}
\chapter{Samenvatting}
%
Gedurende de afgelopen vier eeuwen hebben fysici veel bewijzen verzameld dat
alle interacties tussen verschillende voorwerpen en materialen gereduceerd
kunnen worden tot combinaties van vier fundamentele krachten. 
Van deze natuurkrachten komen de eerste twee het meest aan bod in het dagelijkse
leven. 

De eerste is de {\bf zwaartekracht}. Dit is de kracht waarmee alle
lichamen elkaar aantrekken. Ze is evenredig met het product van de massa's en
omgekeerd evenredig met het kwadraat van de afstand. Deze wet van Newton (1687)
ver\-klaart perfect zaken zoals de baan van kanonskogels,
de beweging van de planeten en het traject van ruimtesondes. Als de lichamen te
snel bewegen ten opzichte van elkaar, 
is de wet van Newton niet nauwkeurig genoeg en moet men de algemene
relativiteitstheorie van Einstein (1916) gebruiken. In deze theorie wordt de
zwaartekracht in verband gebracht met kromming in ruimte en tijd.

De tweede is de {\bf elektromagnetische kracht}. Zoals de naam suggereert
heeft deze kracht twee aspecten. De elektrische kracht is de kracht waarmee twee
elektrisch geladen deeltjes elkaar afstoten (als ze gelijke ladingen hebben) of
elkaar aantrekken (als de ladingen tegengesteld teken hebben). De magnetische
kracht is de kracht die een snel bewegend elektrisch geladen deeltje ondervindt
in een magneetveld. Een magneetveld ontstaat als elektrisch geladen deeltjes
snel bewegen, bijvoorbeeld als ze door de wikkelingen van een spoel stromen.
Maxwell slaagde er in de volledige vergelijkingen voor de elektrische en
magnetische krachtvelden op te schrijven (1873). Het werd een ``ge\"unificeerde''
theorie, die nu elektromagnetisme genoemd wordt.

\begin{figure}[ht]
\begin{center}
\begin{psfrags}
\psfrag{e}[][]{elektron}
\psfrag{f}[][]{foton}
\epsfbox{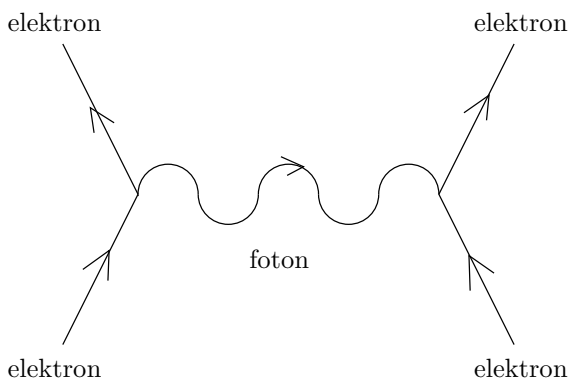}
\end{psfrags} 
\caption{Twee elektronen stoten elkaar af door uitwisseling van een foton.
}
\label{fig:QED}
\end{center}
\end{figure}

Men heeft van deze theorie ook een kwantummechanische versie, waarmee men de
elektromagnetische wis\-sel\-wer\-king tussen twee elementaire deeltjes 
kan berekenen.
Deze versie noemt men kwantumelektrodymanica (QED). In QED is de drager van
de kracht het foton, een deeltje met massa~0 en spin~1. Zie
figuur~\ref{fig:QED} voor een illustratie.

\begin{figure}[ht]
\begin{center}
\begin{psfrags}
\psfrag{1}[][]{$\nu_{\mu}$}
\psfrag{2}[][]{$\bar\nu_e$}
\psfrag{3}[][]{$e^{-}$}
\psfrag{4}[][]{$W^{-}$}
\psfrag{5}[][]{$\mu^{-}$}
\epsfbox{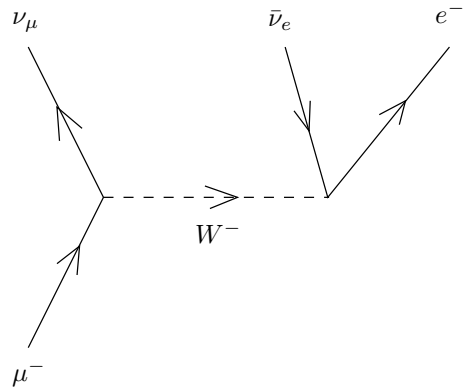}
\end{psfrags} 
\caption{Door de zwakke kracht vervalt een muon ($\mu^{-}$) 
in een elektron ($e^{-}$), een
muon-neutrino ($\nu_{\mu}$) en een anti-elektron-neutrino ($\bar\nu_e$).}
\label{fig:zwak}
\end{center}
\end{figure}

De laatste twee krachten komen in het dagelijkse leven minder aan bod omdat de
reikwijdte van deze krachten beperkt is tot de kern van atomen. Daarom worden
deze twee krachten gezamenlijk de kernkrachten genoemd. De ene kernkracht is de
{\bf zwakke kernkracht}. Deze kracht is er o.a.~verantwoordelijk voor dat vele
deeltjes en ook vele atoomkernen instabiel zijn. De zwakke kracht kan deze
deeltjes doen overgaan in een ander deeltje door een elektron
(of positron) en een neutrino uit te zenden. De dragers van de zwakke kracht 
zijn gevonden
in de jaren tachtig, het $W^{+}$- en het $W^{-}$-deeltje, zie
figuur~\ref{fig:zwak} waar een voorbeeld gegeven wordt. Net als het foton
hebben deze spin~1, maar ze zijn elektrisch geladen en zeer zwaar (vandaar de
korte reikwijdte van deze kracht). Een derde drager, het $Z^0$-deeltje, is
verantwoordelijk voor een iets andere variant van de zwakke kracht.
Ze veroorzaakt bijvoorbeeld elastische 
botsingen tussen neutrino's en andere deeltjes.

De {\bf sterke kernkracht} tenslotte werkt tussen quarks. Deze kracht wordt
uitstekend beschreven door de kwantumchromodynamica (QCD). 
In deze theorie zijn de dragers van de sterke kracht de gluonen. 

De kwantummechanische versie van het elektromagnetisme, de zwakke en sterke
kernkracht wordt het Standaardmodel (rond de jaren 1970) genoemd. Met het Standaardmodel kan
men de elektromagnetische, zwakke en sterke wissel\-wer\-king tussen elementaire
deeltjes perfect berekenen. Dit model is wiskundig elegant: alle krachten
hebben dezelfde wiskundige structuur, de krachtoverbrengende deel\-tjes (foton,
$W^{+}$, $W^{-}$, $Z^0$ en gluonen) zijn alle ijkbosonen in een ijktheorie.
Het Standaardmodel is
uitvoerig getest in deeltjesversnellers zoals die in CERN. Alle experimenten kloppen
uitstekend met het Model\footnote{\emph{Bijna} alle experimenten. Het Standaard
Model begint barstjes te vertonen. De neutrino's blijken niet massaloos te zijn,
maar een zeer kleine massa te hebben. Recente metingen van het magnetisch moment
van het muon wijken 6 cijfers na de komma af van de theoretische
voorspellingen.}. 

Het werk dat vier eeuwen geleden begon is echter nog niet af. Van de eerste
natuurkracht, de zwaartekracht, heeft men nog geen kwantummechanische versie.
De zwaartekrachtswerking tussen twee elementaire deeltjes kan men dus 
momenteel nog niet berekenen tot in zijn volle quantummechanische 
nauwkeurigheid. De
constructie van \emph{kwantumgraviteit} is momenteel het belangrijkste
onder\-zoeks\-onder\-werp binnen de theoretische hoge energie fysica. 
Fysici weten enkel zeker dat het krachtoverbrengend deeltje van de zwaartekracht
een deeltje met spin~2 zal zijn. Men heeft dit nog niet gedetecteerde deeltje 
``graviton'' genoemd.

De meeste fysici
denken dat snaartheorie de grootste kans heeft om een correcte theorie van
kwantumgraviteit te zijn.
\section{Snaartheorie}
Snaartheorie is echter een theorie die nog in volle opbouw is. In het bijzonder
is snaartheorie slechts \emph{perturbatief} gedefinieerd. Dit betekent dat
snaartheorie een formalisme is om verstrooiingsprocessen te berekenen
als storingsreeks in een kleine interactiesterkte. Als we een vergelijking maken
met een theorie van puntdeeltjes, dan komt perturbatieve snaartheorie overeen
met de kennis van de Feynmanregels van een kwantumveldentheorie. 
We gaan nu in op enkele elementen van perturbatieve snaartheorie. 

Het uitgangspunt van snaartheorie is dat alle elementaire
deeltjes verschillende vibratiemogelijkheden van een zeer kleine elastische
snaar zijn. Sommige trillingswijzen geven deeltjes met spin~1. Deze massaloze
deeltjes gedragen zich precies als de ijkbosonen die we tegenkwamen in de vorige
sectie. Het mooie is dat een andere trillingswijze van de snaar spin~2 heeft.
Dit massaloze deeltje gedraagt zich als een graviton, het krachtoverbrengend
deeltje van de zwaartekracht! Dit betekent dus dat snaartheorie 
\emph{automatisch} de
zwaartekracht voortbrengt! Dus de snaartheorie levert niet alleen deeltjes op
van het type dat we ook werkelijk waarnemen, maar de zwaartekracht zit kennelijk
al ingebakken in deze theorie.


Er zijn een aantal verschillende perturbatieve snaartheorie\"en.
Allereerst is er \'e\'en bosonische snaartheorie, waarin alle deeltjes
bosonen zijn. Interne consistentie van bosonische snaartheorie vereist dat \`als de
snaren in een vlakke ruimte-tijdachtergrond bewegen, deze achtergrond
26-dimensionaal moet zijn. Een fundamenteler probleem is dat een van de
vibratiemogelijkheden van de bosonische snaar in een vlakke achtergrond 
een deeltje genereert met negatieve
massa kwadraat. Dit is een tachyon. De aanwezigheid van zo'n deeltje in een
theorie wordt meestal bekeken als teken dat de theorie instabiel is. Omwille
van deze instabiliteit en van het ontbreken van fermionen, is deze theorie
geen correcte theorie van kwantumgraviteit.

Ten tweede zijn er vijf snaartheorie\"en met ruimte-tijdsupersymmetrie, kortweg
supersnaartheorie\"en. In deze
theorie zijn er zowel bosonen als fermionen. De ruimte-tijdsupersymmetrie is
een symmetrie tussen deze bosonen en fermionen.
Deze theorie\"en hebben geen
tachyonen en worden beschouwd als mogelijke kandidaten om de correcte
beschrijving van kwantumzwaartekracht te zijn. 
Interne consistentie eist dat 
een vlakke achtergrond 10-dimensionaal is. 

Ten derde zijn er nog enkele niet-supersymmetrische heterotische
snaartheorie\"en. De meeste van deze theorie\"en bevatten tachyonen en worden
niet zo vaak bestudeerd. Ze komen in dit proefschrift verder niet aan bod.
\section{Problemen in snaartheorie}
Zoals boven vermeld voorspelt supersnaartheorie dat de ruimte-tijd
10-di\-men\-si\-o\-naal is. 
Daarom moeten 6 van de 10 dimensies 
opgerold zijn tot zo'n kleine
afmetingen dat ze nog nooit waargenomen zijn. Dit is mijns inziens het
be\-lang\-rijkste open probleem in snaartheorie. Is de symmetrie tussen die 10
dimensies gebroken? Zijn 6 van de 10 dimensies opgerold? Hoe zijn ze opgerold?
Fysischer geformuleerd is dit probleem
\begin{center}\label{Vraag}
$\mathcal{W}${\it at is het correcte vacu\"um van snaartheorie?}
\end{center}
Hier stuiten we op de inherente beperkingen van perturbatieve snaartheorie. De
vergelijkingen die men moet oplossen om het correcte vacu\"um te vinden zijn
momenteel slechts perturbatief gekend. Ze geven niet genoeg informatie om de
speurtocht te kunnen aanvangen. Omdat men het correcte vacu\"um van snaartheorie
niet kent, kan men ook geen contact leggen met de werkelijkheid. De kennis van
het correcte vacu\"um is een conditio sine qua non om concrete 
fysische voor\-spel\-lin\-gen te
doen. Momenteel kan snaartheorie slechts generieke voorspellingen
doen die gelden voor elke achtergrond.

Ondanks het feit dat snaartheorie slechts een perturbatieve theorie is, is men
er door de grote inwendige symmetrie van snaartheorie toch in geslaagd om ook
\emph{niet-perturbatieve} resultaten te bekomen. Bijvoorbeeld kan men
argumenteren dat bepaalde snaartheorie\"en duaal zijn
t.o.v.~elkaar. Dit betekent dat men bepaalde berekeningen voor grote
koppelingsconstante in \'e\'en theorie kan vertalen naar een berekening in een
kleine koppelingsconstante in een andere snaartheorie. Een belangrijke rol bij
de ontdekking van deze dualiteiten werd gespeeld door $D$-branen (1995). 

\begin{figure}[ht]
\begin{center}
\begin{psfrags}
\epsfbox{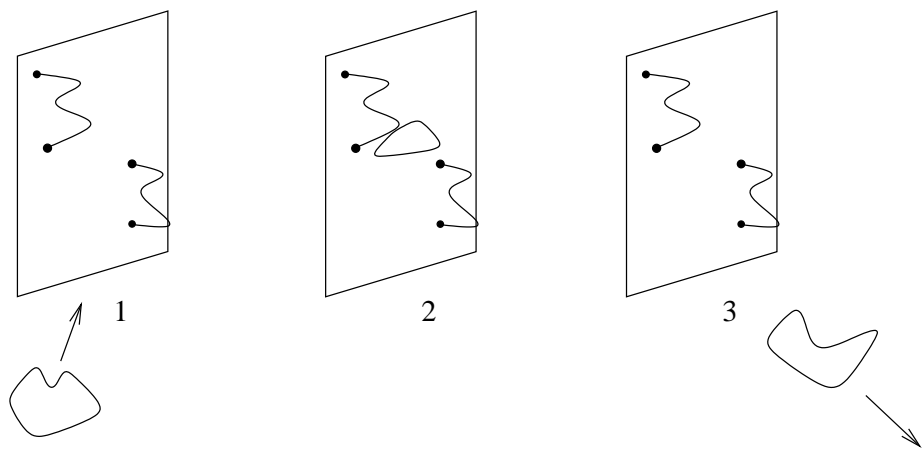}
\end{psfrags} 
\caption{(1) Een gesloten snaar nadert een $D$-braan.
(2) De gesloten snaar interageert met het $D$-braan via de open snaren die eraan
vasthangen.
(3) Na de botsing verwijdert de gesloten snaar zich van het braan.
}
\label{fig:Dbraan}
\end{center}
\end{figure}

$D$-branen maakten hun intrede in snaartheorie als 
hypervlakken waarop open
snaren kunnen eindigen. De massa van $D$-branen is omgekeerd evenredig met de
koppelingsconstante~$g$. Daarom zijn het niet-perturbatieve toestanden. Toch
hebben ze een be\-schrij\-ving in perturbatieve snaartheorie: de dynamica van
$D$-branen wordt beschreven door de open snaren die erop eindigen, zie
figuur~\ref{fig:Dbraan}.

Men heeft tal van aanwijzingen dat hogervermelde dualiteiten correct zijn, echte
bewijzen heeft met echter nog niet. Daarvoor heeft men 
een niet-pertur\-ba\-tieve beschrijving van snaartheorie nodig. Kort
na de ontdekking van snaartheorie heeft men zo'n niet-perturbatieve definitie 
van
snaartheorie proberen construeren in de vorm van snaarveldentheorie\"en. Van
deze snaarveldentheorie\"en is wellicht Wittens open bosonische
veldentheorie de best 
bekende. Zwiebach heeft ook een gesloten snaarveldentheorie geconstrueerd, maar
deze lijkt technisch omslachtig en niet geschikt om concrete berekeningen mee te
doen. Bij de constructie van een open \emph{supersnaar} veldentheorie doken tal
van problemen op. 
Omwille van deze moeilijkheden en omwille van
het feit dat tal van fysici meenden dat snaarveldentheorie een te conservatieve
poging was om snaartheorie niet-perturbatief te defini\"eren, heeft men
snaarveldentheorie gedurende een tiental jaar laten vallen.
Na de ontdekking van hoger vermelde dualiteiten leek 
het erop dat een nog
onbekende 11-dimensionale theorie -- $\mathcal{M}$-theorie -- het juiste 
kader zou zijn
om snaartheorie niet-perturbatief te behandelen. Van deze $\mathcal{M}$-theorie
heeft men zelfs een geconjectureerde exacte beschrijving in de vorm van een
matrix-theorie. De jongste twee jaar echter 
zijn de snaarveldentheorie\"en weer volop
in de aandacht gekomen omdat ze bij uitstek geschikt zijn om de conjecturen van
Sen te controleren.
\section{De conjecturen van Sen}
Zoals tal van recente ontdekkingen in snaartheorie ligt de oorsprong van deze
conjecturen in $D$-branen.
De ontdekking van $D$-branen werpt immers 
een ander licht op het tachyon van de open
bosonische snaartheorie. In 1999 suggereerde Sen dat men deze theorie nu moet 
bekijken als de theorie die een $D25$-braan beschrijft. Inderdaad, na de
ontdekking van $D$-branen realizeerde men zich dat de open snaren van open
bosonische snaartheorie moeten eindigen op een $D$-braan. Omdat de open snaren
van open bosonische snaartheorie door heel de ruimte kunnen bewegen, vult dit
$D$-braan de hele ruimte op -- een $D25$-braan. De ruimte-tijd van bosonische
snaartheorie is niet leeg, maar helemaal opgevuld door het $D25$-braan!
Het bestaan van het open snaar tachyon is dan niets anders dan een perturbatief
gevolg van 
de instabiliteit van dit $D25$-braan. Sen argumenteerde dat de condensatie van 
het tachyon overeenkomt met het verval van het $D25$-braan, m.a.w.~het 
eindproduct van de tachyoncondensatie op het onstabiele $D25$-braan is niets
ander dan de grondtoestand van gesloten snaartheorie. In het bijzonder deed Sen
de volgende drie conjecturen:
\begin{figure}[ht]
\begin{center}
\begin{psfrags}
\psfrag{A}[][]{potentiaal}
\psfrag{B}[][]{$A$}
\psfrag{C}[][]{$\Psi_0$}
\psfrag{D}[][]{$B$}
\psfrag{E}[][]{massa $D25$ braan}
\psfrag{F}[]{$\Psi$}
\epsfbox{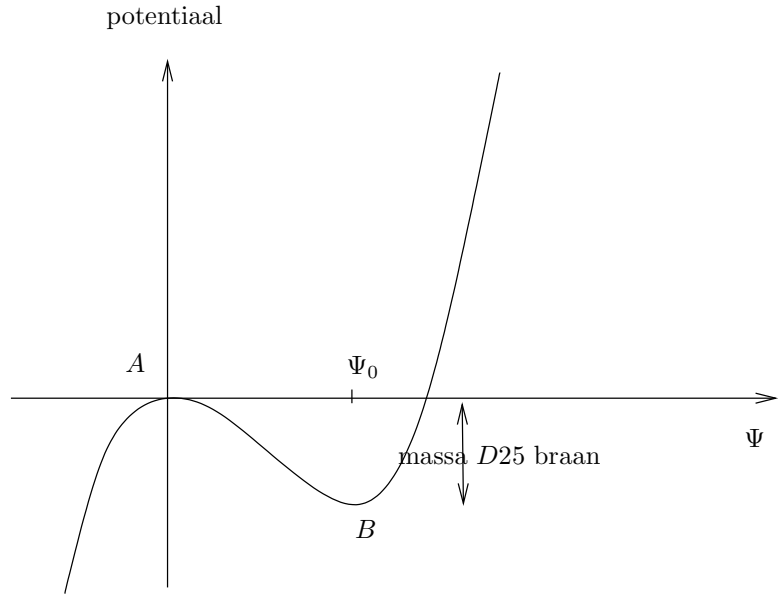}
\end{psfrags} 
\caption{De tachyonpotentiaal in de open bosonische snaar theorie. In punt 
$A$ is het snaarveld $\Psi$ gelijk aan 0. Daar is het onstabiele $D25$-braan nog
aanwezig. Punt $B$ is het gesloten snaartheorie vacu\"um. Volgens de conjectuur
van Sen vervalt het $D25$-braan naar dit vacu\"um. Daarom is het verschil in
potenti\"ele energie tussen deze twee extrema gelijk aan de massa van het
$D25$-braan.}
\label{fig:Senbos}
\end{center}
\end{figure}
\begin{itemize}
\item[(1)] 
Het verschil tussen de waarde van de potentiaal op zijn maximum (dat overeenkomt
met het $D25$-braan) en de waarde op het minimum (dat de stabiele eindtoestand
voorstelt) moet precies gelijk zijn aan de massa van het $D25$-braan, 
zie figuur~\ref{fig:Senbos}.
\item[(2)]
Lager-dimensionale $D$-branen kunnen gerealiseerd worden als 
soliton\-con\-fi\-gu\-ra\-ties van het tachyon en andere snaarvelden.
\end{itemize}
Omdat er na condensatie geen braan meer is waarop open snaren kunnen eindigen, 
leidt dit tot:
\begin{itemize}
\item[(3)]
Het stabiele vacu\"um kan ge\"identificeerd worden met het 
vacu\"um van de ge\-slo\-ten snaartheorie. In het bijzonder zijn er in het 
minimum van de potentiaal geen fysische open snaar toestanden.
\end{itemize}
In supersnaartheorie zijn er ook onstabiele branen, zoals het $D9$-braan in type
IIA snaartheorie. Er zijn analoge conjecturen in dit geval, zie
figuur~\ref{fig:Sensup}. 
\begin{figure}[ht]
\begin{center}
\begin{psfrags}
\psfrag{A}[][]{potentiaal}
\psfrag{B}[][]{$D9$}
\psfrag{C}[][]{$\Psi_0$}
\psfrag{D}[][]{geen $D9$}
\psfrag{E}[][]{massa $D9$ braan}
\psfrag{F}[]{$\Psi$}
\epsfbox{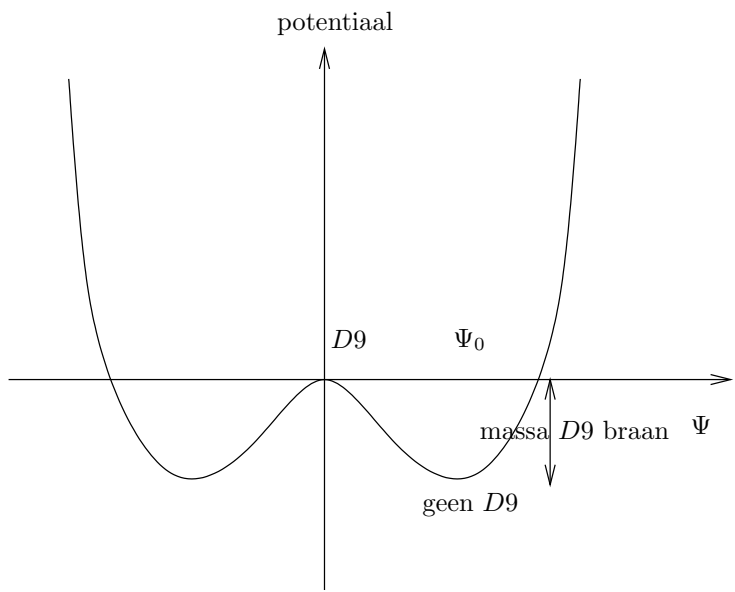}
\end{psfrags} 
\caption{De tachyonpotentiaal op de onstabiele $D9$-braan.}
\label{fig:Sensup}
\end{center}
\end{figure}

De open snaarveldentheorie\"en zijn bij uitstek geschikt om de conjecturen van
Sen te verifi\"eren. Immers, om de eerste conjectuur te controleren moet men
slechts de potentiaal berekenen. Dit is een berekening die uitstekend past in
een veldentheoriecontext. 

\begin{figure}[ht]
\begin{center}
\epsfysize=10.5cm
\begin{psfrags}
\psfrag{a}[][]{(a)}
\psfrag{b}[][]{(b)}
\epsfbox{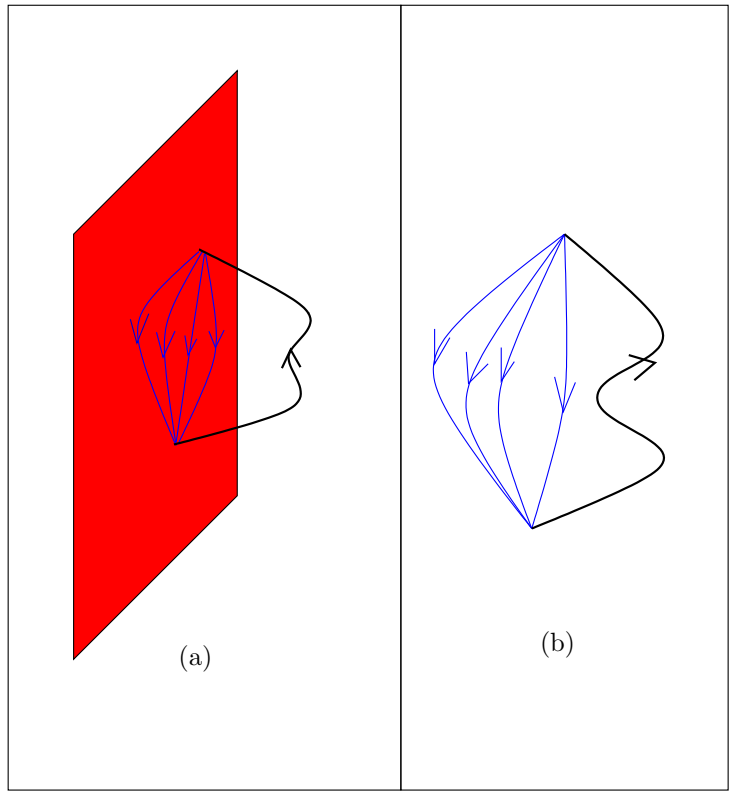}
\end{psfrags} 
\caption{(a) Een open snaar met eindpunten op een niet-BPS $D$-braan voor
tachyoncondensatie. De eindpunten van de open snaar zijn geladen onder een
ijkveld. (b) Deze configuratie na tachyoncondensatie: de open snaar
tesamen met de fluxtube vormen een gesloten snaar.}
\label{fig:flux}
\end{center}
\end{figure}

Het belang van dit onderzoek is het volgende. 
Ten eerste tonen dit soort berekeningen aan dat
snaarveldentheorie\"en inderdaad een niet-perturbatieve be\-schrij\-ving 
geven van open
snaartheorie. Verder is het zeer fascinerend om te speculeren dat open
snaarveldentheorie zelfs een \emph{volledige} niet-perturbatieve definitie geven
van snaartheorie, zowel van de open snaartheorie als van de \emph{gesloten}
snaartheorie. Tot deze speculatie komt men op de volgende manier. Volgens de
derde conjectuur van Sen kan het stabiele vacu\"um ge\"identificeerd worden met
het vacu\"um van gesloten snaartheorie. Daarom kan men hopen dat als $S(\psi)$
de actie is van open snaarveldentheorie en $\psi_0$ het stabiele vacu\"um 
(zie de figuren~\ref{fig:Senbos} en~\ref{fig:Sensup}), $S(\psi+\psi_0)$ dan
de actie is van een gesloten snaarveldentheorie. Men kan ook intu\"itief zien
hoe gesloten snaren kunnen verschijnen na condensatie van een $D$-braan, zie
figuur~\ref{fig:flux}.
\clearpage
Als dit het geval is, kan men in het bosonische geval het lot van het gesloten
snaar tachyon bestuderen. 
En als men zo in het supergeval een gesloten
supersnaarveldentheorie krijgt, zou men eindelijk het correcte vacu\"um van
snaartheorie kunnen bepalen, zie bladzijde~\pageref{Vraag},
en fenomenologische voorspellingen doen.
\section{Overzicht van het proefschrift}
\subsection*{Hoofdstuk 2: Open bosonische snaarveldentheorie.} In 
hoofdstuk~2
bestuderen we het open snaar tachyon in Wittens open bo\-so\-ni\-sche
snaarveldentheorie. Dit is de context waarin Sens conjecturen het meest
bestudeerd zijn. Alhoewel ons eigen werk de berekening van de 
tachyonpotentiaal in
\emph{supersymmetrische} theorie\"en is, behandelen we toch het bosonische geval
om verschillende aspecten van de berekening te kunnen uitleggen in dit
technisch eenvoudigere geval. In sectie~\ref{s:gluing} argumenteren we
dat interacties tussen snaarvelden beschreven worden door het aan elkaar plakken
van \emph{wereldlakens}. Dit kan men vergelijken met de interacties tussen
puntdeeltjes. Daar worden \emph{wereldlijnen} aan elkaar geplakt
d.m.v.~interactievertices in Feynman diagrammen. In sectie~\ref{s:star}
bespreken we het belangrijkste ingredient van Wittens open snaarveldentheorie:
het sterproduct. Na deze twee inleidende secties kunnen we Wittens actie
behandelen. Deze actie heeft de vorm van een Chern-Simons actie. Daarom geven we
een kort overzicht van deze actie in sectie~\ref{s:CS}. In sectie~\ref{s:SFT},
het centrale deel van dit hoofdstuk, behandelen we Wittens actie. We fixeren de
ijkinvariantie in
sectie~\ref{s:Siegel}. Daarna
gebruiken we Wittens actie om de instabiliteit van het open snaar tachyon te
bespreken in sectie~\ref{s:tachpot}. We sluiten het hoofdstuk af met drie
secties. In sectie~\ref{s:Neumann} geven we een concrete realisatie van Wittens
interactievertex. In sectie~\ref{s:proofs} geven we enkele bewijzen die we in
vorige secties hebben overgeslagen. In de laatste sectie tenslotte vermelden we
enkele interessante onderzoeksrichtingen.
\subsection*{Hoofdstuk 3: Berkovits' supersnaarveldentheorie.} 
In dit hoofdstuk bestuderen we het open supersnaar tachyon in Berkovits'
supersnaarveldentheorie. Dit hoofdstuk valt uiteen in twee delen.
\begin{description}
\item[1. De definitie van Berkovits' actie.] Berkovits' actie lijkt formeel op een
Wess-Zumino-Witten (WZW) actie. Daarom geven we in sectie~\ref{s:WZW} een kort
overzicht van deze actie. In sectie~\ref{s:BPS} construeren we Berkovits' actie
voor een BPS $D9$-braan gebaseerd op de WZW-actie. In sectie~\ref{s:nonBPS}
behandelen we Berkovits' actie voor een niet-BPS $D9$-braan.
\end{description} 
In het tweede deel bestuderen we de tachyonpotentiaal. Dit tweede deel bestaat
voornamelijk uit eigen werk. 
\begin{description}
\item[2. Studie van de tachyonpotentiaal.] 
In sectie~\ref{s:level4} berekenen we de tachyonpotentiaal tot en met
niveau\footnote{Deze notatie wordt uitgelegd in sectie~\ref{s:level4}}~$(2,4)$ en
verifi\"eren we Sens eerste conjectuur met een nauwkeurigheid van $89\%$. We
willen meer niveaus gebruiken om een nauwkeurigere overeenkomst te hebben.
Echter, de berekeningsmethode die we in sectie~\ref{s:level4} gebruikten is te
omslachtig voor deze hogere niveaus. Daarom werken we
in sectie~\ref{s:cons} een effici\"entere rekenmethode uit. Deze methode is
gebaseerd op het gebruik van behoudswetten. Ze was in de 
li\-te\-ra\-tuur gegeven voor
de bosonische snaar. Wij breiden deze methode uit tot de supersnaar.
In sectie~\ref{s:high} gebruiken we deze methode om de 
tachyonpotentiaal te berekenen tot en met niveau~$(4,8)$ en 
verifi\"eren we Sens eerste conjectuur tot op $94\%$.
\end{description}
In de laatste sectie tenslotte bespreken we andere berekeningen die gedaan zijn
in Berkovits' snaarveldentheorie en geven we enkele richtingen aan voor verder
onderzoek.  
\subsection*{Hoofdstuk 4: Een speelgoedmodel.}
Het is waarschijnlijk dat we een exacte vorm moeten hebben van het stabiele
vacu\"um $\psi_0$ om te weten te komen hoe gesloten snaren verschijnen in dat
vacu\"um. De constructie van een exacte vorm is echter zeer moeilijk omwille van
twee redenen. De eerste reden is dat er oneindig veel ruimte-tijdvelden
verwachtingswaarden krijgen. De tweede reden is de gecompliceerde vorm van de
interactie in open snaarveldentheorie. Daarom construeren we in hoofdstuk~4
een eenvoudiger model van Wittens open snaarveldentheorie en zoeken daarin een
exacte oplossing. In dit eenvoudiger model zijn er geen spoken en in plaats van
de oneindige verzameling oscillatoren
$\alpha^\mu_n$ met $\mu: 1,\ldots,26$  en  $n: 1,\ldots,+\infty$ die voldoen
aan de commutatierelaties
$[\alpha^\mu_m,\alpha^\nu_n] = m\ \eta^{\mu\nu}\delta_{m+n}$,
nemen we slechts \'e\'en oscillator $a$ met $[a,\crea]=1$.
\subsection{Besluit}
Door de berekeningen die gepresenteerd zijn 
in dit proefschrift heeft men meer
vertrouwen gekregen in snaarveldentheorie\"en als niet-perturbatieve
beschrijvingen van open snaartheorie\"en. 
In het bosonische geval is men er nu zeker van dat het
$D25$-braan inderdaad vervalt naar het vacu\"um van gesloten snaartheorie. 
In het
supergeval heeft dit soort berekeningen er voor gezorgd dat men Berkovits'
supersnaarveldentheorie nu beziet als de correcte supersnaarveldentheorie. 

Als open snaarveldentheorie\"en ons iets kunnen leren over
niet-perturbatieve gesloten snaren, dan zou men het correcte vacu\"um van
snaartheorie kunnen bepalen. Omdat dit zo'n belangrijk probleem is, is
snaarveldentheorie een veelbelovend onderzoeksdomein.

\selectlanguage{english}

\printindex
\cleardoublepage
\thispagestyle{empty}
\selectlanguage{dutch}
\thispagestyle{empty}
\setlength{\parindent}{.7cm}
\parskip 5pt plus 1pt minus 1pt
\noindent {\centerline{\large\bf Dankwoord}}
\ \\[1cm]
Mijn carri\`ere als ``snarentheoreet'' is begonnen met het lezen van
Polchinski's boeken; ik worstelde me met Jeanne De Jaegher en Joris
Raeymaekers door de hoofdstukken en oefeningen heen. Hen wil ik van 
harte bedanken voor
de aangename discussies en verhelderende berekeningen. Het onderzoek in deze
thesis heb ik verricht in samenwerking
met Joris. Ik wil hem erg danken voor zijn orde en
grondigheid waardoor het onderzoek vlot verliep. De berekeningen die we 
gedaan
hebben onder tijdsdruk van concurrenten op andere 
continenten, crashende computers,
discussies over subtiliteiten in onduidelijk Mathematica-jargon, 
e-mails over fermionen
en tekens zijn onlosmakend verbonden met de boeiendste weken die 
ik beleefd heb op het
instituut.

Verder bedank ik mijn promotor Walter Troost.
Hij maakte altijd tijd vrij om te luisteren naar mijn vragen en om 
verhelderende antwoorden te geven. Tenslotte bedank ik het leescomit\'e voor de
tijd die ze gestoken hebben in het lezen van dit proefschrift: 
D\'esir\'e Boll\'e, 
Mark Fannes, 
Raymond Gastmans,
Joseph Minahan,
Walter Troost,
Toine Van Proeyen en
Andr\'e Verbeure.

\end{document}